\documentstyle[aaspp4,natbib]{article}
\def\eq#1{equation~(\ref{eqn:#1})}
\def\Eq#1{eq.~(\ref{eqn:#1})}
\def\gsim{{{}_>\atop{}^{{}^\sim}}}
\def\lsim{{{}_<\atop{}^{{}^\sim}}}

\def\kms{{\rm km}\,{\rm s}^{-1}}
\def\kpc{{\rm kpc}}
\def\dls{{D_{\rm LS}}}
\def\te{{t_{\rm E}}}
\def\tp{t_{\rm p}}
\def\re{{r_{\rm E}}}
\def\umin{u_{0}}

\def\days{{\rm days}}
\def\day{{\rm day}}
\def\au{{\rm AU}}
\def\ob#1#2{OGLE-19#1-BUL-#2}
\def\mb#1#2{MACHO #1-BLG-#2}
\def\dos{D_{\rm S}}
\def\dls{D_{\rm LS}}
\def\dol{D_{\rm L}}
\def\thetae{\theta_{\rm E}}
\def\thetap{\theta_{\rm p}}
\def\thetas{\theta_*}
\def\muas{\mu {\rm as}}
\def\murel{\mu_{\rm rel}}
\def\pirel{\pi_{\rm rel}}
\def\drel{D_{\rm rel}}
\def\dchit{\Delta\chi^2_{\rm thresh}}
\def\dchim{\Delta\chi^2_{\rm min}}

\def\dchi{\Delta\chi^2}
\def\elz{{\epsilon}_{{\rm LZ},i}} 
\def\vperp{v}
\def\vtilde{\tilde v}
\def\fb{F_{\rm B}}
\def\fs{F_{\rm S}}
\def\rhos{\rho_{*}}
\def\vmin{(V-I)}
\def\vmi0{(V-I)_0}
\def\i0{I_0}

\def\u0{u_0}
\def\t0{t_0}
\def\du0{\delta\u0/\u0}
\def\amax{A_{\rm max}}
\def\massp{m_{p}}
\def\msun{M_{\odot}}
\def\mjup{M_{\rm Jup}}
\def\ap{r_{p}}
\def\ave#1{\left<#1\right>}
\def\acfmax{A^{\rm cf}_{\rm max}}
\def\accmax{A^{\rm cc}_{\rm max}}
\def\accmobs{A^{\rm cc}_{\rm max,obs}}
\def\acfmobs{A^{\rm cf}_{\rm max,obs}}
\def\teobs{t_{\rm E, obs}}
\def\qmin{q_{\rm min}}
\bibpunct[,]{(}{)}{;}{a}{}{,}
\begin{document}

\title{Microlensing Constraints on the
Frequency of Jupiter-Mass Companions: Analysis of Five Years of PLANET Photometry}

\author{
B. S. Gaudi\altaffilmark{1,2,3},
M. D. Albrow\altaffilmark{4,5}, 
J. An\altaffilmark{1},
J.-P. Beaulieu\altaffilmark{6}, J. A. R. Caldwell\altaffilmark{7},
D. L. DePoy\altaffilmark{1}, M. Dominik\altaffilmark{8},
A. Gould\altaffilmark{1},J. Greenhill\altaffilmark{9}, K. Hill\altaffilmark{9},
S. Kane\altaffilmark{9,10}, R. Martin\altaffilmark{11},
J. Menzies\altaffilmark{7}, R. M. Naber\altaffilmark{8},
J.-W. Pel\altaffilmark{8},R. W. Pogge\altaffilmark{1},
K. R. Pollard\altaffilmark{4,12}, P. D. Sackett\altaffilmark{8},
K. C. Sahu\altaffilmark{5},P. Vermaak\altaffilmark{7},
P. M. Vreeswijk\altaffilmark{8,13}, R. Watson\altaffilmark{9},
A. Williams\altaffilmark{11}\\
The PLANET Collaboration}

\altaffiltext{1}{Ohio State University, Department of Astronomy, Columbus, 
OH 43210, U.S.A.}
\altaffiltext{2}{Institute for Advanced Study, Einstein Drive, Princeton, NJ 08540, U.S.A.}
\altaffiltext{3}{Hubble Fellow}
\altaffiltext{4}{Univ.\ of Canterbury, Dept.\ of Physics \& Astronomy, 
Private Bag 4800, Christchurch, New Zealand}
\altaffiltext{5}{Space Telescope Science Institute, 3700 San Martin Drive, 
Baltimore, MD. 21218, U.S.A.}
\altaffiltext{6}{Institut d'Astrophysique de Paris, INSU CNRS, 98 bis 
Boulevard Arago, F-75014, Paris, France}
\altaffiltext{7}{South African Astronomical Observatory, P.O. Box 9, 
Observatory 7935, South Africa}
\altaffiltext{8}{Kapteyn Astronomical Institute, Postbus 800, 
9700 AV Groningen, The Netherlands}
\altaffiltext{9}{Univ. of Tasmania, Physics Dept., G.P.O. 252C, 
Hobart, Tasmania~~7001, Australia}
\altaffiltext{10}{School of Physics \& Astronomy, University
of St.\ Andrews, North Haugh, St.\ Andrews, Fife KY16 9SS, UK}
\altaffiltext{11}{Perth Observatory, Walnut Road, Bickley, Perth~~6076, Australia}
\altaffiltext{12}{Physics Department, Gettysbrug College, 300 North Washington Street, Gettysburg, PA 17325, U.S.A.} 
\altaffiltext{13}{Astronomical Institute ``Anton Pannekoek'', University of 
Amsterdam, Kruislaan 403, 1098 SJ Amsterdam, The Netherlands}

\begin{abstract}

We analyze five years of PLANET photometry of microlensing events
toward the Galactic bulge to search for the short-duration deviations
from single lens light curves that are indicative of the presence of
planetary companions to the primary microlenses.  Using strict event
selection criteria, we construct a well defined sample of 43
intensively monitored events.  We search for planetary perturbations
in these events over a densely sampled region of parameter space
spanning two decades in mass ratio and projected separation, but find
no viable planetary candidates.  By combining the detection
efficiencies of the events, we find that, at 95\% confidence, less
than $25\%$ of our primary lenses have companions with mass ratio
$q=10^{-2}$ and separations in the lensing zone, $[0.6-1.6]\thetae$,
where $\thetae$ is the Einstein ring radius.  Using a model of the
mass, velocity and spatial distribution of bulge lenses, we infer that
the majority of our lenses are likely M dwarfs in the Galactic bulge.
We conclude that $<33\%$ of M-dwarfs in the Galactic bulge
have companions with mass $\massp = \mjup$ between 1.5 and 4$~\au$,
and $<45\%$ have companions with $\massp = 3\mjup$ between 1 and
7$~\au$, the first significant limits on planetary companions to
M-dwarfs.  We consider the effects of the finite size of the source
stars and changing our detection criterion, but find that these do not
alter our conclusions substantially.

\end{abstract}

  
\setcounter{footnote}{0}
\section{Introduction}

The discovery in 1995 of a massive planet orbiting 51 Peg
\citep{may95}, followed by the discovery of many more planets
orbiting nearby dwarf stars using the same radial velocity technique
(\citealt{mcm2000} and references therein) has focussed both public
and scientific attention on the search for extrasolar planets and the
experimental and theoretical progress being made in developing other
viable detection techniques.

Due to their small mass and size, extrasolar planets are difficult to
find.  Proposed detection methods can be subdivided into direct and
indirect techniques.  Direct methods rely on the detection of the
reflected light of the parent star, and are exceedingly challenging
due to the extremely small flux expected from the planet, which is
overwhelmed by stray light from the star itself \citep{aandw1997}.
Some direct imaging searches have already been performed
\citep{boden1998b}, but the future of this method lies in the
construction and launching of space-based instrumentation
\citep{wanda1998}.

Astrometric, radial velocity, and occultation measurements can be used
to detect the presence of a planet indirectly.  Astrometric detection
relies on the measurement of the positional wobble of the stellar
centroid caused by the motion of the star around the center of mass of
the planet-star system and yields the mass ratio and orbital
parameters of the planet-star system.  Many attempts to find
extrasolar planets in this way have been made, but the measurements
are difficult and the detections remain controversial; planned
space-based missions astrometric missions such as the Full-Sky
Astrometric Mapping Explorer (FAME), the Space Interferometry Mission
(SIM), and the Global Astrometric Interferometer for Astrophysics
(GAIA) are expected to be substantially more successful.  Occultation
methods use very accurate photometry of the parent star to detect the
small decrease in flux ($\la 1\%$) caused by a planet transiting the
face of the star \citep{bands1984,handd1994}.  Many occultation
searches are currently being conducted \citep{deeg, bandc2000}, with
important new limits being placed on planetary companions in 47 Tuc
\citep{gilliland2000}.  Recently, one of the extrasolar planets
detected via radial velocity surveys was also found to transit its
parent star, yielding a measurement of the mass, radius, and density
of the companion \citep{charbon2000, henry2000}.  Spaced-based
missions are being planned to increase the sensitivity to low-mass
planets (COROT, \citealt{corot}; KEPLER, \citealt{kepler}).  By far
the most successful indirect method for discovering planets has been
the Doppler technique, which employs precise radial velocity
measurements of nearby stars to detect Doppler shifts caused by
orbiting planets.  Several teams have monitored nearby stars with the
aim of detecting the Doppler signal of orbiting planets
\citep{mcmilan1993,may95,butler1996,cochran1997,noyes1997,vogt2000}.
To date these groups combined have discovered over 50 extrasolar planets, with
new planetary companions being announced every few months.  Several
exciting discoveries using the radial velocity technique include the
first detection of extrasolar planetary systems \citep{butler1999, marcy2001a,marcy2001b, fischer2002}
and the detection of planets with masses below that of Saturn
\citep{mbv2000}.

These detection techniques are complementary to one another both 
in terms of their sensitivity to planetary mass and orbital separations 
and the specific physical quantities of the planetary system that they measure.  
All share two distinct advantages: 
the experiments are repeatable and, due to their reliance on 
flux measurements of the parent star or the planet itself, they are 
sensitive to stars in the solar neighborhood where 
follow-up studies can be easily pursued.  For example, 
spectroscopic follow-up studies may enable the detection of molecules 
commonly thought to be indicative of life, such as water, carbon dioxide, 
and ozone \citep{wanda1998}.
This advantage is linked to a common drawback: 
most of the searches can be conducted only on a limited number of nearby stars,
and are thus unable to address questions about the nature of 
planetary systems beyond the immediate solar neighborhood.  
In addition, most of the methods (astrometry, radial velocity and
occultation) can only probe companions with orbital periods smaller
than the duration of the experiment.  Furthermore, most are
fundamentally restricted to massive planets, for example, radial
velocity searches probably have an ultimate limit of $\sim 1~{\rm
m~s^{-1}}$ due to random velocity variations intrinsic to the parent
stars \citep{sbm1998}. Of these methods, 
only space-based interferometric imaging and transit searches 
are expected to be sensitive to Earth-mass planets.

Microlensing is a relatively new method of detecting extrasolar
planets that overcomes many of these difficulties.  
Galactic microlensing occurs when a massive, compact object (the lens) passes near
the observer's line of sight to a more distant star (the source).  If the
observer, lens, and source are perfectly aligned, then the lens images
the source into a ring, called the Einstein ring, which has angular radius
\begin{equation}
\thetae=\sqrt{{{4 G}\over c^2} {M \over \drel}}\simeq 320\muas\left({M \over 
0.3~\msun}\right)^{1/2}, 
\label{eqn:thetae}
\end{equation}
where $M$ is the mass of the lens, $\drel$ is defined by,
\begin{equation}
{1 \over \drel} \equiv {1 \over \dol} - {1 \over \dos},
\label{eqn:pirel}
\end{equation}
and $\dol$ and $\dos$ are the distances to the lens and source,
respectively. The lens-source relative parallax is then
$\pirel=\au/\drel$.  Note that $\thetae$ corresponds to a physical distance
at the lens of
\begin{equation}
\re = \thetae \dol \simeq 2~\au \left({M \over 
0.3~\msun}\right)^{1/2}.
\label{eqn:re}
\end{equation}
If the lens is not perfectly aligned with the line of sight to the source, then the
lens splits the source into two images.  The separation of these
images is $\sim 2\thetae$ and hence unresolvable.  However, the
source is also magnified by the lens, by an amount that depends on the angular
separation between the lens and source in units of $\thetae$.  Since the observer, lens, and
source are all in relative motion, this magnification is a function of
time: a `microlensing event.'  The characteristic time scale for such an event is 
\begin{equation}
\te={\thetae \over \murel}\simeq 20~\days \left({M\over
0.3~M_{\odot}}\right)^{1/2},
\label{eqn:te}
\end{equation}
where $\murel$ is the  lens-lens relative proper motion, which we
have assumed to be typical of events toward the Galactic bulge, $\murel=25~\kms~\kpc^{-1}$.

If the primary lens has a planetary companion, and the position of
this companion happens to be near the path of one of the two images
created during the primary event, then the planet will perturb the
light from this image, creating a deviation from the primary
light curve.  The duration  $\tp$ of the deviation is roughly the time it
takes the source to cross the Einstein ring of the planet, $\thetap$.
From \eq{thetae}, $\thetap=(\massp/M)^{1/2}\thetae$, where $\massp$ is
the mass of the planet.  Therefore, from \eq{te},
$\tp=(\massp/M)^{1/2}\te$, or 
\begin{equation}
\tp = \sqrt{q} \te
\label{eqn:tep}
\end{equation}
where $q\equiv \massp/M$ is the mass ratio of the system.
For a Jupiter/Sun mass ratio ($q\simeq10^{-3}$), the
perturbation time scale is ${\cal O}({\rm day})$.
Since the perturbation time scale is considerably less than
$\te$, the majority of the light curve will be indistinguishable from a
single lens.  Hence the signature of a planet orbiting the primary
lens is a short-duration deviation imposed on an otherwise normal single lens curve.

Because microlensing relies on the mass
(and not light) of the system, planets can be searched for around
stars with distances of many kiloparsecs. 
Also, the sensitivity can, in principle, be extended down to Earth-mass planets \citep{bandr1996}.
Finally, orbital separations of many AU can be probed immediately, without having to
wait for a full orbital period.  The primary disadvantages of
microlensing searches for planets are that the measurements are not
repeatable and there is little hope for follow-up study of discovered planetary
systems.  

\citet{mandp1991} first suggested that microlensing might be used to find
extrasolar planets.  Their ideas were expanded upon by
\citet{gandl1992}, who in particular noted that if all stars had
Jupiter-mass planets at projected separations of $\sim\re$, then $\sim20\%$ of all
microlensing events should exhibit planetary perturbations and that
the detection probability will be
highest for planets with projected separations lying within $[0.6-1.6]\thetae$ of the primary,
the ``lensing zone.''  Since
these two seminal papers, the theoretical basis of planetary microlensing
has developed rapidly.  Numerous authors have studied detection
probabilities and observing strategies incorporating a variety of new
effects \citep{bandf1993, bandr1996,  peale1997, sackett1997,
gands1998, gns, dands1999a, dands1999b, vermaak2000,  handk2001, peale2000}.  Notably, \citet{bandr1996}
found that the detection probability for Earth-mass planets could be
appreciable ($\sim2\%$), and \citet{gands1998} found that for high magnification
events the detection probability can be nearly 100\% for Jovian
planets in the lensing zone.   \citet{gandg1997}, \citet{me1998} and
\cite{meands2000} all discussed extracting information from observed
microlensing events.  In particular, \cite{meands2000} developed a
method to calculate the detection efficiency of observed datasets to
planetary companions; this method is employed extensively here.
Planetary microlensing has been placed in the global context of binary
lensing by \citet{martin1999}, and studied via perturbative analysis
by \citet{bozza1999,bozza2000a,bozza2000b}.  

On the observational front, progress has been somewhat slower.  This
is primarily because the survey collaborations that
discover microlensing events toward the Galactic bulge, EROS \citep{derue1999}, MACHO
\citep{alcock1997a}, and OGLE \citep{udalski2000}, have sampling periods
that are of order or smaller than the planetary perturbation
time scale, $\tp$.  However, soon after these searches commenced, these
collaborations developed the capability to recognize microlensing
events in real time \citep{alcock1996,ews}, thus allowing 
publically available alerts of ongoing events.  In response to this potential,
several ``follow-up'' collaborations were formed: GMAN
\citep{pra96,alcock1997b}, PLANET \citep{albrow1998} and MPS \citep{mpssmc},
with the express purpose of intensively monitoring alerted events 
to search for deviations from the standard point-source point-lens
(PSPL) light curve, and in particular the short duration signatures of
planets.  The feasibility of such a monitoring campaign was
demonstrated in the 1995 pilot season of PLANET
\citep{albrow1998}, during which we achieved $\sim 2$~hour sampling and
few percent photometry on several concurrent bulge microlensing events.   

The MPS collaboration used observations of
the high-magnification event \mb{98}{35} to rule out Jovian
companions to the primary microlens for a large range of separations \citep{mps}.
We performed a similar study of \ob{98}{14} \citep{albrow2000b},
demonstrating that companions with mass $>10~\mjup$ were ruled out
for separations $1-7~\au$.  Our detection efficiency for this event
was $\sim60\%$ for a companion with the mass and separation of
Jupiter, thereby demonstrating that a combined analysis of many
events of similar quality would place interesting constraints on
Jovian analogs.  A similar analysis was performed for events
\ob{00}{12} and MACHO~99-LMC-2 by the MOA collaboration
\citep{bond2001}.

\citet{bennett1999} claimed to detect a planet
orbiting a binary microlens \mb{97}{41}.  As we discuss in
\S\ref{sec:esel}, we exclude binaries with mass ratios $q>10^{-2}$
from our search because of the difficulty of modeling binaries and
therefore of making an unambiguous detection of planetary
perturbations amongst the wealth of other perturbations that can occur
in these systems.  Indeed,
\citet{albrow2000a} found that all available data for
this event were explained by a rotating binary without a planet.

\citet{mps} claimed ``intriguing evidence'' for a planet with
mass ratio $4 \times 10^{-5} \le q \le 2 \times 10^{-4}$ in event
\mb{98}{35}. This perturbation had a reduced $\Delta\chi^2\sim 21$,
far below our threshold of 60.  As can be seen from Figure
\ref{fig:detdist}, our data set contains many perturbations with
$\Delta\chi^2\la 50$.  As we show in \S\ref{sec:detections}, based on
studies of constant stars, we find that systematic and statistical
noise can easily give rise to deviations in our data with
$\Delta\chi^2 \la 60$.

	\citet{bond2001} reanalyzed all available data for
\mb{98}{35} including the then unpublished PLANET data that
are now presented here.  They found fits for 1--3 planets all with
masses $q<3\times 10^{-5}$, with $\Delta\chi^2=60$.  This mass range
is below our search window, primarily because our sensitivity to it is
quite low (see \S\ref{sec:ulimits}).  In our view,
planetary detections in this mass range should be held to a very
rigorous standard, a standard not met by $\Delta\chi^2=60$ which would
be just at our threshold.

Thus, none of these claimed detections \citep{bennett1999,
mps,bond2001} would have survived our selection criteria even if they
had been in our data.  Therefore, they pose no conflict with the fact that
we detect no planets among 43 microlensing events, and are not in
conflict with the upper limits we place on the abundance of planets among bulge stars.

Despite the excellent prospects for detecting planets with
microlensing, and after more than five years of intensive monitoring of
microlensing events, no unambiguous detections of 
Jupiter-mass lensing companions have been made.
These null results broadly imply that such planetary companions must not be very common.  
In the remainder of this paper we quantify this conclusion by analyzing five years of PLANET
photometry of microlensing events toward the bulge for the presence of
planets orbiting the primary microlenses.  We use strict event
selection criteria to construct a well defined subsample of events.  
Employing analysis techniques presented in
\citet{meands2000} and applied in \citet{albrow2000b}, we search for
the signals of planets in these events.  We find no planetary microlensing signals.  Using this
null result, and taking into account the detection efficiencies to
planetary companions for each event, we derive a statistical upper
limit to the fraction of primary microlenses with a companion.   Since
most of the events in our sample are likely due to normal stars in the
Galactic bulge, we therefore place limits on the fraction of stars in
the bulge with planets.

We describe our observations, data reduction and post-processing in
\S\ref{sec:data}.  In \S\ref{sec:gconsid}, we describe and categorize
our event sample. We define and apply our event selection criteria in
\S\ref{sec:esel}; this section also includes a description of how our
events are fitted with a PSPL model.  We summarize the characteristics
of our final sample of events in \S\ref{sec:echars}.  In
\S\ref{sec:sdes},  we describe our algorithm for searching for
planetary perturbations (\S\ref{sec:algorithm}) as well as various
nuances in its implementation
(\S\S\ref{sec:cofdata}-\ref{sec:threshold}).  We describe our detections
(or lack thereof) in \S\ref{sec:detections} and our detection
efficiencies in \S\ref{sec:des}.  Our method of correcting for finite
source effects is discussed in \S\ref{sec:fs}, and we derive our upper
limits in \S\ref{sec:ulimits}.  We interpret our results in
\S\ref{sec:iandd}, compare our results with other constraints on
extrasolar planets in \S\ref{sec:discuss}, and conclude in \S\ref{sec:conclude}.
Appendix~\ref{sec:app1} lists our excluded anomalous events, and 
Appendix~\ref{sec:parallax} discusses parallax contamination.

This paper is quite long, and some of the discussion is technical and
not of interest to all readers.  Those who want simply to
understand the basic reasons why we conclude there are no planets  
and understand our resulting upper limits on companions should 
read \S\ref{sec:gconsid}, and \S\S\ref{sec:ulimits}-\ref{sec:conclude}.
Those who want only the upper limits and their implications should
read \S\S\ref{sec:discuss} and \ref{sec:conclude}, especially focusing on 
Figures~\ref{fig:ulam} and \ref{fig:lzulam}.
A brief summary of this work is given in \citet{albrow2001a}.

\section{Observations, Data Reduction, and Post-Processing\label{sec:data}}

Details of the observations, detectors, telescopes, and primary data reduction
will be presented elsewhere \citep{albrow2001c}.  Here we will
summarize the essential aspects of the observations and
primary data reduction, and discuss only our post-processing in
detail.  

The photometry of the microlensing events presented and analyzed here
was taken over five bulge seasons starting from June of 1995 and ending in
December 1999, with a few scattered baseline points taken in early in
2000.   These data were taken with six different telescopes:  the CTIO
0.9m, Yale-CTIO 1m, and Dutch/ESO 0.91m in Chile, the SAAO 1m in South
Africa, the Perth 0.6m near Perth, Australia, and the Canopus 1m in
Tasmania.  Measurements were taken in the broadband filters $V_J$ and $I_C$ using a total of 11
different CCD detectors.  

The data are reduced as follows.  
Images are taken and flat-fielded in the usual way;  these
images are then photometered using the DoPHOT package \citep{dophot}.  A
high-quality image is chosen for each field, which is then used to
find all the objects on the frame.  From this ``template'' image,
geometrical transformations are found for all the other frames.
Fixed-position photometry is then performed on all the objects in all
the frames.  The time-series photometry of all the objects found on
the original template image is then archived using specialized
software designed specifically for this task.  This software enables
photometry relative to an arbitrarily chosen set of reference 
stars.  We treat each
light curve for each site, detector and filter as independent.  The
number of independent light curves for each event ranges from one to
twelve.  For the majority of the
events, the $V$-band data are reduced using the source positions
identified with the $I$-band template image,
since, in general, the signal-to-noise is considerably higher in
$I$-band and more objects are detected. This improves the subsequent
photometry relative to what can be achieved using a $V$-band template.

Once the photometry of all objects in the microlensing target fields
are archived, we perform various post-reduction procedures to
optimize the data quality. The light curves of the microlensing source stars are extracted
using reference stars chosen in a uniform manner.  
Four to 10 reference stars are chosen that are close to the
microlensing source star (typically within $30''$) and exhibit no
detectable brightness variations.  We require that the ratio of the mean
DoPHOT-reported error in the measurements of each reference star to
standard deviation all of the measurements of the star is
approximately unity, with no significant systematic trend over the
entire set of observations.  Generally, the mean DoPHOT-reported error
in a single measurement of a reference star is 0.01~mag.  Reference
stars are selected for each independent light curve, although
typically the set of reference stars is similar for all observations
of a particular event.  Only those points on the microlensing event light curve with
DoPHOT types\footnote{DoPHOT types rate the quality of the
photometry.  DoPHOT type 11 indicates an object consistent with a point source
star, whereas DoPHOT type 13 indicates a blend of two close stars.
From our experience, all other DoPHOT types often provide
unreliable or suspect photometry.} 11 or 13, and 
DoPHOT-reported errors $<0.4$~mags are
kept.  Further data points are rejected based on unreliable reference
star photometry as follows.  For each reference star, the
error-weighted mean is determined and the point that deviates
most ($>3\sigma$) from the mean is removed.  The errors of the
remaining points are scaled to force the $\chi^2$ per degree of
freedom (d.o.f.) for the reference star light curve to unity.  The
error-weighted mean is then recomputed, and
the entire process repeated until no $>3\sigma$ outliers remain.
The outliers are reintroduced with error scalings determined
from their parent light curves.
Then, for each data point in the microlensing light curve, the
$\chi^2$ of all the reference stars are summed.  If this $\chi^2$ is larger
than four times the number of reference stars, the data point is discarded.
After this procedure, individual light curves are then examined, and light curves
for which the microlensing target was
too faint to be detected on the template image were eliminated.  In addition, 
individual light curves with less than 10 points are eliminated.
Since at least three parameters are needed to fit each light curve (see \S\ref{sec:esel}),
light curves with fewer than 10 points contain very little information.   
Finally, a small number ($\la 10$ over the entire dataset) of
individual data points were removed by hand.  These data points were
clearly highly discrepant with other photometry taken nearly
simultaneously, and were typically taken under extreme seeing and/or
background conditions, or had obvious cosmic ray strikes near the
microlensing target.  Since there are only a handful of such points,
their removal has a negligible effect on the overall sensitivity.  Furthermore, these
points cannot plausibly be produced by a real planetary signal, but would 
lead to spurious detections if not removed.

\section{General Considerations\label{sec:gconsid}}

During the 1995-1999 seasons, PLANET relied on alerts from three
survey teams, EROS (1998-99), MACHO (1995-99), and OGLE (1995;
98-99).  During these five years, several hundred  events were
alerted by the three collaborations combined.  Often, there are too many to
follow at one time, and PLANET must decide real-time
which alerts to follow and which to ignore.  Since the event
parameters are typically poorly known at the time of the alert, and survey team data
are sometimes unavailable, it is impossible to set forth
a set of rigid guidelines for alert selection.
The entire process is necessarily organic: decisions are made primarily by 
one (but not always the same) member of the collaboration, and secondarily by the observers at
the telescopes, and are based on considerations such 
as the predicted maximum magnification and time scale of the event,
the brightness and crowding of the source, and the number and quality of other
events currently being followed.  
Our final compilation of events does not therefore represent a well defined
sample.  Some
selection effects are present both in the sample of events alerted by the survey teams and the
sample of events we choose to follow.  Although these
selection effects could in principle bias our conclusions, in
practice their effects are probably quite minor, since the reasons that  an
event was or was not alerted and/or monitored
(i.e.\ crowding conditions and/or brightness of the source, number of
concurrent events, maximum magnification) are not related to
the presence or absence of a planetary signal in the light curve.   
The one exception to this is the microlensing time scale, which as we
show in \S\ref{sec:echars}, is typically twice as long in our sample
as in the parent population of microlensing events.   One might
imagine that, since our sample is biased toward longer time scale
events, we are probing higher mass lenses.  In fact, as
we show in \S\ref{sec:iandd},  it is likely that we are primarily
selecting slower, rather than more massive, lenses.  Thus the bias toward
more massive primaries is small.
This is not necessarily a bias, per se, as long as we
take care to specify the population of primary lenses around which we
are searching for planets.  
Thus, provided that any a posteriori cuts we make are also not
related to the presence or absence of planetary anomalies in the
light curves, our sample should be relatively unbiased.

We would like to define a sample of events in which we can search for
and reliably identify planetary companions to the primary lenses.  
The events in this sample must have sufficient data quality and
quantity that the nature of the underlying lensing system can be
determined.  Also, our method of searching for planetary perturbations
is not easily adapted to light curves arising from non-planetary anomalies, 
such as those arising from parallax or equal mass binaries.
Therefore, such events must be discarded.  The remaining events represent the well-defined sample, which can then be search for planetary companions.  In the next section, we describe
our specific selection criteria designed to eliminate these two categories of events 
and the implementation of these criteria used to define our sample.
However, for the most part, our events could be placed cleanly into these categories by eye, 
without the need of detailed modeling or analysis.  Examination of our full sample of light curves 
reveals that the events generally fall into three heuristic categories:
\begin{description}
\item[{(1)}] Poor-quality events.  
\item[{(2)}] High-quality events which are obviously deviant from the PSPL form for a large fraction of the data span, or are deviant from the PSPL form in a manner that is unlikely to be planetary. 
\item[{(3)}] High-quality events which follow the PSPL form, with no obvious departures from the PSPL form.
\item[{(4)}] High-quality events which exhibit a short-duration
deviation superimposed on an otherwise normal PSPL light curve. 
\end{description}
Events in the first category are the most plentiful:  they consist of events
with either a very small number of points ($\la 20$), poor photometric precision, and/or incomplete light curve coverage.  Events in the second category are those with high-quality data, in terms of photometric precision, coverage, and sampling.  They typically consist of anomalies recognized real-time, and are comprised of both events that deviate from the PSPL form in a way not
associated with binary lensing (i.e.\ finite source effects, parallax, and binary source events), and
events arising from roughly equal-mass (mass ratio $\gsim 0.1$) binary lenses.
Events in the third category are high-quality, apparently normal events that follow the PSPL form
without obvious deviations.  Events in the last category are planetary candidates.

The first two categories correspond to events that should be 
removed from the sample; events 
in the last two categories make up the final event sample, and should be analyzed in detail for planetary companions.
Of course, some cases are more subtle, and the interpretation of the event is not so clear.  
In general, however, other deviations from the PSPL form are easily distinguishable from
planetary deviations, with two caveats.  First, there is no clear
division between ``roughly equal mass ratio'' and ``small mass ratio''
binary lenses: if the mass ratio distribution of binary lenses were, e.g,
uniform between $q=10^{-5}$ and $q=1$, one would expect
grossly deviant light curves, light curves with short-duration
deviations, and everything in between.  In practice, however, this
does not appear to be the case, as we discuss below.  Second, there
exists a class of binary-source events that can mimic the
short-duration deviations caused by planetary companions
\citep{me1998}.  Detections of short-duration anomalies must therefore
be scrutinized for this possibility.

All of the 126 Galactic bulge\footnote{We exclude events toward the
Magellanic Clouds.} microlensing events for which PLANET has acquired data
during the 1995-1999 seasons are listed in Table~\ref{tab:tab01}.  
A cursory inspection of these events reveals that $\sim 40\%$ clearly belong in category (1), 
$\sim 11\%$ clearly belong in category (2), and $\sim 25\%$ clearly belong in category (3).
 The remaining $\sim 24\%$ are marginal events that could be placed in either category
(1) or (3).  However, no events clearly belong to the last category,
i.e., there are no events that have anomalies that are clearly consistent with a low
mass-ratio companion.  Since we do not see a continuous distribution in the time scale of deviations with
respect to the parent light curve time scale, this implies that 
either the mass ratio distribution
is not uniformly distributed between equal mass and small mass
ratios or our detection efficiency to companions drops 
precipitously for smaller mass ratios.  In fact, as we show in \S\ref{sec:des}, our efficiencies are substantial for mass ratios $\gsim 10^{-3}$, implying that massive planetary
companions are probably not typical.  For the remainder of the paper,
we will use strict event selection criteria and sophisticated methods of
analysis to justify and quantify this statement.  

\section{Event Selection\label{sec:esel}}

The goal of our selection criteria is to provide a clean sample of events for which we can
reliably search for planetary deviations and robustly quantify the detection
efficiency of companions.  Such criteria are also
necessary so that future samples of events (and possibly future
detections) can be analyzed in a similar manner, and thus combined 
with the results presented here. 
Our selection criteria roughly correspond to the categorization
presented in \S \ref{sec:gconsid}.  Note that any arbitrary rejection
criterion is valid, as long as the criterion is not related the
presence or absence of a planetary signal in the light curve.

We first list our adopted rejection criteria, and then describe
the criteria, our reasons for adopting them, and the procedure to implement them.
The three rejection criteria are:
\begin{description}
\item[{(1)}] Non-planetary anomalies (including parallax, finite
source, binary sources, and binaries of mass ratio $>0.01$). 
\item[{(2)}] Events for which no individual light curve has 20 points or more.
\item[{(3)}] Events for which the fractional uncertainty in the fitted impact
impact parameter, $\u0$, is $>50\%$.
\end{description}
The original sample of 126 events along with an indication of which
events were cut and why is tabulated in
Table~\ref{tab:tab01}.
The first criterion eliminates 19 events, the second 32 events, and the third
32 events, for a final sample of 43 events.  

As stated previously, criterion (1) is necessary because
we do not have an algorithm that can systematically search for planetary companions
in the presence of such anomalies.  
We are confident that the anomalies in the events that we have rejected by
criterion (1) are, in fact, non-planetary in origin, based on our own analyses, 
analyses in the published literature, and a variety of secondary
indicators.  Descriptions of each of
these events and the reasons why we believe the anomaly to be
non-planetary in origin are given in Appendix \ref{sec:app1}.

We fit the observed flux $F_l$ of observatory/band
$l$ and time $t_k$ to the microlensing-event model,
\begin{equation}
F_l(t_k)=F_{{\rm S},l} A(t_k) + F_{{\rm{ B}},l} +\eta_l [\theta(t_k)-\theta_{0,l}]
\label{eqn:foft}
\end{equation}
where $A(t_k)$ is the magnification at time $t_k$; $F_{{\rm S},l}$ and
$F_{{\rm B},l}$ are the source and blend fluxes for
light curve $l$.  The last term is introduced to account for the
correlation of the flux with seeing that we observe in almost all of our
photometry (see \citealt{albrow2000b}).  Here $\eta_l$ is
the slope of the seeing correlation, $\theta(t_k)$ is the full width
at half maximum (FWHM) of the
point spread function (PSF) at time $t_k$, and $\theta_{0,l}$ is the error-weighted mean FWHM of
all observations in light curve $l$.  For a single lens, the
magnification is given by \citep{ref64,pac1986}.
\begin{equation}
A_0[u(t)]={ {u^2(t)+2}\over u(t)
\sqrt{u^2(t)+4}}; \qquad u^2(t)= \tau^2
+\u0^2 ,
\label{eqn:aoft}
\end{equation}
where $\tau$ is the ``normalized time,''
\begin{equation}
\tau \equiv { {t-\t0}\over \te}.
\label{eqn:tau}
\end{equation}
Here $\t0$ is the time of maximum magnification, $\te$ is the
characteristic time scale of the event, and $\u0$ is the minimum
angular separation (impact parameter) between the lens and source in
units of $\thetae$.  A single lens fit to a multi-site, multi-band
light curve is thus a function of $3+3 N_{l}$ parameters: $\te$,
$\u0$, $\t0$, and one source flux $F_{{\rm S},l}$, blend flux $F_{{\rm
B},l}$, and seeing correlation slope $\eta_l$ for each of $N_{l}$
independent light curves.  For a binary lens, three additional
parameters are required: the mass ratio of the two components, $q$,
the binary separation $d$ in units of $\thetae$, and the angle of the
source trajectory with respect to the binary axis, $\alpha$.  Thus for
an event to contain more information than the number of free
parameters, at least one observatory must have at least 9+1=10 data
points.  In order for the fit to be well-constrained, considerably
more data points than fit parameters are needed.  We therefore impose
criterion (2): if no independent light curve has at least 20 data
points, the event is rejected.  The number 20 is somewhat arbitrary,
however the exact choice has little effect on our conclusions: a
natural break exists such that the majority of events are well above
this criterion, and those few events that are near the cut have little
sensitivity to planetary perturbations.

All events that pass criterion (2) are fit to a PSPL model 
[eqs.\ (\ref{eqn:foft}) and (\ref{eqn:aoft})].  At this stage, we also
incorporate MACHO and/or OGLE data into the fit, when 
available\footnote{MACHO data are available for those events alerted
by MACHO in 1999, along with a few events that were originally alerted by OGLE
in 1999.
OGLE data is available for events alerted by OGLE in 1998-99, along
with a few events that were originally alerted by MACHO during these years.}.   To fit the PSPL model, we
combine the downhill-simplex minimization routine AMOEBA
\citep{cookbook} with linear least-squares fitting.  Each
trial combination of the parameters ($\te, \te, \u0$)
immediately yields a prediction for $A_0(t)$ [eqs.~(\ref{eqn:aoft})
and (\ref{eqn:tau})].  The flux is then just a 
linear combination of $F_{{\rm S},l}$,  $F_{{\rm B},l}$ and $\eta_l$ [eq.\ (\ref{eqn:foft})].
The best fit parameters $a_i=(F_{{\rm S},1}, F_{{\rm B},1},
\eta_1, F_{{\rm S},2}, F_{{\rm B},2}, \eta_2, ...)$ can then be found by
forming,
\begin{equation}
b_{ij}\equiv\sum_k {{1}\over{\sigma^2_k}}{{\partial
F(t_k)}\over{\partial a_i}}{{\partial F(t_k)}\over{\partial
a_j}}, \qquad c=b^{-1}, \qquad d_i = \sum_k {{F_k}\over{\sigma^2_k}}  {{\partial
F(t_k)}\over{\partial a_i}}, 
\label{eqn:bij}
\end{equation}
where the index $k$ refers to a single observation, the sum is over
all observations, and $\sigma_k$ is the photometric error in the observed flux $F_k$.   
The parameter combination $a_i$ that minimizes $\chi^2$ is then,
\begin{equation}
a_i=\sum_j c_{ij} d_{j}.
\label{eqn:alphas}
\end{equation}
Occasionally, the values of $F_{{\rm B},l}$ obtained from this procedure
are negative.  If $F_{{\rm B},l}$ is negative by more than
its uncertainty, we apply a constraint to $c_{ij}$ to force $F_{{\rm B},l}=0$.
We then use AMOEBA to find the values of ($\te, \te, \u0$) that
minimize $\chi^2$.  Note that since neither MACHO nor OGLE report
seeing values, we do not correct their data for seeing correlations. 

We know from experience \citep{albrow1998,albrow2000b} that 
DoPHOT-reported photometric errors are typically underestimated by a factor of
$\sim1.5$.  Naively adopting the DoPHOT-reported errors would thus
lead one to underestimate the uncertainty on fitted parameters, and overestimate 
the significance of any detection.   However, simply scaling all
errors by a factor to force $\chi^2/{\rm d.o.f.}$ to unity is also not
appropriate, as we find that our photometry usually contains significantly more large ($>
3\sigma$) outliers than would be expected from a Gaussian distribution
\citep{albrow2000b, albrow2000c}.  Furthermore, independent light
curves from different sites, detectors, and filters typically have different error
scalings.  Therefore we adopt the following
iterative procedure, similar to that used by \citet{albrow2000b}.  We
first fit the entire dataset for a given event to a PSPL model in the manner explained
above.  We find the largest $>3\sigma$ outlier, and reject it.
We then renormalize the errors on each individual light curve to force
$\chi^2/{\rm d.o.f.}$ to be equal to unity for that light curve.
Next, we 
refit the PSPL model, find the largest $>3\sigma$ outlier, etc.  This
process is repeated until no $>3\sigma$ outliers are found.  
The outliers are then reintroduced, with error scalings
appropriate to their parent light curve.  We typically find 3 to 6
outliers $>3\sigma$ in the PLANET data and OGLE data, and a larger
number for MACHO data (which contain significantly more data points).
The median error scaling for PLANET data is 1.4, with $90\%$ of our
data having scalings between 0.8 and 2.8.  The errors reported by OGLE
are typically quite close to correct (scalings of $\sim 1.1$),
while MACHO errors are typically overestimated (scalings of $\sim 0.8$).

Once the best-fit PSPL model is found, we determine the uncertainties
on the model parameters by forming $c_{ij}$ as in \eq{bij}, except
that now the parameters are $a_i=(\t0, \te, \u0, F_{{\rm S},1},
F_{{\rm B},1}, \eta_1, F_{{\rm S},2}, F_{{\rm B},2}, \eta_2, ...)$,
i.e., we have included $\te$, $\t0$, and $\u0$.  The uncertainty in
parameter $a_i$ is then simply $\delta a_i=(c_{ii})^{1/2}$.  Note that
we include the outliers to determine the uncertainties.  As discussed
by \cite{gands1998} the sensitivity of a light curve to planetary
companions is strongly dependent on the path of the source trajectory
in the Einstein ring, such that trajectories that pass closest to the
primary lens, i.e.\ events with small $\u0$, will have larger
sensitivity than events with larger $\u0$.  Thus, in order to
accurately determine the detection efficiency to a given binary lens,
the source path in the Einstein ring, $u(t)$, must be
well-constrained; poor knowledge of $u(t)$ translates directly
into poor knowledge of the sensitivity of the event to planets
\citep{meands2000}.  
The values of $u(t)$ for a given dataset are
determined from the mapping between flux and magnification, which
depends on the source and blend fluxes, and the mapping between the
magnification and time, which depends on $\u0$, $\te$, and $t_0$.  In
blended PSPL fits, all these parameters are highly correlated.  Thus,
a large uncertainty in $\u0$ implies a large uncertainty in other
parameters.  Thus the uncertainty in $\u0$ in a PSPL fit
can be used as an indication of the uncertainty in $u(t)$, and thus
the uncertainty in the detection efficiency.    
Furthermore, for a planetary perturbation, the projected separation
$d$ is a function of the observables $(t_{0,{\rm p}}-t_0)/\te$, where
$t_{0,{\rm p}}$ is the time of the planetary perturbation, while the
mass ratio is $q \sim {{\tp}/ \te}$ \citep{gandl1992,gandg1997}, where
${t_p}$ is the duration of the perturbation.  Therefore the detection
of a planet in an event with poorly constrained $\te (\u0)$ would be
highly ambiguous, as the neither the projected separation $d$ nor the
mass ratio $q$ would be well-constrained.  We therefore impose a cut 
based on the fractional uncertainty in the fitted value of $\u0$.

\begin{figure*}[t]
\epsscale{1.0}
\centerline{\plotone{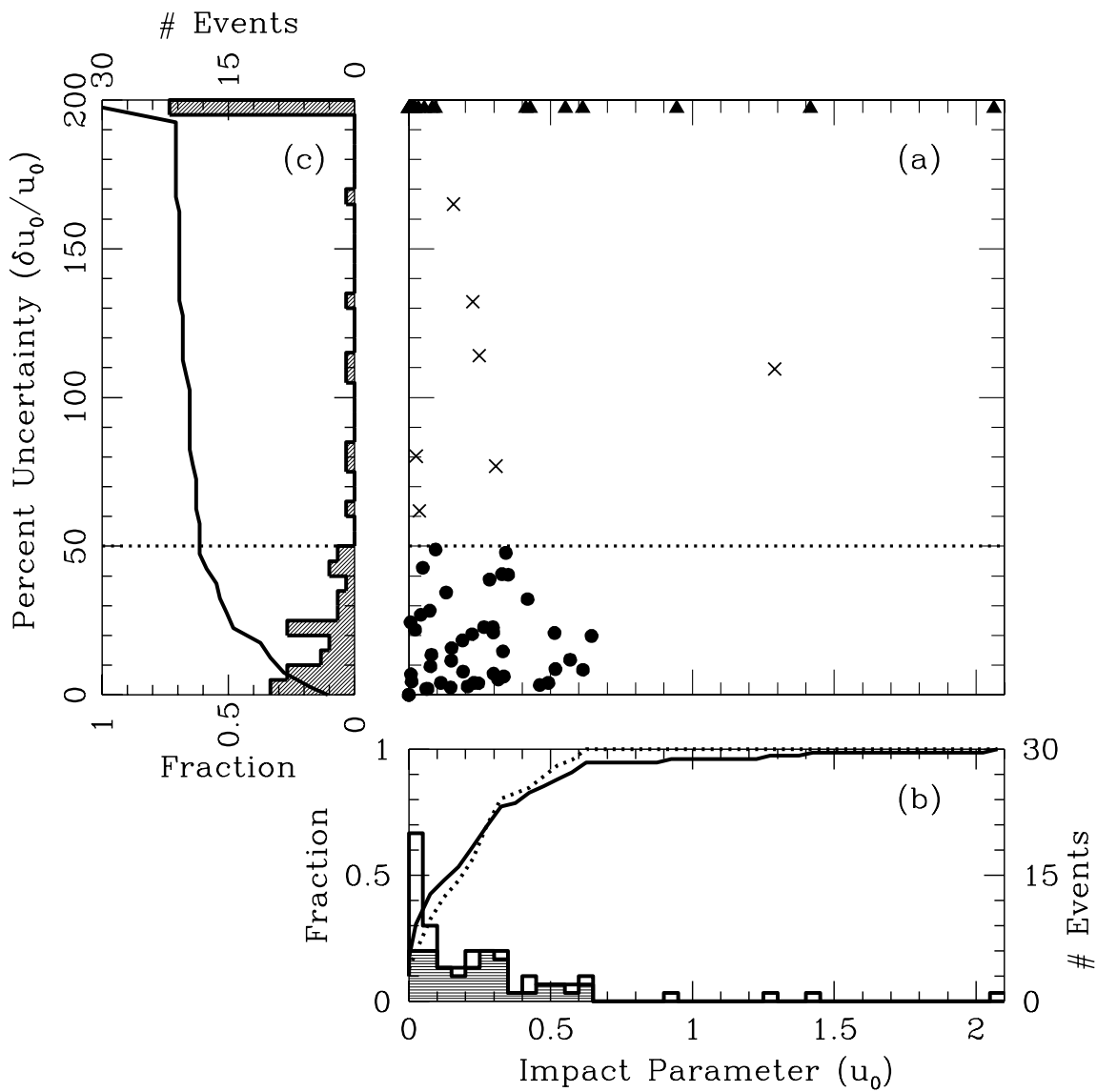}}
\caption{
\footnotesize
(a) The percentage uncertainty $\du0$ in the impact parameter  is
plotted versus $\u0$ for all events that pass our data
quantity cut. The $\u0$ for those events with $\du0>200\%$ are
plotted as triangles.  The dotted line indicates our cut on the
fractional uncertainty, $\du0 =50\%$. Events with $\du0\le50\%$ are included
in the final event sample and are shown as solid circles, while events
with $\du0 >50\%$ are discarded (crosses and triangles). (b) The lines
show the cumulative distribution of $\u0$ for all events (solid) and
those events that pass our cut (dashed).  The histograms show the
differential distributions of $\u0$ for all events (unshaded) and
those events that pass our cut (shaded).  The left axis refers to the 
cumulative distributions, while the right axis refers to the
differential distributions.  (c)  The line shows the cumulative
distribution of $\du0$ (bottom axis).  The histogram shows the
differential distribution (top axis). 
}
\label{fig:u0dist}
\end{figure*}

Figure \ref{fig:u0dist} shows the fractional uncertainty  $\delta \u0/\u0$ in the impact
parameter versus $\u0$ for all events that passed
selection criteria (1) and (2).  Examination of the distribution of fractional
uncertainty in $\u0$ for these events reveals a large clump of events with small fractional
uncertainty; many scattered, smoothly distributed events with larger uncertainties, and a natural
break in the distribution at $\delta \u0/\u0 \approx 50\%$.  We
therefore adopt $\delta \u0/\u0 = 50\%$ for our final event cut.  The
exact choice for the cut on $\delta \u0/\u0$ has little effect on our
conclusions; as we discuss in \S\ref{sec:des}, events with $\delta
\u0/\u0 \ga 30\%$ typically have low detection efficiencies.  
Four classes of events have  poorly-constrained $\u0$.  These are events: for which the data
cover only one (usually the falling) side of the event;
for which no baseline information is available; that are highly
blended; with an
intrinsically low maximum magnification.  Thus by
imposing a cut on $\delta \u0/\u0$, we eliminate all low magnification
events; the event with largest impact parameter in our final sample
has $\u0=0.61$.  
Note that the majority of events that fail the cut on $\delta
\u0/\u0$ fall into the first two classes, which 
emphasizes the need for coverage of the peak and baseline
information.  In particular, without MACHO and OGLE data, many more events
would not have passed this last cut, and our final sample would
have been considerably smaller. 

\begin{figure*}[t]
\epsscale{0.7}
\centerline{\plotone{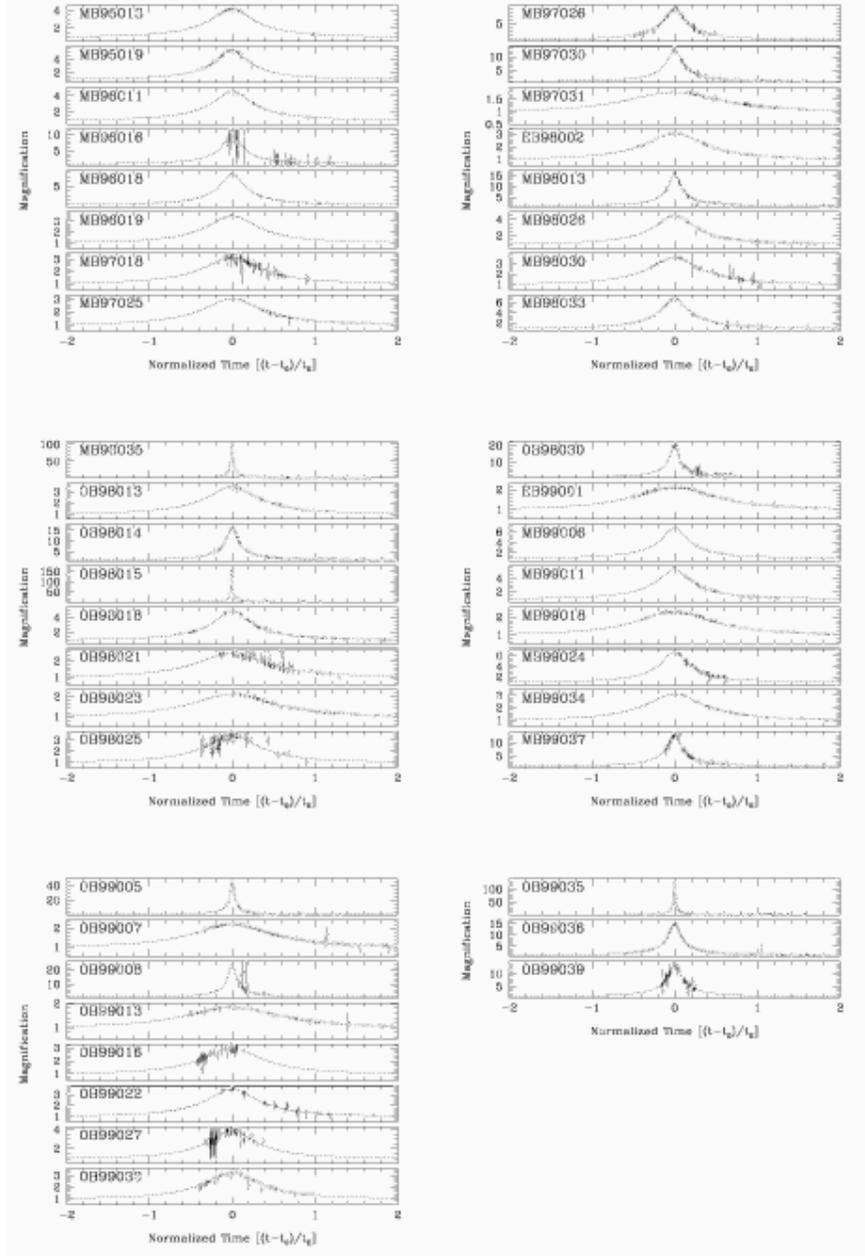}}
\caption{
\footnotesize
PLANET data for the events that pass our selection criteria
(\S ~\ref{sec:esel}).  The magnification, $(F-\fs)/\fb$, is plotted as a function of
normalized time, $\tau=(t-\t0)/\te$, for the ``cleaned''
light curves, i.e. with seeing systematics removed and rescaled errors.
See \S ~\ref{sec:data}.
}
\label{fig:lca}
\end{figure*}

After imposing cuts 1 (non-planetary anomalies), 2 (data quantity), and
3 (uncertainty in the impact parameter), we are left with a sample of 43 events.
The light curves for these events are shown in
Figure~\ref{fig:lca}.  In order to display all independent
light curves (which in general have different $\fs$, $\fb$, and
$\eta$), we plot the magnification, which is
obtained by solving \eq{foft} for $A_0(t)$.  Rather than show the
magnification as a function of true time, we show the magnification as
a function of normalized time $\tau$ [\Eq{tau}].  When
plotted this way, perturbations arising from a given 
$q$ would have the same duration on all plots [\Eq{tep}].
Thus the sensitivity of different light curves to companions can be compared
directly. In the next section, we describe the properties of these events,
paying particular attention to those properties relevant to the
detection of planetary anomalies.

\section{Event Characteristics\label{sec:echars}}

The parameters $\t0$, $\te$, and $\u0$ and their respective $1\sigma$ uncertainties  
for the final event sample are tabulated in Table~\ref{tab:tab02},
along with the percent uncertainty in $\u0$.  The sensitivity of an event to planetary companions
depends strongly on $\u0$ \citep{gandl1992,gands1998,meands2000}, and thus the exact distribution
of $\u0$ influences the overall sensitivity of any set of light curves.   The time scale is
important in that the population of lenses
we are probing is determined from the distribution of $\te$.  In addition, we
use $\te$ in \S\ref{sec:fs} to estimate the effect of finite sources on planetary
detection efficiencies and therefore the effect on our final
conclusions.  For the current analysis, the parameter $\t0$ is of
no interest.  

\begin{figure*}[t]
\epsscale{1.0}
\centerline{\plotone{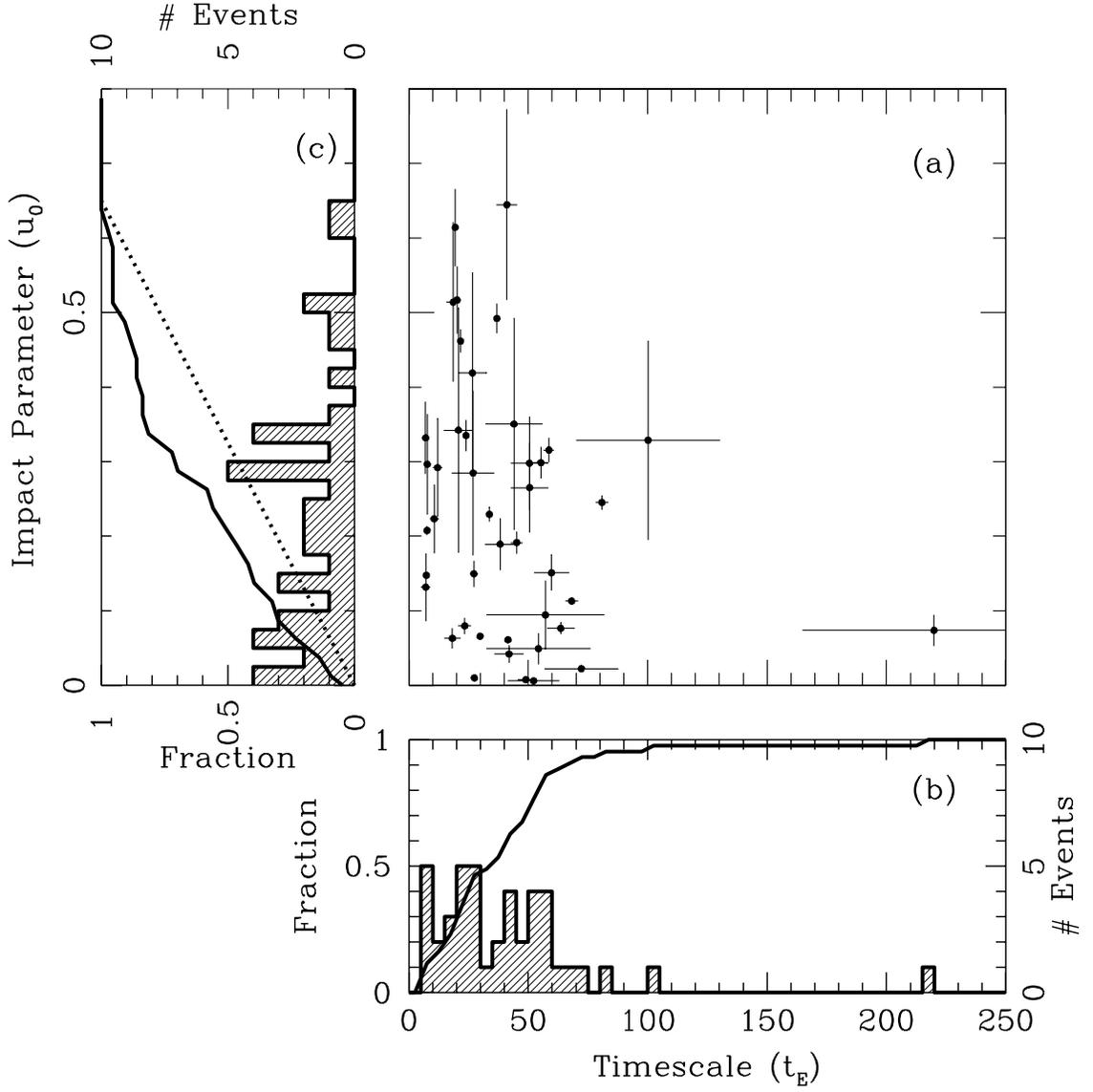}}
\caption{
\footnotesize
(a) The fitted impact parameter, $\u0$, is
plotted versus the fitted time scale, $\te$, for those events that pass all our cuts. 
(b) The line shows the cumulative distribution of $\te$ (left axis), while the histogram shows the
differential distribution (right axis). (c)  The
line shows the cumulative distribution of $\u0$ (bottom axis), while
histogram shows the differential distribution (top axis).  The dotted line is for a uniform distribution in $\u0$.
}
\label{fig:u0vte}
\end{figure*}

In Figure~\ref{fig:u0vte}, we plot $\u0$ against $\te$ for our event
sample, revealing no obvious correlation between the two.
This lack of correlation between $\te$ and $\u0$ implies that the
lenses that give rise to the events with the most sensitivity  to
planets (i.e., those with small $\u0$) comprise a sample that is unbiased with respect to the entire
sample of lenses.  Given this, we can then inspect the distributions
of $\u0$ and $\te$ independently.   

Both the differential and cumulative distributions of $\te$ are shown
in Figure~\ref{fig:u0vte}.  The median time scale of our events is
$\sim 40~\days$, about a factor of two higher than the median time
scale for events found by the MACHO and OGLE teams toward the Galactic
bulge \citep{alcock1997a, udalski2000}.  This is almost certainly a
selection effect caused by the fact that longer time scale events are
more likely to be alerted before peak magnification, and thus are more
likely to be chosen by us as targets for follow-up photometry.  This
is compounded by the fact that, for short time scale events, we are
less likely to get good coverage of the peak, even if they are alerted
pre-peak.  Events with poor or no peak coverage will often fail our
selection criterion of $<50\%$ fractional uncertainty in $\u0$. In
principle, this deficiency could be partially alleviated by including
MACHO and/or OGLE data.  However, in practice, we often stop observing
the event altogether if we do not get good peak coverage.  As we
discuss in \S\ref{sec:iandd}, the primary effect of this selection is
a bias toward slower lenses.

We also show in Figure~\ref{fig:u0vte} the differential and cumulative
distributions of $\u0$.  The median $\u0$ is $\sim0.2$, and the fraction
of high-magnification ($\amax>10$) events is $\sim 30\%$.
As it is a purely random quantity, the
intrinsic distribution of $\u0$ should be uniform.  The observed distribution 
of $\u0$, however, is clearly not uniform.  This is due to a combination of
various selection effects.  First, faint events are more likely to be
detected (and hence alerted) by the survey teams if they have a larger
maximum magnification \citep{alcock1997a, udalski2000}.  Since there are
more faint stars than bright stars, this results in a bias toward
smaller impact parameters with respect to a uniform distribution.
Second, since events with smaller impact parameters are also more
sensitive to planets, we preferentially monitor high-magnification
events.  This bias does not affect our
conclusions, since the value $\u0$ is unrelated to the presence or absence of
a planetary companion.  However, as emphasized by \citet{meands2000},
it does imply that in order to determine accurately  the overall
sensitivity of an ensemble of light curves to planetary companions, the
actual distribution of observed $\u0$ must be used.

\begin{figure*}[t]
\epsscale{1.0}
\centerline{\plotone{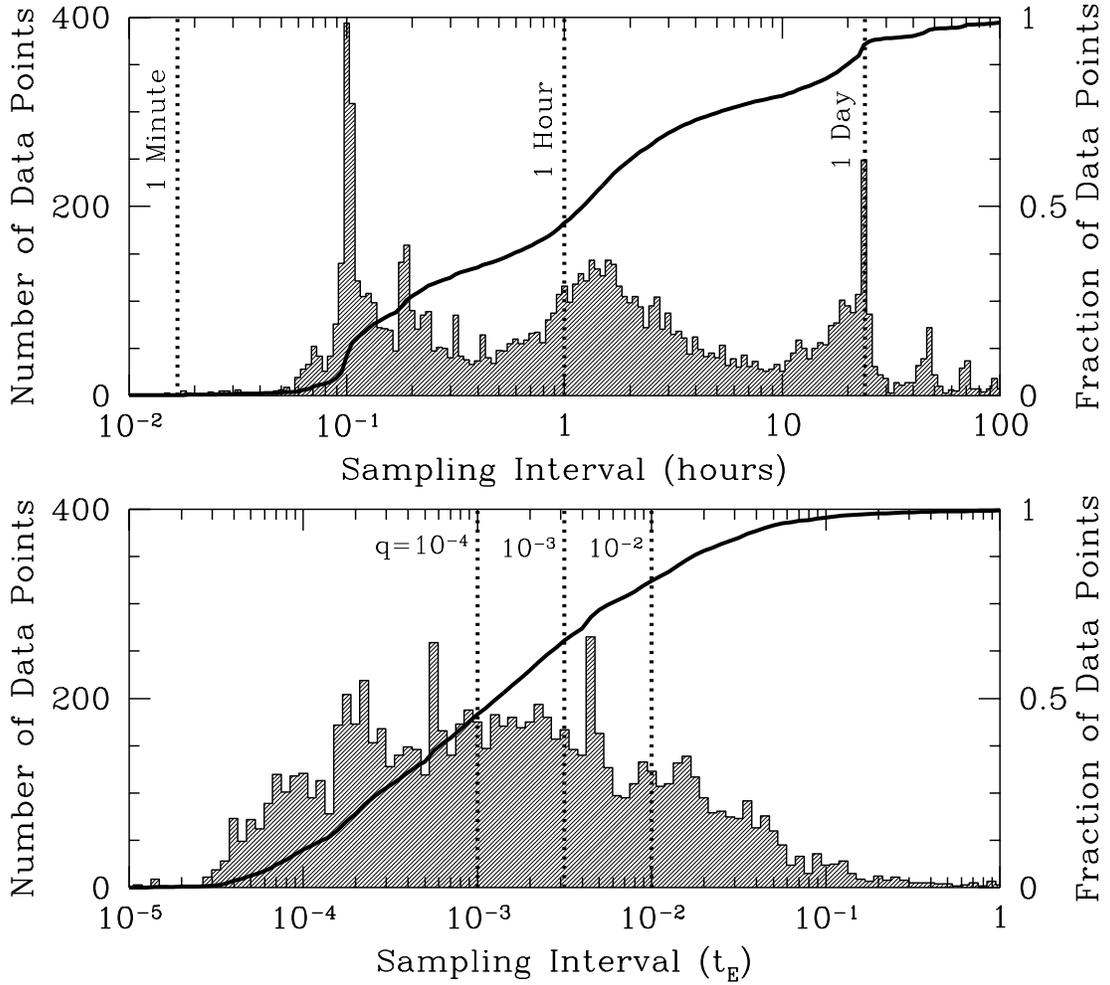}}
\caption{
\footnotesize
(a) The histogram shows the differential distribution of sampling intervals (in hours) for
our final event sample (left axis).  The solid line shows the
cumulative distribution (right axis).  (b) Same as (a), except in
units of $\te$.   The vertical dotted lines indicate the approximate minimum sampling rates
necessary for detection of companions of the indicated mass ratios.
}
\label{fig:dtdist}
\end{figure*}

Since one of the primary
goals of PLANET is to obtain very dense sampling of microlensing
events, it is interesting to examine how well this goal has been achieved.  In
Figure~\ref{fig:dtdist}, we show the distribution of sampling
intervals, that is, the time between successive exposures of a given
event.  Three peaks are evident.  The first at $\sim 6$~minutes is
our typical $I$-band exposure time of 5 minutes plus 1 minute of overhead time;
this peak is dominated by events that are followed continuously and
also pairs of $I$-$V$ data points.  The second peak at $\sim
1.5$~hours represents our fiducial sampling interval.  The third peak at 1~day
arises primarily from sampling of the wings and baselines of
light curves.   The median sampling interval is $\sim 1.5$~hours, with 90\%
of all data taken between 5~minutes and 1~day of one
another for a given event.  What is of particular relevance to the detection of planets
is the sampling interval in units of $\te$, which is shown in
the lower panel of Figure~\ref{fig:dtdist}.  Assuming that at least 10 data points are
needed on a planetary perturbation for detection, the sampling
interval needed to detect a companion of mass ratio $q$ is approximately,
\begin{equation}
\Delta t = 3\times 10^{-3} \te \sqrt{{ q \over 10^{-3}}}.
\label{eqn:sample}
\end{equation}
Using this formula and comparing to Figure~\ref{fig:dtdist}, we find
that (80\%, 65\%, 45\%, 25\%) of our data have sufficient sampling to detect
companions of mass ratio ($10^{-2}, 10^{-3}, 10^{-4}, 10^{-5}$).  Thus
we expect the majority of our data to have sufficient sampling to
detect companions with mass ratios $\ga 10^{-4}$.   This is not an accident, since PLANET observations are planned to have sensitivity to Jovian mass planets orbiting main sequence stars \citep{albrow1998}.

\begin{figure*}[t]
\epsscale{1.0}
\centerline{\plotone{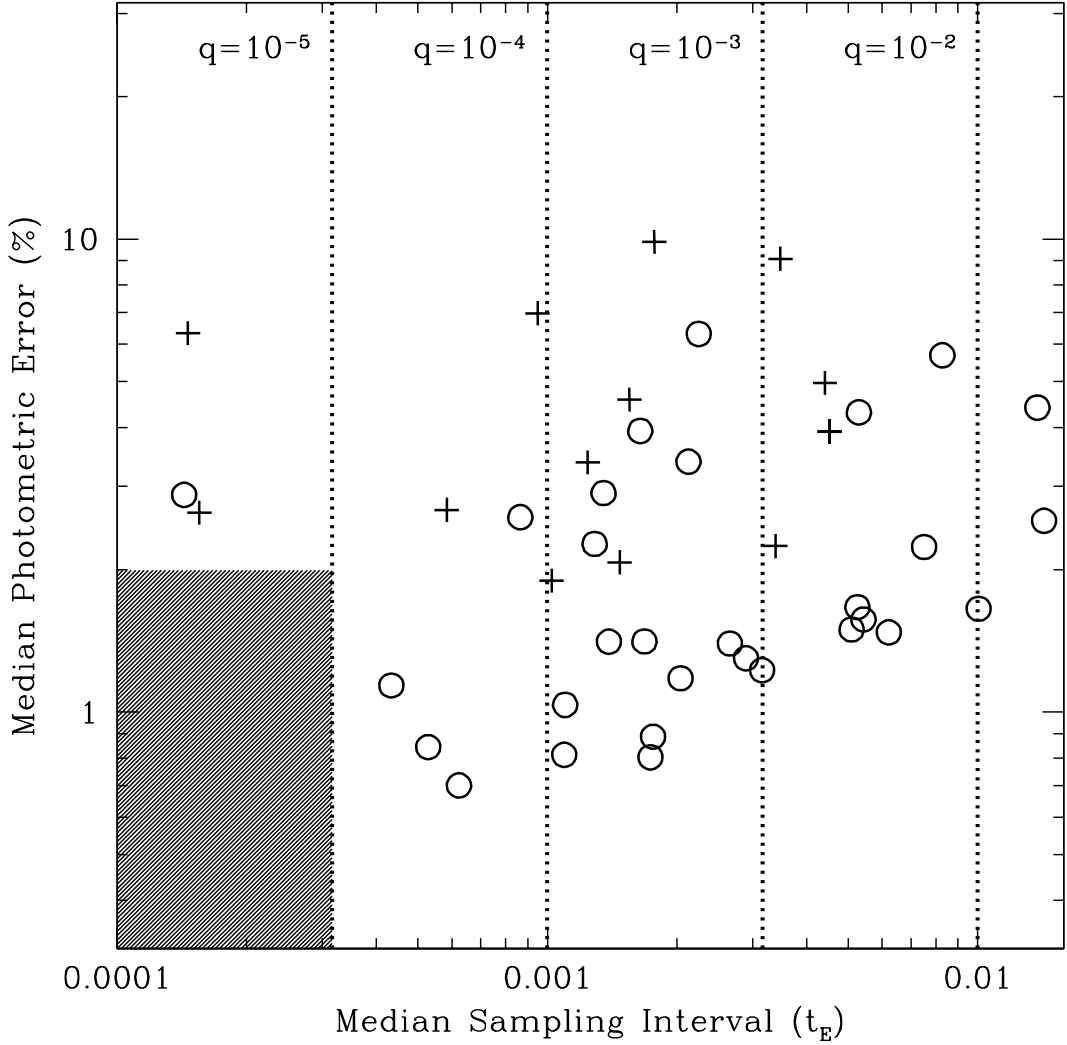}}
\caption{
\footnotesize
The median photometric error is plotted versus the median sampling
interval in units of $\te$ for our final event sample.  The plus signs
indicate high-magnification events ($\u0<0.1$ or $\amax>10$).  The
vertical dotted lines indicate the approximate minimum sampling rates
necessary for detection of companions of the indicated mass ratios.
The shaded box indicates approximately the median error and sampling needed
to have significant sensitivity to $10^{-5}$ mass ratio companions.
}
\label{fig:sigvdt}
\end{figure*}

The sensitivity of a given light curve to planetary companions is
primarily determined by three factors:  photometric errors,
temporal sampling, and impact parameter.  
In Figure~\ref{fig:sigvdt}, we plot the median photometric error,
$\sigma_{\rm med}$, 
versus the median sampling interval, $\Delta t_{\rm med}$ for all events; 
high-magnification ($\amax>10$) events are indicated.  These are also tabulated
in Table~\ref{tab:tab03}. High magnification events
that occupy the lower left quadrant of Figure~\ref{fig:sigvdt} will have the highest
sensitivity to planetary companions.  Of the 13 high-magnification
events, all have sufficiently small median sampling intervals to
detect $q=10^{-2}$ companions; we therefore expect our sensitivity to
such to companions to be quite high.  Two high-magnification events
have sufficient sampling rates to detect companions with $q=10^{-5}$;
however, for companions as small
as this, excellent photometry ($\la 2\%$) along with
excellent sampling is required  to obtain significant efficiency for
detection \citep{bandr1996}. No events satisfy both of these
requirements ($\sigma_{\rm med}<2\%$ and $\Delta t_{\rm med}/\te
<10^{-3.5}$).  We therefore restrict our attention to $q\ge10^{-4}$.

Considering the large number of high-magnification events, 
and the dense sampling and precise photometry, our
sample should be quite sensitive to planetary companions,
especially those with $q \ga 10^{-3}$.  This fact, combined with the
fact that no planetary-like perturbations are clearly evident in the
light curves, is an indication that such planetary companions are probably
not common.  In the following sections, we strengthen and quantify this statement.

\section{Search for Detections and Calculation of Detection Efficiencies\label{sec:sdes}}

Although a cursory inspection of Figure~\ref{fig:lca} reveals no
obvious candidate planetary perturbations, such perturbations could
be quite subtle, and thus missed by eye.  Furthermore, the
significance of the lack of planetary perturbations must be
quantified.  Specifically, the frequency with which planetary companions of given 
$d,q$ could be detected in individual light curves, the detection efficiency, must be
determined.  We simultaneously search for planetary
signatures in and determine the detection efficiency of individual
events using the method suggested by \citet{meands2000} and applied to microlensing
event \ob{98}{14} by \citet{albrow2000b}.  We briefly review the
algorithm here, but point the reader to these two papers for a more
thorough discussion of the method and its application.

\subsection{Algorithm\label{sec:algorithm}}

Of the $6+3N_l$ parameters (see \S\ref{sec:esel}) in a point source binary
microlensing fit, $3+3N_l$ have analogs in the PSPL fit:
$\te$, $\u0$, $\t0$ and one $F_{{\rm S},l}$, $F_{{\rm B},l}$, and $\eta_l$ for each of $N_l$
independent light curves.  The
parameters $\te$, $\u0$ and $\t0$ have different meanings in the
binary-lens model than in the PSPL model, and depend on the choice of the origin of the binary-lens
and the reference mass.  For small mass-ratio binaries, however, if
one chooses the origin to be the location of the primary lens, and
normalizes to the mass of the primary, then the values of these
parameters will be quite similar in a binary-lens and single-lens fit
to a light curve.  Three parameters are not included in the PSPL fit: 
the mass ratio $q$, the projected separation $d$, and the angle $\alpha$ of the
source relative to the binary-lens axis.  While $q$ and $d$
 are related to the physical nature of the
planet-star system, the angle $\alpha$ is a nuisance parameter which is of 
no physical interest. It is a random
geometric parameter and therefore uniformly distributed.  However, the
value of $\alpha$ does have a significant effect on the amplitude and
duration of the planetary perturbation.  Thus, some values of $\alpha$
lead to detectable perturbations to the PSPL model, while others do
not.  Marginalization over $\alpha$ for a given binary lens specified by
($q,d$) therefore determines the geometric detection efficiency
$\epsilon_i(d,q)$ for event $i$ and such a binary system.
Repeating this process for all $(d,q)$ pairs 
of interest yields the efficiency for all systems. This is the basis of the method of
determining the detection efficiency for individual events suggested
by \citet{meands2000}.

Operationally, the procedure to search systematically for planetary
signatures and determine $\epsilon_i$ for each event is as follows:
\begin{description}
\item[{(1)}] Fit event $i$ to the PSPL model, obtaining $\chi^2_{\rm
PSPL}$ (\S\ref{sec:esel}).
\item[{(2)}]  Holding  $d$ and $q$ fixed, find the binary-lens model
that best fits light curve $i$ for source
trajectory $\alpha$, leaving the $3+3N_l$ parameters
($\te, \umin, t_0, [\fs, \fb, \eta]N_l$) as free parameters.
This yields $\chi^2 (d,q,\alpha)$. 
\item[{(3)}] Repeat step (2) for all source trajectories $0\le \alpha
< 2\pi$. 
\item[{(4)}] Evaluate the difference in $\chi^2$ between the binary
and PSPL fits: $\dchi(d,q,\alpha)\equiv\chi^2(d,q,\alpha)-\chi^2_{\rm
PSPL}$.  Compare this to some threshold value $\dchit$:
\begin{description}
\item[{(a)}] If $\dchi(d,q,\alpha) < -\dchit$, then we tentatively
conclude we have a detected a planet with parameters $d,q,$ and $\alpha$. 
\item[{(b)}] If $\dchi(d,q,\alpha) > \dchit$ then the geometry
$(d,q,\alpha)$ is excluded.
\end{description}
\item[{(5)}]The detection efficiency $\epsilon_i(d,q)$ of event $i$ for the assumed
separation and mass ratio is then
\begin{equation}
\epsilon(d,q)\equiv{1 \over 2\pi} \int_0^{2\pi} {\rm d}\alpha\,\, \Theta[\dchi(d,q,\alpha)
- \dchit],
\label{eqn:ebq}
\end{equation}
where $\Theta[x]$ is a step function.
\item[{(6)}] Repeat steps (2)-(5) for a grid of $(d,q)$ values.  
This gives the detection efficiency $\epsilon_i(d,q)$ for event $i$ 
as a function of $d$ and $q$,  and also yields all binary-lens
parameters $(d,q,\alpha)$ that give rise to significantly better fits
to the event than the PSPL model.
\item[{(7)}] Repeat steps (1)-(6) for all events in the sample.
\end{description}

In step (2), we find the parameters ($\te, \umin, t_0, [\fs, \fb, \eta]N_l$ ) that minimize $\chi^2$ in the same way
as the PSPL fit:  we choose trial values of ($\te, \u0, t_0$) which
(along with the values of $d,q,\alpha$) immediately yield the
binary-lens magnification\footnote{For an explanation of how to
calculate the binary-lens magnification, see \citet{witt1990}.} as a
function of time, $A_{\rm B}(t)$.  This
is used to find the least-squares solution for the other
parameters, and the resultant $\chi^2$.  A downhill-simplex routine is then
used find the combination of parameters ($\te, \u0, t_0$) that
minimize $\chi^2$ (see \S\ref{sec:esel}).  The procedure is slightly
more complicated for those events for which MACHO and/or OGLE data was used
for the PSPL fit, as we discuss in \S\ref{sec:maog}. 

\begin{figure}
\epsscale{1.0}
\centerline{\plotone{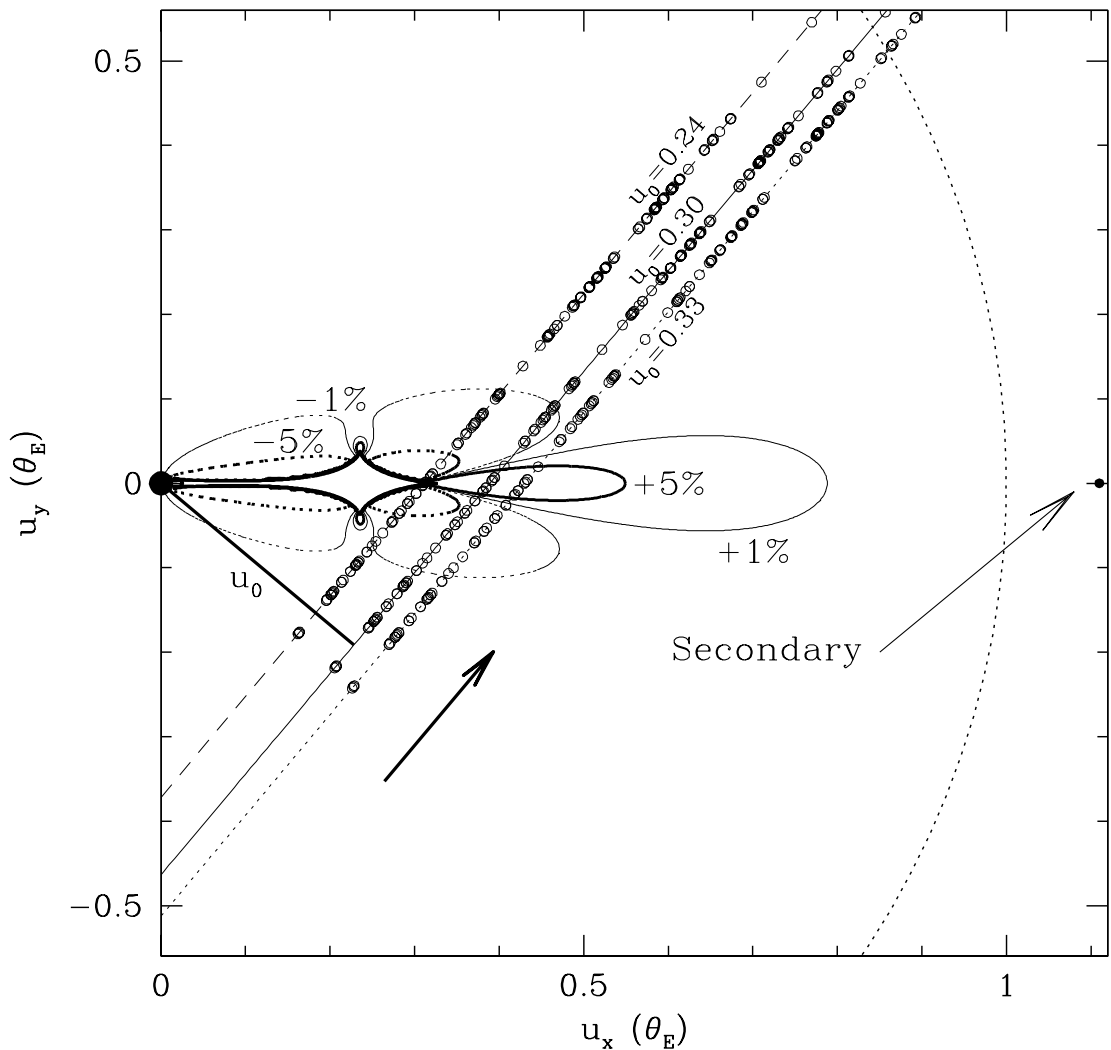}}
\caption{
\footnotesize
The vector positions in the source plane of the data points for event
\ob{98}{13}, assuming $\alpha=50^\circ$. 
 We plot these for the best-fit $\u0$ as determined from the PSPL fit,
$\u0=0.30$, as well as for the $\pm 4\sigma$ bounds on $\u0$.  The
arrow shows the direction of motion of the source with respect to the
lens.  The solid line connects the origin to the trajectory with
$\u0=0.30$ at time $t=\t0$.  Also shown are contours of constant
fractional deviation $\delta$ from the PSPL magnification for a mass ratio $q=10^{-3}$ and projected separation of $d=1.11$.  The solid
contours are $\delta=\infty, +5\%, +1\%$ (heaviest to lightest),
while the dotted contours are $\delta=-5\%, -1\%$ (heaviest to
lightest).  The solid black dots show the positions of the masses, the
large dot is the primary lens, the small dot the secondary.
}
\label{fig:u0const}
\end{figure}

Due to the perturbative nature of the planetary companion, for the
appropriate choice of the origin of the binary and the total mass of
the system, the majority of structure of the $\chi^2$ hypersurface
with respect to the parameters ($\te, \u0, t_0$) will be very similar
in the PSPL and the binary lens cases.  The two hypersurfaces will
only deviate significantly in some localized region of the ($\te, \u0,
\t0$) parameter space where the planetary perturbation from the PSPL
form is large. Consider a set of parameters $(d,q,\alpha)$ for which
the characteristic size of such a region in ($\te, \u0, \t0$) space is
much smaller than the intrinsic uncertainty of these parameters.
Since we find the binary-lens fit that minimizes $\chi^2$, rather than
integrating over the whole $\chi^2$ surface, our algorithm will find
best-fit parameters ($\te, \u0, \t0$) for the binary-lens model that
avoids this region without significantly increasing the $\chi^2$ with
respect to the single lens.  Thus we will always underestimate the
detection efficiency.  The amount the detection efficiency is
underestimated depends on how well $\te, \u0$, and $\t0$ are constrained.  
For events with poorly-constrained 
parameters, the efficiency can be underestimated by a significant amount
\citep{meands2000}.  This is illustrated in Figure~\ref{fig:u0const},
using event \ob{98}{13} as an example.  The fractional uncertainty in
$\u0$ for this event is $\sim 7\%$.  We show the vector positions in
the source plane of the data points for this event for the best-fit
$\u0$ as determined from the PSPL fit, along with the $\pm 4\sigma$
bounds on $\u0$\footnote{Note that the bounds on $\u0$ were calculated
by projecting the $\Delta\chi^2$ surface on $\u0$, rather than by the
linearized covariance matrix, as in Table~\ref{tab:tab02}.  In
general, the former method gives asymmetric bounds on $\u0$ due to the
$F_B\ge 0$ constraint, whereas the latter gives symmetric bounds by
definition. }  The data are more ``compressed'' in the Einstein ring
for values of $\u0$ smaller than the best-fit value because $\te$ is
anti-correlated with $\u0$, and thus smaller $\u0$ implies larger
$\te$.  For reference, we also show contours of constant fractional
deviation from a single lens for a binary with $q=0.001$ and $b=1.11$.
It is clear that the difference in $\chi^2$ between the binary-lens
and single-lens fits will differ substantially between these three
fits.  Our algorithm will always choose the one that minimizes
$\chi^2$, and thus will underestimate the efficiency.  This could in
principle be avoided by integrating over $\u0, \t0$, and $\te$, rather
than evaluating $\chi^2$ at the best-fit parameters. However, for the
large number of binary-lens geometries we test (see \S\ref{sec:grid}),
this is not computationally feasible.  These underestimated detection
efficiencies could be a serious problem if planetary deviations were
detected, as they would lead to an overestimate of the true number of
planets.  However, as we show in \S\ref{sec:detections}, we do not
detect any planetary deviations.  Thus, the underestimated
efficiencies represent conservative upper limits.

\subsection{Implementation of the Algorithm}
Although the algorithm described in \S\ref{sec:algorithm} is conceptually simple and appears
straightforward, there are some subtle details that must be addressed before
implementation.  Specifically, in the following subsections we discuss
photometric errors, the inclusion of MACHO/OGLE photometry, the grid size and spacing for the
binary-lens parameters $d$, $q$, and $\alpha$, the method by which the
binary-lens magnification is evaluated, and the choice of the
detection threshold $\dchit$. 

\subsubsection{Photometric Errors\label{sec:cofdata}}

As we discussed in \S\ref{sec:esel}, the errors reported by DoPHOT
are typically underestimated by a factor of $\sim 1.5$; adopting such
errors would both overestimate the significance of any planetary
detections, and overestimate the detection efficiency.  Furthermore,
since events can have error scaling
factors that differ by a factor of three, even the relative significances
for different events would not be secure.  Ideally, one would like to
determine the magnitude of the photometric errors without reference to
any model.  Unfortunately, this is not possible in general, primarily because the
error depends strongly on the local crowding conditions of the microlensing
source object in a manner that is impossible to access a priori.  
Therefore, in order to put all events
on the same footing and to arrive at the best possible estimate of
the significance of planetary detections and detection efficiencies, we
adopt the error scaling factors as determined in the PSPL fit
(see \S\ref{sec:esel}).
We typically find that, after scaling in this way, the error distributions 
are nearly Gaussian, with the exception of a small handful of large
$>3\sigma$ outliers \citep{albrow2000b}.

If the PSPL model is truly the ``correct'' model, this procedure is 
valid, and does not bias the results in any way.  However, if the
light curve actually deviates from the PSPL model, this procedure will
overestimate the error scaling factors, and thus {\it underestimate}
the significance of the anomaly.  Assuming that binary-lens model is
correct, it is straightforward to show that
the true difference in $\chi^2$, which we will label  $\dchi_0$, is related to the $\dchi$
evaluated assuming the PSPL fit is correct by,
\begin{equation}
\dchi_0=\dchi \left(1 - {\dchi\over {\rm d.o.f.}}\right)^{-1},
\label{eqn:dchiover}
\end{equation}
where ${\rm d.o.f.}$ is the number of degrees-of-freedom of the event. Thus
for an event with $\sim 300$ data points and $\dchi=60$, using the
errors determined from the PSPL fit would lead us to underestimate
the ``true'' $\dchi_0$ by $20\%$.  For events with a small number of
d.o.f., this underestimate can formally be as large as $100\%$.
This would seem to argue that the values of $\chi^2$ computed in all
fits (PSPL and binary) should be renormalized by the best-fit model
(PSPL or binary).  However, there are several reasons we feel this is not
appropriate.  First, for any fit, $\chi^2$ is {\it not}
dominated by the number of ${\rm d.o.f.}$: instead, typically only a handful
of large outliers contribute a significant fraction of the
evaluated $\chi^2$.  Thus, in reality ${\rm d.o.f.}$ should be
replaced by $\chi^2_{\rm binary}$ in \eq{dchiover}, which is typically
larger by $\sim 100$, thus reducing the underestimate considerably.
Furthermore, renormalizing $\chi^2$ in this way would give extra
weight to binary-lens models that ``succeed'' by fitting isolated
large-$\sigma$ outliers, particularly for events with a small number
of data points, where $\chi^2$ is dominated by such outliers.  The
smaller the number of data points, the more difficult it is to
objectively judge the reality of such fits.  Although some of these
biases could in principle be calibrated by Monte Carlo techniques,
i.e.\ by inserting many artificial planetary signals into the
light curves, and then repeating the algorithm on all of these
artificial datasets, in practice the large number of fits required
(see \S\ref{sec:grid}) makes this computationally
prohibitive. Furthermore, it is difficult to address the effects of
large-$\sigma$ outliers in this way.  We will therefore adopt the conservative
and simpler choice of using the errors determined with reference to
the PSPL model in order to avoid the danger of detecting spurious
planets in data with isolated outliers in sparse datasets.

\subsubsection{Including MACHO/OGLE Data\label{sec:maog}}

As discussed in \S\ref{sec:esel}, we include
MACHO and/or OGLE data for some events in order to better constrain $\u0$. This is
necessary in order to robustly determine $\epsilon_i$ for 
events for which our data are poorly sampled near the peak or do not
have baseline information.  However, as we do not have access to these
raw data, nor do we know the details of the data reduction procedures,
we have no way of
independently judging the quality of the MACHO or OGLE photometry.
Furthermore, we do not have access to the seeing values for these data,
and hence cannot correct for the seeing correlations that can often
mimic low-amplitude planetary deviations.  Thus any planetary
``signal'' discovered using this photometry would be difficult
to interpret, and the reality of the signal impossible to determine.
Therefore, while we use these data to constrain the
global parameters $\te$, $\t0$ and $\u0$, we do not use these
data in either the search for planetary signatures or the calculation of the
planet detection efficiency.  We accomplish these goals in the
following manner.  

All information on the parameters $\te$, $\t0$, $\u0$ and their 
covariances with other parameters is contained within the covariance matrix
$c_{ij}$ and the vector $d_{i}$ as determined from the PSPL fit
with all parameters [see \S\ref{sec:esel} and \Eq{bij}]. Therefore,
we simply need to extract the information provided by
the MACHO/OGLE data and apply it to the binary-lens fit with only
PLANET data.  First we calculate the
covariance matrix $c_{ij}$ of the best-fit parameters $a_i=(\t0,
\te, \u0, F_{{\rm S},1}, F_{{\rm B},1}, \eta_1, F_{{\rm S},2}, F_{{\rm B},2}, \eta_2, ...)$
as determined by the PSPL fit to all (MACHO+OGLE+PLANET) data.  Note
that this is identical to the procedure used in \S\ref{sec:esel} to
calculate the uncertainties of $a_i$.  We then restrict $c_{ij}$ and $a_{i}$
to the parameters $F_{{\rm S},l}, F_{{\rm
B},l}, \eta_l$ for PLANET data.  We call these
restricted quantities $c_{ij}^{\rm MOP}$ and $a_{i}^{\rm MOP}$.  
We calculate the covariance matrix $c_{ij}^{\rm P}$ of the best-fit parameters $a_i^{\rm P}$
determined from the PSPL fit to only PLANET data, again restricting
these quantities to the parameters $F_{{\rm S},l}, F_{{\rm
B},l}, \eta_l$.  Next, we form
the matrix and vector,
\begin{equation}
b^{\rm MOP}\equiv\left(c^{\rm MOP}\right)^{-1}\qquad d_i^{\rm MOP} \equiv \sum_j
b_{ij}^{\rm MOP} a_{j}^{\rm MOP},
\label{eqn:MOP}
\end{equation}
and similarly for $b_{ij}^{\rm P}$ and $d_i^{\rm P}$.  Finally, we
calculate,
\begin{equation}
b_{ij}^{\rm MO}=b_{ij}^{\rm MOP}-b_{ij}^{\rm P}\qquad d_i^{\rm
MO}=d_i^{\rm MOP}-d_i^{\rm P}.
\label{eqn:MO}
\end{equation}
The resultant matrix $b_{ij}^{\rm MO}$ and vector $d_i^{\rm MO}$
contain only the information on $\t0, \te, \u0$ and the parameters $F_{{\rm S},l}, F_{{\rm
B},l}, \eta_l$ for PLANET data provided by the MACHO/OGLE data.  We
then use these two quantities to constrain the binary-lens fits using
PLANET data only in the following manner.  For each trial $\t0, \te, 
\u0$, we compute $b_{ij}$ and $d_i$ for the quantities $F_{{\rm S},l}, F_{{\rm
B},l}, \eta_l$ using only PLANET data.  We add to these the
constraints from MACHO/OGLE by forming
\begin{equation}
b_{ij}^{\rm cons}=b_{ij}+b_{ij}^{\rm MO}\qquad d_i^{\rm cons}=d_i+d_i^{\rm MO},
\label{eqn:bconst}
\end{equation}
which are then used to find the best-fit parameters 
$a_i=(F_{{\rm S},1}, F_{{\rm B},1}, \eta_1, F_{{\rm S},2}, F_{{\rm
B},2}, \eta_2, ...)$ via \eq{alphas}.  The $\chi^2$ of the resultant fit is then evaluated. 
We add to this $\chi^2$ a contribution,
\begin{equation}
\chi^2_{\rm MO}\equiv\sum_{ij} \delta a_i b_{ij}^{\rm MO} \delta a_j
\label{eqn:chi2pen}
\end{equation}
where $c^{\rm MO}=(b^{\rm MO})^{-1}$ and 
\begin{equation}
\delta a_i = a_i - a_i^{\rm MO}, \qquad a_i^{\rm MO}=\sum_j c_{ij}^{\rm MO} d_j^{\rm MO}.
\label{eqn:deltaa}
\end{equation}
The contribution $\chi^2_{\rm MO}$ to $\chi^2$ is a penalty for violating the constraints from MACHO/OGLE
data.  The remainder of the fitting procedure is as before:  this
$\chi^2$ is then used by the downhill-simplex routine AMOEBA \citep{cookbook} to find
the parameters $\t0$, $\u0$ and $\te$ that minimize $\chi^2$ for the
particular $d,q,\alpha$ binary-lens geometry.

\subsubsection{Grid of Binary-Lens Parameters\label{sec:grid}}

Several factors dictate our choice of grid size and
spacing in $d,q,\alpha$ parameter space.  First, the grid spacing
must be dense enough to avoid missing possible
planetary signals and prevent sampling errors from dominating the uncertainty in 
$\epsilon_i$.  Second, the grid
must cover the full range of parameter space for which we have significant
sensitivity.  Finally, the computation must be performed in
a reasonable amount of time.  

We restrict our attention to $10^{-4} \le q \le 10^{-2}$.  The upper
end of this range is dictated by the fact that we are primarily
interested in planetary companions, and also because our procedure for
finding binary-lens fits fails for events that are grossly deviant
from the PSPL form.  In fact, finding all satisfactory fits to such binary-lens
light curves is quite difficult (see \citealt{mandi1995,diandp1998,albrow1999b}).
We do detect binaries well fit by $q>0.01$.  Incorporating such
binaries into the analysis would entail finding {\it all} possible fits
to these observed binaries {\it and} calculating the efficiency of all
other events.  Although such a study is interesting in its own right,
it would be quite an undertaking, well beyond the scope of this paper.
The lower end of the range of mass ratios we test is dictated by the
fact that we are unlikely to have significant sensitivity below
$q=10^{-4}$ (\S\ref{sec:echars}). We sample $q$ at equally
spaced logarithmic intervals of 0.25. 

Numerous studies \citep{gandl1992,diandm1996, bandr1996, gands1998, mps, albrow2000b} 
have shown that the planetary detection probability is largest in the
``lensing zone,'' $0.6 \le d \le 1.6$, and is negligible for 
$d\la 0.1$ and $d \ga 10$.  Furthermore planetary perturbations exhibit a $d \rightarrow d^{-1}$
symmetry \citep{gandg1997,gands1998, martin1999}.  Therefore, we
sample $d$ at $0.1, 0.2,..., 0.9, 1.0$, and also the inverse
of these values, for a full range of $0.1 \le d \le 10$. 

In order to avoid missing any possible planetary signals, we
choose a variable step size for $\alpha$ that depends on $q$.  The
size of the region of significant perturbation is $\sim q^{1/2}$, and
thus a perturbation at the Einstein ring radius would cover an
opening angle with respect to the center of the primary lens of $\sim q^{1/2}$.
Therefore in order to sample the perturbed region at least twice, we
choose a step size of
\begin{equation}
\Delta \alpha = { \sqrt{q} \over 2}.
\label{eqn:alphastep}
\end{equation}
For every $d,q$ pair, we thus find the best-fit binary-lens model for a total of 
$4 \pi q^{-1/2} \sim 400 (q/10^{-3})^{-1/2}$ choices of $\alpha$.

\subsubsection{Magnification Maps\label{sec:maps}}

With the grid size and spacing described in \S\ref{sec:grid}, 
we perform a total of $8.8 \times 10^{4}$ binary-lens fits
to each event, for a grand total of $3.8 \times 10^{6}$ fits for all 43 events .
Each fit requires at least 50 evaluations of the binary-lens
magnification light curve to converge, for a total of more than $10^8$
binary-lens light curve evaluations. Given this large number
of evaluations, re-evaluating the magnification for each data point of each
event is both
prohibitive and inefficient.  We therefore first create
magnification maps for each of the $d,q$ grid points, and interpolate
between these maps to evaluate the binary-lens magnification.  Maps are generated for 
source positions $-2 \le x \le 2$ and $-2 \le y \le 2$ (in units of $\thetae$).  
For source positions outside this range, we use the PSPL
magnification.  For a binary with $q\ll 1$ and $d \ne 1$, there are two sets of caustics.  The
``central caustic'' is always located at the position of the primary, i.e.\ $x=0,~y=0$.  The 
 ``planetary caustic(s)'' are separated from the primary by an amount $|d^{-1}-d|$.  Therefore by only evaluating the binary-lens magnification for source positions in the ranges above, 
we are implicitly assuming
that we are not sensitive to the
planetary caustics of companions with separations $d \la 0.4$ and $d \ga
2.4$, although we are still sensitive to such planets via the central
caustic.  This assumption is essentially correct since the vast majority ($\sim 95\%$) of the 
data was taken within $\le 2\te$ of the peak.
To generate the maps, the source position is sampled at intervals of 
$2 \times 10^{-3}\thetae$, the typical sampling interval of our events (\S\ref{sec:echars}). 
We have performed numerous tests comparing fits using these maps and fits
using the exact binary-lens magnification, and find that using the
maps introduces an error of $\dchi \la 2$, which is far below any of
our thresholds $\dchit$.   Typically, efficiencies determined using
these maps are in error by $\la 1\%$.  We have also inserted planetary
deviations into selected light curves, and confirm that these
``detections'' are recovered when the maps are used to evaluate the
magnification. 

\subsubsection{Choice of Detection Threshold\label{sec:threshold}}

Ideally, one would like to choose the detection threshold $\dchit$ a priori,
without reference to the results of the binary-lens fits.
Specifically, one would like to be able to determine the probability
$P(\ge \dchi)$ of obtaining a given $\dchi$ or larger by chance, and then choose a
probability threshold for detection, say $P=0.01$.  Naively, one might
expect that the probability of getting a certain value of $\dchi$ or larger by chance is
given by, 
\begin{equation}
P(\ge \dchi)= \left(2\pi\right)^{-1/2}\int_{\dchi}^\infty {\rm d} x~x^{1/2} {\rm e}^{ -x/2},
\label{eqn:pgauss}
\end{equation}
for the three extra binary parameters ($d,q,\alpha$), assuming they
are independent and have Gaussian distributed uncertainties.  However,
this formula fails for several reasons.  First, most events contain
large outliers that are not described by Gaussian statistics.  Second,
and more importantly, such a naive calculation fails to take into
account the fact that many independent trial binary-lens fits to the
datasets are being performed, thereby effectively increasing the
difference in the number of
degrees-of-freedom between the binary and single lens models.  In
other words, while the success of a single binary lens model is given
by \eq{pgauss} in the limit of Gaussian errors, the success of {\it
any} binary-lens model is not.  Unfortunately, the effect of this increase in the
effective number of degrees-of-freedom on the probability cannot be
assessed analytically, and must be determined via a Monte Carlo simulation.
This would entail generating many different realizations of synthetic
events with sampling and errors drawn from the sampling and error distributions  
of each of the 43 events in our sample.  The algorithm in \S\ref{sec:algorithm}
would then need to be performed on each of these synthetic events, in
order to determine the mapping $P(\ge \dchi)$ for each event.  Given that each event
requires $\sim 10^{5}$ binary-lens fits, this is clearly
impossible.  Furthermore, as we demonstrate \S\ref{sec:detections}, it is likely that 
unrecognized systematics exist in the data which give
rise to temporal correlations in the fluxes of 
observed light curves.  These systematics will result in false detections.  The rate of  
such false detections cannot
be recovered with Monte Carlo simulations of synthetic light curves unless the actual temporal correlations (which are not understood) are introduced in these light curves.
We therefore use the distribution of $\dchi$ from the actual
events to choose $\dchit$, as described in the next section.

\subsection{Detection Threshold and Candidate Detections\label{sec:detections}}

\begin{figure*}[t]
\epsscale{1.0}
\centerline{\plotone{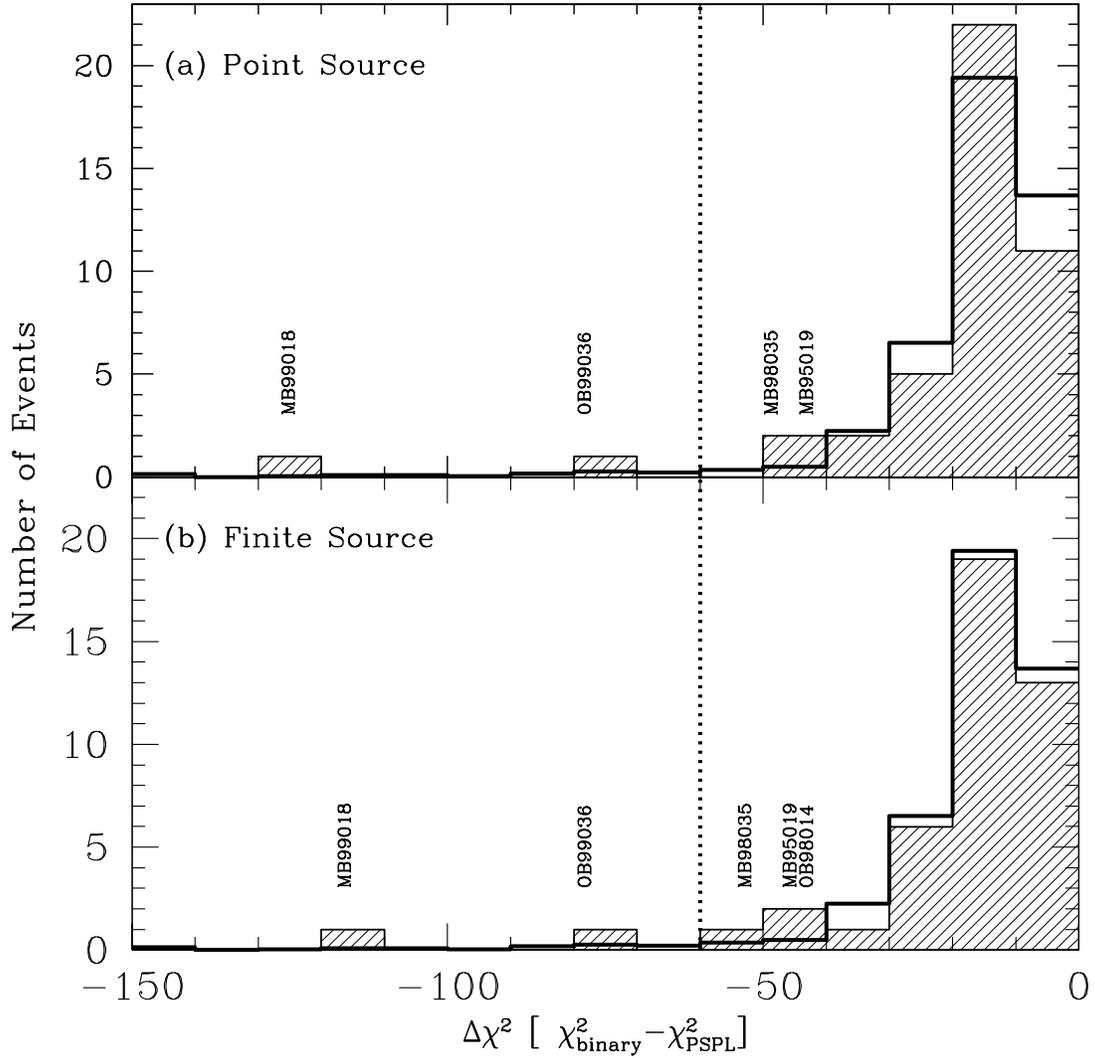}}
\caption{
\footnotesize
The shaded histogram shows the distribution of the difference in
$\chi^2$ between the best-fit binary-lens model in the range
$q=10^{-2}-10^{-4}$, and the point-source point-lens fit.  Events with
$\dchi<-40$ are labelled.  The dotted line is our adopted
detection threshold, $\dchit=60$.  The unshaded, bold histogram is
the distribution of $\dchim$ found from a Monte Carlo analysis of constant 
light curves.  See \S\ref{sec:detections}. 
(a) Binary-lens models in which the source is assumed to be
point-like. (b) Binary lens models in which the source is assumed to have
the dimensionless size $\rho_*$ given in Table~\ref{tab:tab04}.
}
\label{fig:detdist}
\end{figure*}

We have applied the algorithm presented in \S\ref{sec:algorithm}
for all 43 events in our final sample.  For each event, we find the
absolute minimum $\dchim$ from this procedure.  The distribution of
these $\dchim$ is shown in Figure~\ref{fig:detdist}.  If all the events
harbored planets, we would expect a continuous distribution in
$\dchi$ extending to very large negative values.  If some fraction of events
harbored planets, then we would expect a large ``clump'' of small
$\dchim$ obtained from single events through statistical fluctuations,
and then a few scattered instances of large $\dchim$ from those events
with companions.  In fact most of the
events have $\dchim \ga -60$, with only two events, \mb{99}{18} and \ob{99}{36},
having $\dchim \le -60$.  We therefore 
interpret the binary-lens fits with $\dchim>-60$ to be arising from statistical
fluctuations or unrecognized low-level systematics, and choose $\dchit=60$ as a reasonable threshold for
detections.

To establish the plausibility of our choice of $\dchit$, we
perform a simplistic Monte Carlo simulation.  For one observatory and filter,
we extract 1000 light curves of stars in the field of a typical
microlensing event.  These stars span a large range of brightness
and local crowding conditions.  The overwhelming majority of these
stars have constant brightness, although a handful are almost
certainly variables.  We reduce and post-process these light curves in
the same manner as the microlensing events (\S\ref{sec:data}), using a
constant flux model with seeing correlation correction to rescale the
errors.  Outliers ($>3\sigma$) are included, but not used to determine
the error scaling.  We then fit each of these light curves to the model
designed to mimic the deviation induced by a planetary companion:
\begin{equation}
F(t_k)=F_{\rm S} \left[1 + \delta_0 \exp( -\tau_k^{2})\right] + \eta
[\theta(t_k)-\theta_{0}],\qquad \tau_k=(t_k-\t0)/\tp.
\label{eqn:falmodel}
\end{equation}
This model has a deviation from constant flux with a maximum amplitude
of $\delta_0$ at a time $\t0$, and a characteristic duration $\tp$.  We
vary $\delta_0$ in 80 steps $\delta_0=-20\%$ to $20\%$, 
$\t0$ in 30 steps between the minimum and maximum time of observations,
and $\tp$ in 30 logarithmic steps between $10^{-1}$ and $10^{-4}$ of the total duration of the
observations, for a total of $7.2\times 10^4$ trial combinations.  This is similar to the number of binary-lens fits performed for each microlensing event. 
For each $\delta_0, \t0,$ and $\tp$, we find the
best-fit values of $F_{\rm S}$ and $\eta$, and calculate $\chi^2$.
This is repeated for all sampled values of ($\delta_0, \t0, \tp$) and
the minimum $\dchim$ between the best fit to the model in
\eq{falmodel} and the constant flux model determined for each of the
1000 light curves.  In Figure~\ref{fig:detdist}, we show the resulting
distribution of $\dchim$, normalized to 43 events.  The similarity to
the distribution of $\dchim$ of the microlensing events is remarkable.
We conclude that it is quite likely that 
the binary-lens fits with $\dchim>-60$ arise from statistical
fluctuations or unrecognized low-level systematics, and that our choice of $\dchit$
is reasonable.

Based on this choice of $\dchit=60$, we tentatively conclude that we have
detected anomalies consistent with planetary deviations in events \mb{99}{18} and \ob{99}{36}.  We have
examined both events individually, and find other, more likely, explanations for their anomalous behavior
which we now describe in some detail. 

The light curve for \ob{99}{36} shows an overall asymmetry will respect to the
time of maximum magnification.  This asymmetry is well fit by the
distortion to the overall light curve created by a
planetary companion to the primary lens with $q=0.003$.  However, such
a distortion requires a special geometry, specifically 
$\alpha \sim 0$ or $180^{\circ}$, i.e.\ a source trajectory nearly
parallel to the planet-star axis.  All other values of $\alpha$
produce either no asymmetry or a planetary ``bump.''  
Asymmetries like that of \ob{99}{36} are a generic feature of low-amplitude parallax effects
\citep{gbm1994}; indeed the event is equally well-fit by a parallax model.
Typically, parallax effects are only significant in long time scale events ($\te \gsim 100~{\days}$),
and thus it would seem unlikely that, for typical lens masses and distances,  
such effects should be detectable in the light curve of 
\ob{99}{36}, which has $\te \sim 30~{\days}$.  However, as we describe in
Appendix~\ref{sec:parallax}, the parameters we derive are reasonable: the asymmetry is quite small, and only detectable due to the excellent data quality of the event.
Since both models fit the data equally well, we conclude that we
cannot reliably distinguish between them, although we favor the
parallax interpretation based on the fact that the planetary fit
requires a special geometry and a parallax signal must be present at some level 
in all light curves due to the motion of the earth around the sun.  
We therefore conclude that we cannot robustly
detect a planet from an asymmetry that is equally well-fit by parallax.  
This in turn implies that all
planetary perturbations consistent with such an overall asymmetry should be
ignored in the efficiency calculation for all events.  Although we have not done
this, we have performed simulations which demonstrate that by not
doing so, we overestimate our efficiencies by only a few percent,
which is small compared to our statistical uncertainties.  The
parallax and planetary fits to \ob{99}{36}, as well as a detailed account
of these simulations are presented in Appendix~\ref{sec:parallax}.

The light curve of \mb{99}{18} displays a $\sim 15\,$day
anomaly of amplitude $\sim 2\%$.  Such an anomaly is longer than that
expected from planets with $q\la 0.01$, and we therefore
systematically explored binary-lens fits with $0.01\le q \le 1$.  This
uncovered a fit with $q\sim 0.2$ that is favored over the best-fit
planet ($q = 0.01, d =0.8$) by $\Delta\chi^2=22$.
Clearly we cannot claim detection of a planet when a roughly 
equal-mass binary model provides a substantially better fit.
However, since $\Delta\chi^2=22$ is below our normal threshold 
($\Delta\chi^2=60$), we must estimate the probability that in
excluding  MACHO~99-BLG-18 from the analysis, we have inadvertently
thrown out a real planetary detection.  Naively, this probability
is $\exp(-\Delta\chi^2/2)\sim 10^{-5}$, but we have already seen
that unknown systematic effects generate a whole range of planet-like
perturbations at the $\Delta\chi^2\la 50$ level.  An upper limit to 
the probability that a planetary light curve has been corrupted to look like an equal mass
binary can be estimated directly from the data. It is
$P \le f_{\rm apriori}f_{22}$ where $f_{22}\sim 20\%$ is the
fraction of events with $\Delta\chi^2<-22$, and 
$f_{\rm apriori}$ is the {\it a priori} probability that the event contains
a planet that is being corrupted by systematic effects into a $q >0.01$ binary, 
rather than a true $q> 0.01$ binary.  This last quantity
is unknown, but since we detect of order 10 other binaries and no
other planets, $f_{\rm apriori}$ is certainly less than 50\%. 
Thus $P \la 10\%$.
This probability is smaller than the statistical errors on our
resultant limit on planetary companions from the entire sample of
events.  Thus, excluding MACHO~99-BLG-18 as a binary causes us to
overestimate our sensitivity to planets, but by an amount that is
smaller than our statistical errors.
 
Thus, out of an original sample of 43 events, we are left with 42
events (rejecting \mb{99}{18}), and no viable planet candidates.  Given
this lack of detections, we can use the individual event detection
efficiencies $\epsilon_i$ to determine a statistical upper limit to
the fraction of lenses with a companion in the range of $d,q$
parameter space that we explore.

\subsection{Detection Efficiencies\label{sec:des}}

\begin{figure*}[t]
\epsscale{0.7}
\centerline{\plotone{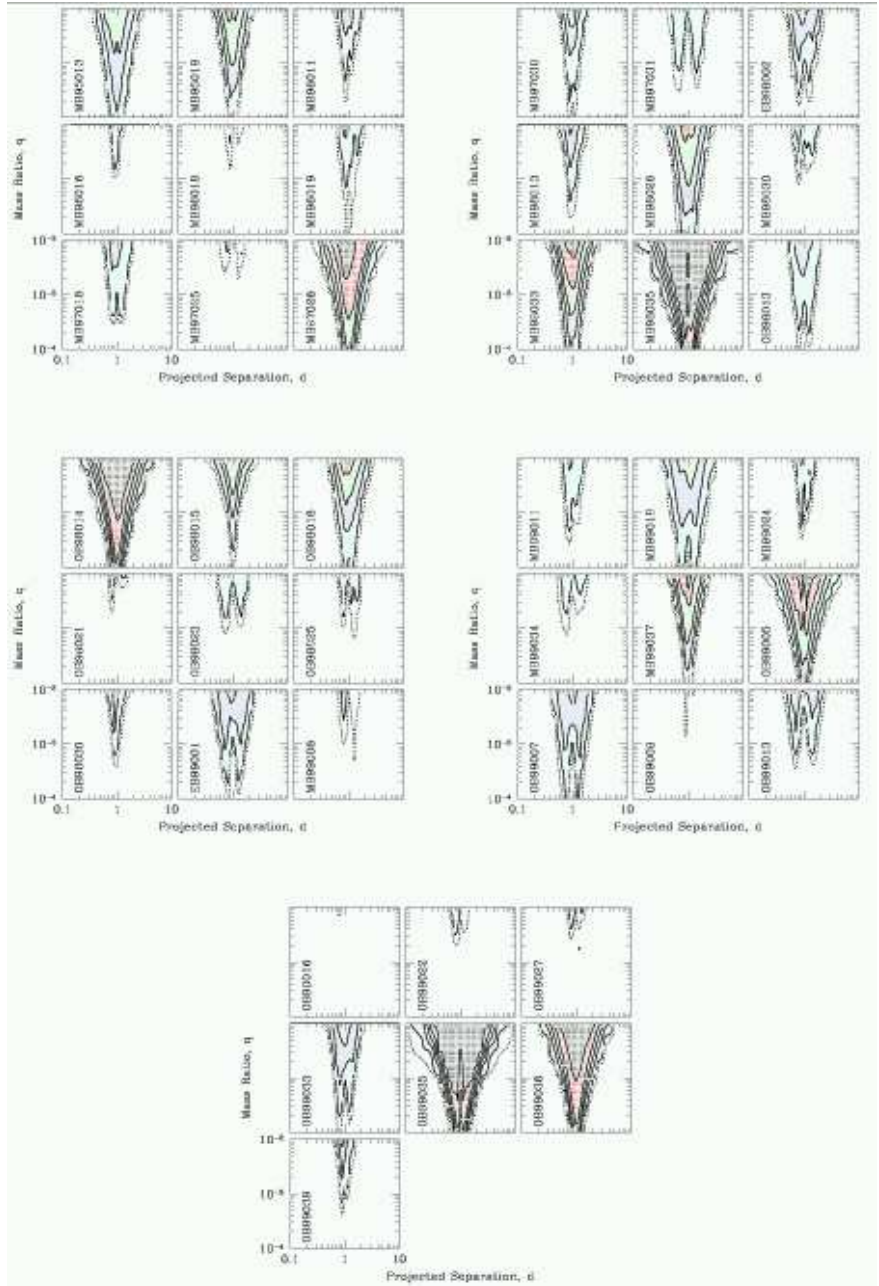}}
\caption{
\footnotesize
Black lines are contours of constant detection efficiency,
$\epsilon(d,q)$, shown for projected separations $d$ 
between the primary and companion in units of
the Einstein ring radius, of $-1 < \log{(d)} \le 1$, and mass ratios
between the primary and companion, $q$, of $-2 > \log(q) >-4$. 
Contours mark $\epsilon=1\%$(outer contour; dotted), $5\%$, $25\%$, $50\%$, $75\%$,
and $95\%$ (inner contour).  Each panel is for a separate event; the
abbreviated event name is indicated in each panel.  The
``wiggly'' nature of the outer contours apparent in some events is an
artifact of the $(d,q)$ sampling and the plotting routine. Point sources have been assumed here.
}
\label{fig:des}
\end{figure*}

The detection efficiency $\epsilon_i(d,q)$ is the probability that a
companion with mass ratio $q$ and projected separation $d$ would
produce a detectable deviation (in the sense of $\dchi\le-\dchit$) in the
observed light curve of event $i$.  Figure~\ref{fig:des} shows $\epsilon_i(d,q)$ for
our fiducial threshold $\dchit=60$ 
and all our events in the parameter range we searched for companions, $0.1
\le d \le 10$ and $10^{-4} \le q \le 10^{-2}$.  

We have plotted
$\epsilon$ as a function of $\log{d}$, which clearly reveals the $d
\rightarrow d^{-1}$ symmetry inherent in planetary perturbations
\citep{gands1998, martin1999}.   For low magnification events ($\u0
\ge 0.1$), the efficiency exhibits a ``two-pronged'' structure as a
function of $d$, such that the efficiency has two distinct maxima, one
at $d_{\rm \epsilon,max} < 1$ and one at $d^{-1}_{\rm \epsilon,max}$,
and a local minimum at $d=1$.  The approximate locations of these maxima
can be found by determining the separations at
which the perturbation due to the planetary caustic occurs at the peak of the light curve,  
\begin{equation}
d^{\pm 1}_{\rm \epsilon,max} \approx {1 \over 2} \u0 \mp {1 \over 2} \sqrt{\u0^2 + 4}.
\label{eqn:dmax}
\end{equation}
For planetary separations $d_{\rm \epsilon,max} < d < d^{-1}_{\rm
\epsilon,max}$, the caustics produced by the companion are within a
radius $\u0$ of the primary lens, and are thus not well probed by the event.
For high magnification events, $\epsilon$ is maximized near $d=1$.
This is not only a consequence of \eq{dmax}, but also because the
central caustic is being probed by the event.  As expected, the detection
efficiency to companions with any $q$ and $d\la 0.2$ or $d\ga 5$ is negligible
for nearly all events. 

Of the 43 events, $13$ have very little detection efficiency: for these events,
$\epsilon(d,q)$ is larger than 5\% for only the most massive
companions, and never gets larger that 25\%. For the most part, these
low efficiencies are due to poorly constrained $\u0$.   Eight events, notably all 
high-magnification events, have excellent
sensitivity to companions and exhibit $\epsilon(d,q)>95\%$ for
a substantial region in the $(d,q)$ plane.  Our
resultant upper limits on small mass ratio $q\la 10^{-3}$ companions (\S\ref{sec:ulimits}) are
dominated by these 8 events.  For the
remainder of the events, the efficiency is substantial ($\ga 25\%$) for some regions
of parameter space.  These events contribute significantly to the upper limits for mass ratios $q\ga 10^{-3}$.

\begin{figure*}[t]
\epsscale{0.4}
\centerline{
\plotone{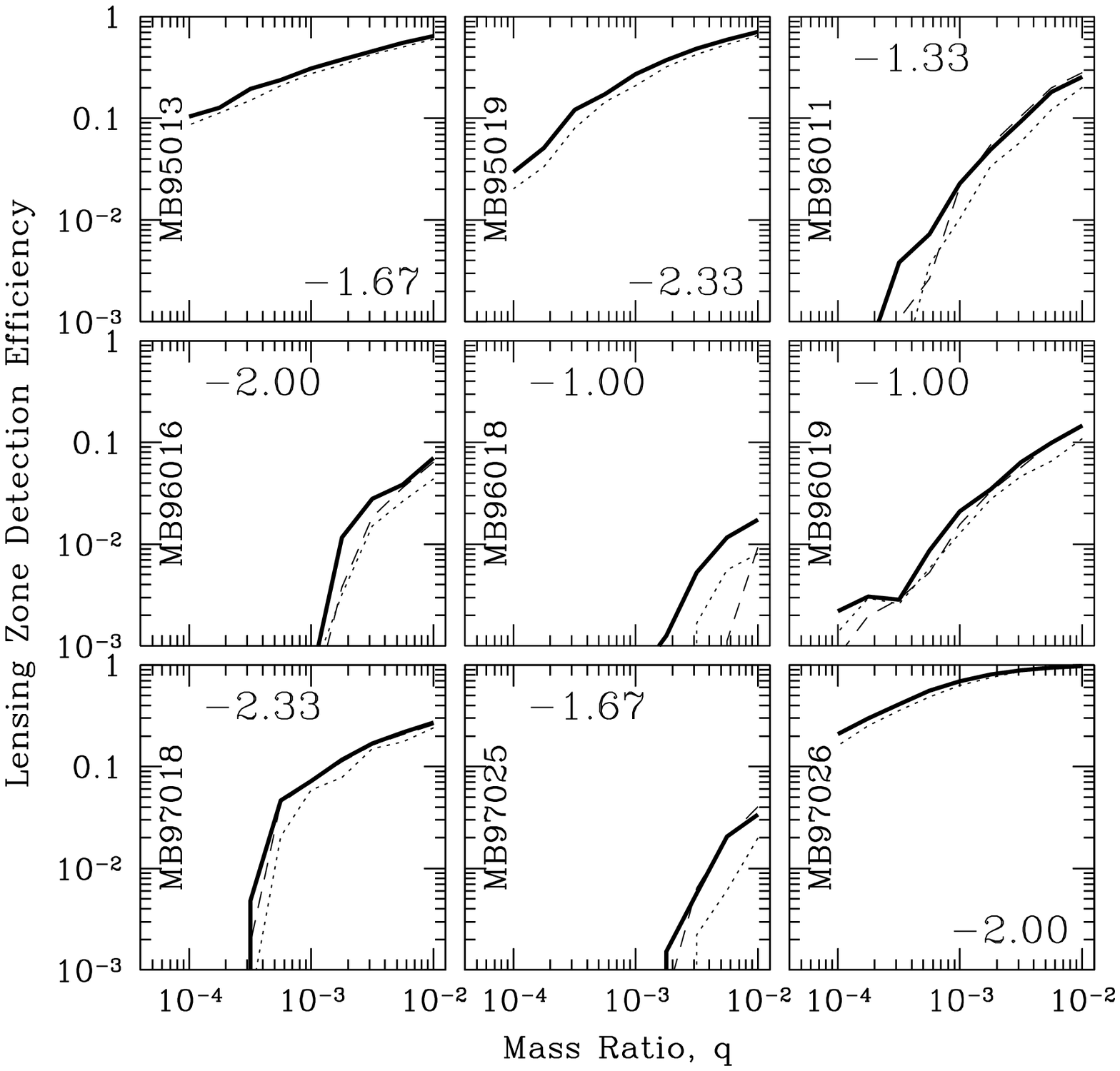}
\plotone{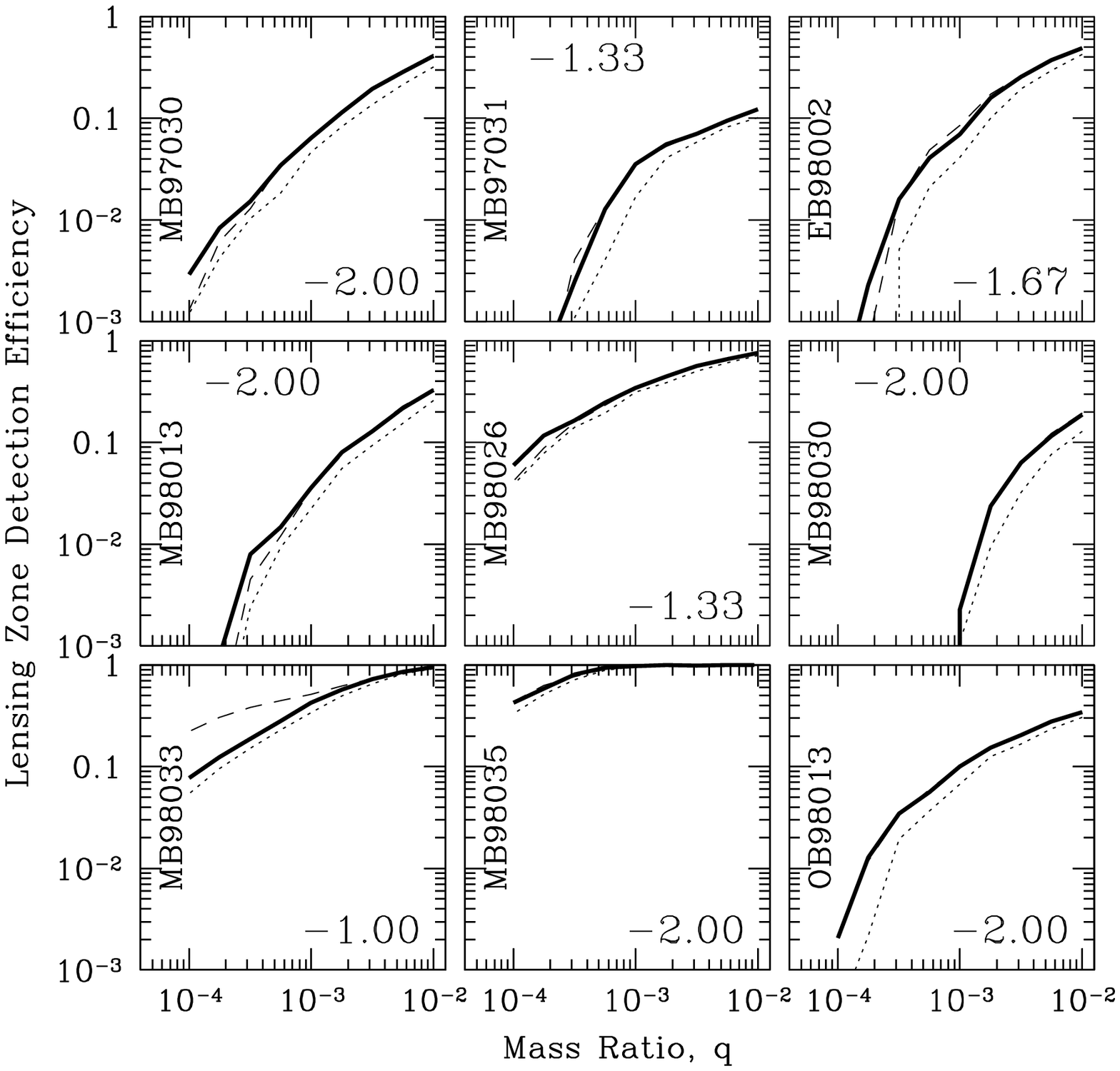}}
\centerline{
\plotone{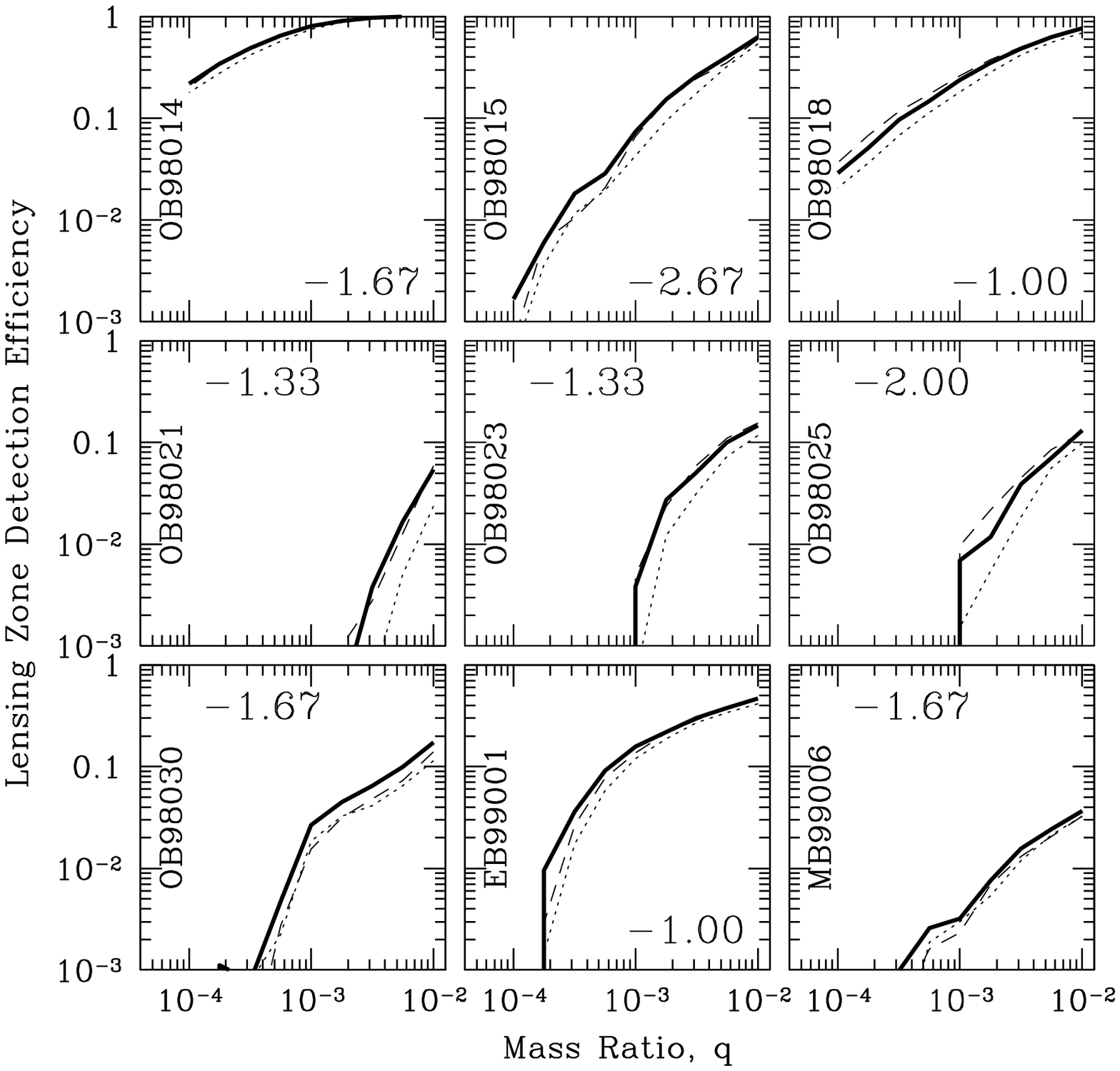}
\plotone{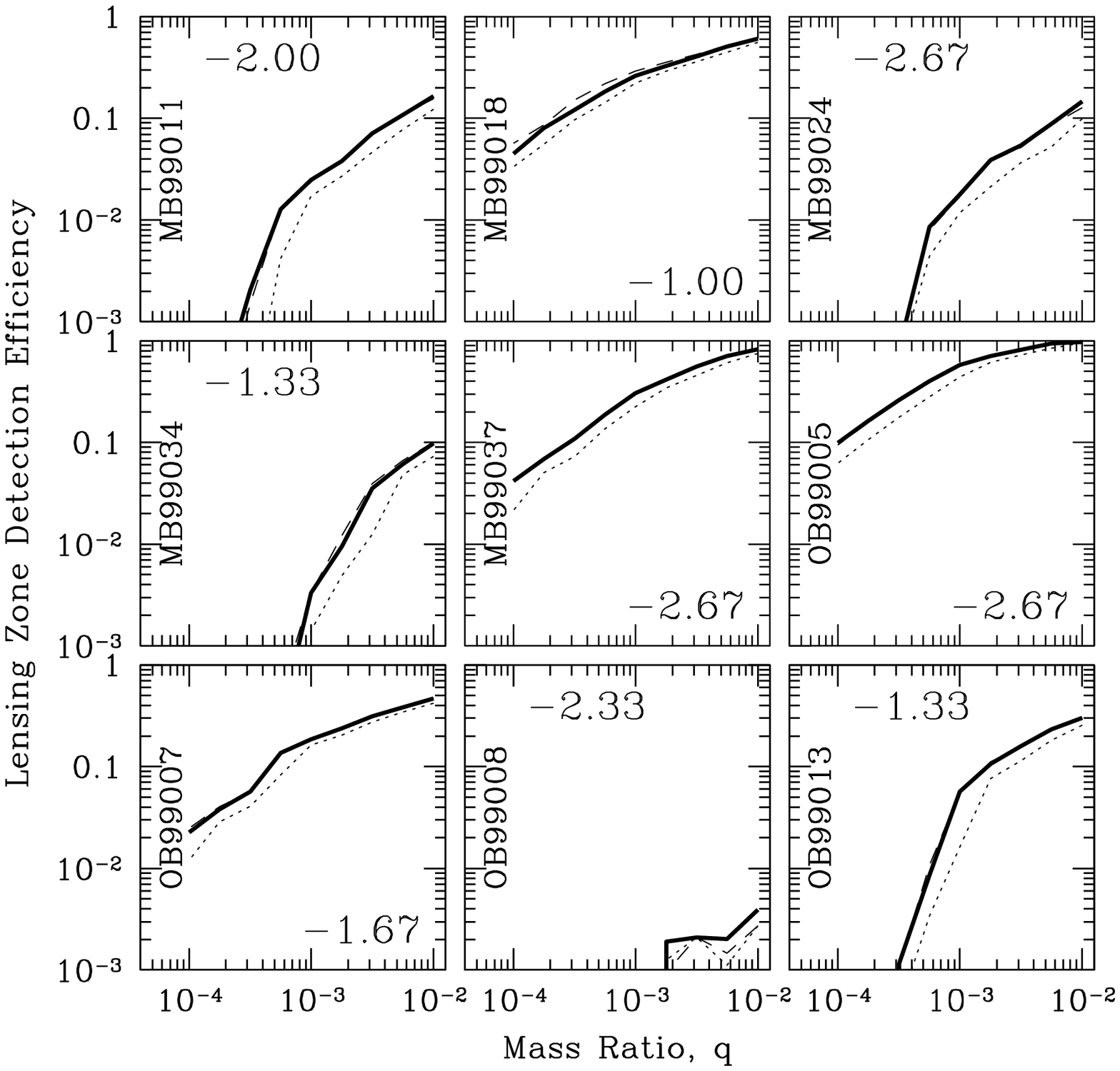}}
\centerline{
\plotone{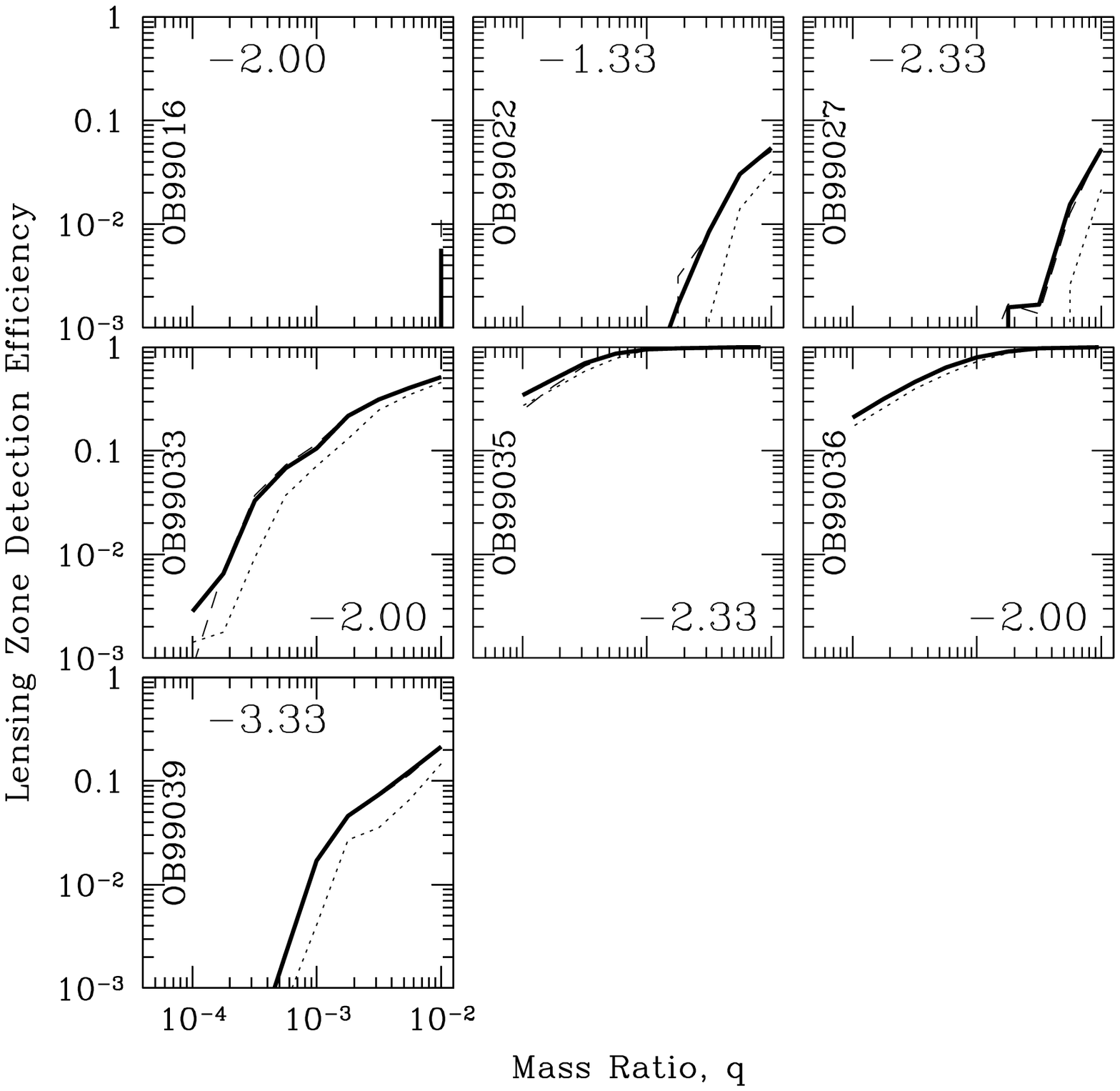}}
\caption{
\footnotesize
Heavy solid lines are point-source detection efficiencies averaged over the
lensing zone ($0.6 \le d \le 1.6$) as a function of the mass ratio of
the companion for a threshold of $\dchit=60$.  Dotted lines are for a
threshold of $\dchit=100$.  Dashed lines are the lensing zone
detection efficiencies for $\dchit=60$ assuming a finite source of size $\rhos$
in units of the angular Einstein ring radius.  
Each panel is for a separate event; the abbreviated event name and $\log{\rhos}$ are
indicated.
}
\label{fig:lze}
\end{figure*}

In Figure~\ref{fig:lze}, we show the efficiency averaged over the
lensing zone (where the detection efficiency is the highest), 
\begin{equation}
\elz (q) \equiv \int_{0.6}^{1.6}  \epsilon_i (d,q)~{\rm d}d~,
\label{eqn:elenszone}
\end{equation} 
as a function of the logarithm of the mass ratio.  For a model in
which companions have projected separations $d$ distributed uniformly 
in the lensing zone, $\elz (q)$ is the probability that a planet
of mass ratio $q$ would have been detected in light curve $i$.  Also
shown is $\elz$ for a detection threshold of $\dchit=100$.  For this
more conservative threshold, the efficiencies are $5-40\%$ lower, though the 
threshold level is most important where the efficiency is smallest.

\section{Finite Source Effects\label{sec:fs}}

The results in \S\ref{sec:detections} and \S\ref{sec:des} were
derived under the implicit assumption that the source stars of the
microlensing events could be treated as point-like.  
Numerous authors have
discussed the effect of the finite size of the source on the 
deviation from the PSPL curve caused by planetary companions
\citep{bandr1996, gandg1997, gands1998, meands2000, vermaak2000}.  Finite sources
smooth out the discontinuous jumps in magnification that occur when
the source crosses a caustic curve, and generally lower the amplitude
but increase the duration of planetary perturbation.  Finite sources
also increase the area of influence of the planet in the Einstein
ring.  Thus finite sources have a competing influence on the detection
efficiency: significant point source deviations can be suppressed below the
detection threshold, while trajectories for which the limb of the
source grazes a high-magnification area can give rise to detectable
perturbations when none would have occurred for a point-source. 
Which effect dominates depends on many factors, including the size of
the source relative to the regions of significant deviation from the
single-lens form, the photometric precision, and the sampling rate.
For large sources and small mass ratios, finite source effects can
significantly alter the detection efficiency \citep{meands2000}. 
Since in principle the results presented in \S\S\ref{sec:detections} and
\ref{sec:des} could  be seriously compromised by
ignoring these effects,  we evaluate the magnitude of the
finite source effect explicitly.

In order to access the magnitude of the
finite source effect, we must estimate the angular radius of the source in units of $\thetae$, 
\begin{equation}
\rhos = {{\thetas}\over{\thetae}}={{\thetas}\over{\murel\te}}
\simeq 0.02 \left(\thetas \over
6~\muas\right) \left({\murel} \over {12.5~\kms\kpc^{-1}}\right)^{-1}
\left({\te \over 40~\days}\right)^{-1},
\label{eqn:rhos}
\end{equation}
where $\thetas=6~\muas$ for a clump giant at $8~\kpc$.   
For deviations arising from the planetary caustic, finite source
effects become important when $\thetas$ is of order or smaller the planetary Einstein ring radius, $\thetap$, i.e, when
\begin{equation}
\rhos \ga \sqrt{q}\qquad{\rm (Planetary~Caustics)}.
\label{eqn:fslimpc}
\end{equation}
The size of the central caustic is $u_{\rm c}\sim q d (d-1)^{-2}$
\citep{gands1998}. Thus finite sources will affect the magnification due to
the central caustic when $\rhos \ga q d(d-1)^{-2}$.  However, in order
for the central caustic to be probed at all, the event must have an
impact parameter $\u0 \la u_{\rm c}$.  Thus finite source will affect deviations arising from
the central caustic if 
\begin{equation}
\rhos \ga \u0 \qquad{\rm (Central~Caustic)}.
\label{eqn:fslimcc}
\end{equation}

The difficulty in assessing the effect of finite sources on the
detection efficiency lies not in evaluating the effect for a given
$\rhos$, but rather in determining the appropriate $\rhos$ for a given
event.  This is clear from \eq{rhos}: of the three parameters that
determine $\rhos$, $\te$ is known from the PSPL fit, $\thetas$ can be
estimated based on the color and magnitude of the source, but $\murel$
is unknown. \citet{meands2000} suggested several possible methods of
dealing with this difficulty.  The simplest is to assume for all events
a proper motion equal to the mean
proper motion $\left<\murel\right>$, adopting the
$\rhos$ given by  \eq{rhos} with $\murel=\left<\murel\right>$.  A more accurate, but
also more complicated and time-consuming, method is to integrate over a
distribution of $\murel$ given by a Galactic model; this would imply
calculation of finite source effects for many different values of
$\rhos$.  Here we adopt the first approach and determine $\rhos$ assuming
$\murel=\left<\murel\right>=12.5~\kms\kpc^{-1}$.  This value of
$\left<\murel\right>$ is a factor of two lower than the expected mean relative
proper motion for all lenses toward the bulge \citep{handg1995}, and
reflects the fact that our median $\te$ is a factor of two larger than
the median of all microlensing events toward the bulge and our belief 
that the larger time scales reflect the fact that we are
preferentially selecting slower (rather than more massive or closer)
lenses.  We justify this assertion in \S\ref{sec:iandd}.  To the extent that 
the masses and distances of the lenses contribute somewhat to this
larger median time scale, our adopted value of 
$\left<\murel\right>$ is likely an overestimate.  Therefore, the
resulting values of $\rhos$ are likely overestimates, so that we are
conservatively computing {\it upper limits} to the effect of the 
finite source sizes on our conclusions.

\subsection{Estimating the Source Sizes\label{sec:angular}}

\begin{figure*}[t]
\epsscale{1.0}
\centerline{\plotone{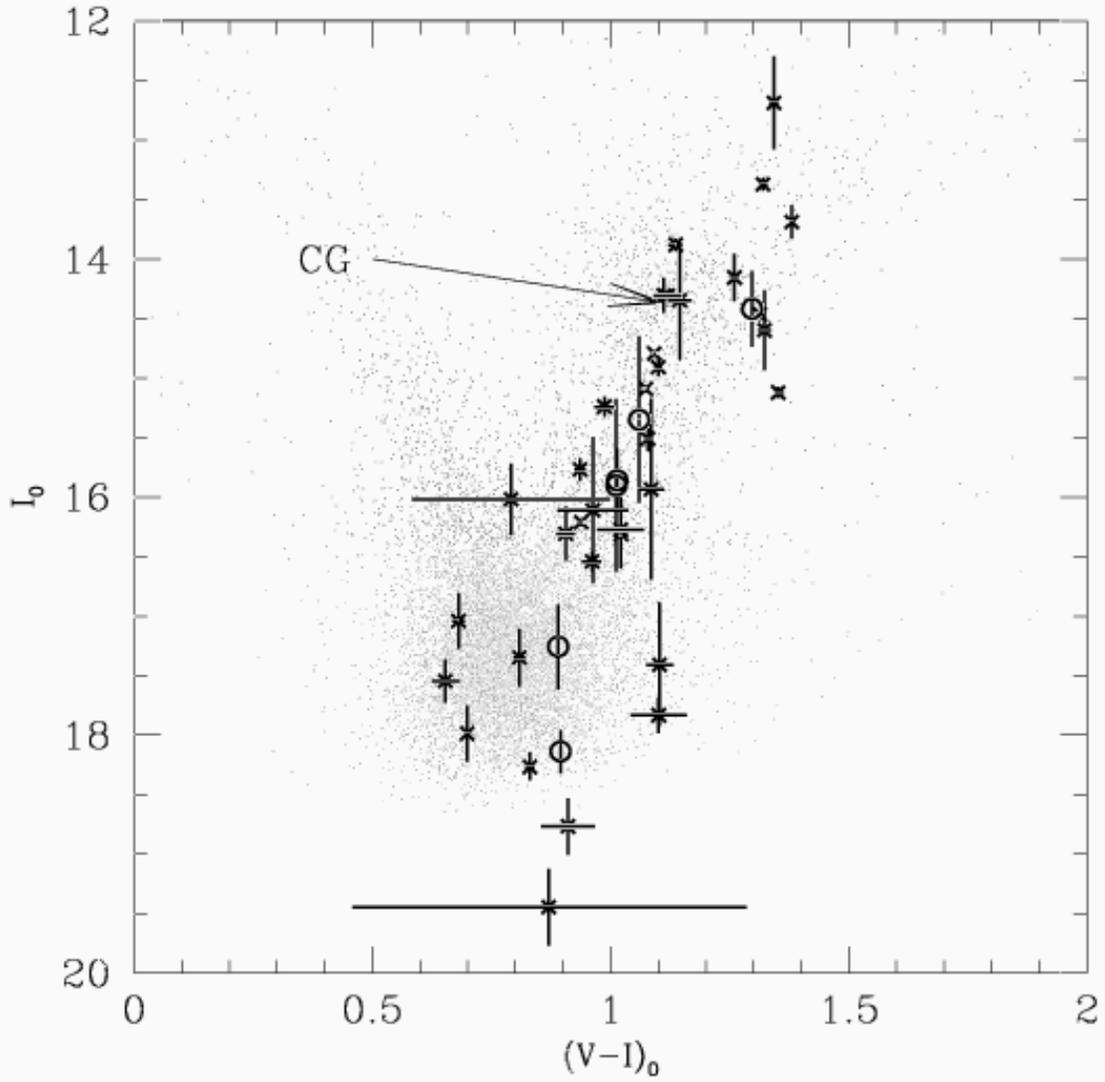}}
\caption{
\footnotesize
The dereddened ``clump-calibrated'' color-magnitude diagram
for the source stars.  Crosses indicate sources for which separate
$\i0$ and $\vmi0$ determination was possible; circles indicate
events for which the $\vmi0$ was assumed to be that typical of stars with
the same $\i0$ as measured for the event.  The center of the clump is
indicated with an arrow. Also shown is the CMD of a typical field
(small dots).
}
\label{fig:cmd}
\end{figure*}

The angular size $\thetas$ of a given source can be estimated from its
dereddened $\vmi0$ color and magnitude $\i0$.  
From the PSPL fits to the $I$ and $V$ photometry, we know the $I$ and $V$ 
fluxes of the sources in instrumental units (see \S\ref{sec:esel}).  
We assume that the dereddened color
$(V-I)_{\rm cl,0}$ and magnitude $I_{\rm cl,0}$ of the clump is
invariant for all our fields, adopting the determination by \citet{packy1999},
\begin{equation}
(V-I)_{\rm cl,0}=1.114 \pm 0.003, \qquad  I_{\rm cl,0}=14.43 \pm 0.02.
\label{eqn:clump}
\end{equation}
We form instrumental color-magnitude diagrams (CMDs) for each of our
fields, and determine the position of the clump in instrumental
units by finding the local maximum in the density of sources.  The
difference between this position and the intrinsic position [\Eq{clump}]
gives the offsets $\Delta\vmin$ and $\Delta I$ for all the stars in
the field (except foreground stars, which have less reddening than the
calibrating clump).  Note that these offsets include both the calibration from
instrumental to true fluxes, and also the correction for the mean
reddening of the field.  Thus we do
not assume a constant redenning law from field to field. 
We apply these offsets to the instrumental $\vmin$ and $I$ of our source
stars, finally arriving at the $\vmi0$ and $\i0$ for all our sources.
These are shown in Figure~\ref{fig:cmd} and listed in Table~\ref{tab:tab04}.
The error bars on $\vmi0$ and $\i0$ are those derived from
modeling uncertainties; we estimate there to be an additional
calibration error of $\sim 5\%$ in both
$\vmi0$ and $\i0$ based on the typical
dispersion of the clump.  Note that the majority ($\sim 70\%$) of our
sources are giants. 

Using these colors and magnitudes, the angular size of the
sources are derived using a modified version of the empirical color-surface
brightness relation derived by \citet{vbelle} and given in
\citet{albrow2000a}.
The resulting $\thetas$ for all of our sources are shown in
Table~\ref{tab:tab04}.  The average uncertainty in 
$\thetas$ is ${\cal O}(20\%)$, combining the
uncertainty in the color and magnitude of the source due to both
modeling and calibration uncertainty and the uncertainty in the
underlying \citet{vbelle} relation.  We do not determine the uncertainty on
$\thetas$ for individual sources because the uncertainty in $\rhos$ (the
parameter in which we are primarily interested) is dominated by the
uncertainty in $\left<\murel\right>$.  Seven of our events have
insufficient $V$-band data to determine the instrumental $\vmin$ of
the source.  For these events, we assumed the source to have the median
$\vmin$ of all sources in the field with similar $I$ magnitudes.  (For these
events, we do not quote uncertainties on $\vmi0$.)  Finally, four events
had either no $V$-band data at all, or the position of the clump was
impossible to determine from the CMD of the field.  For these events, we
simply adopt the conservative assumption that the sources are clump
giants, with $\thetas=6~\muas$.

\begin{figure*}[t]
\epsscale{1.0}
\centerline{\plotone{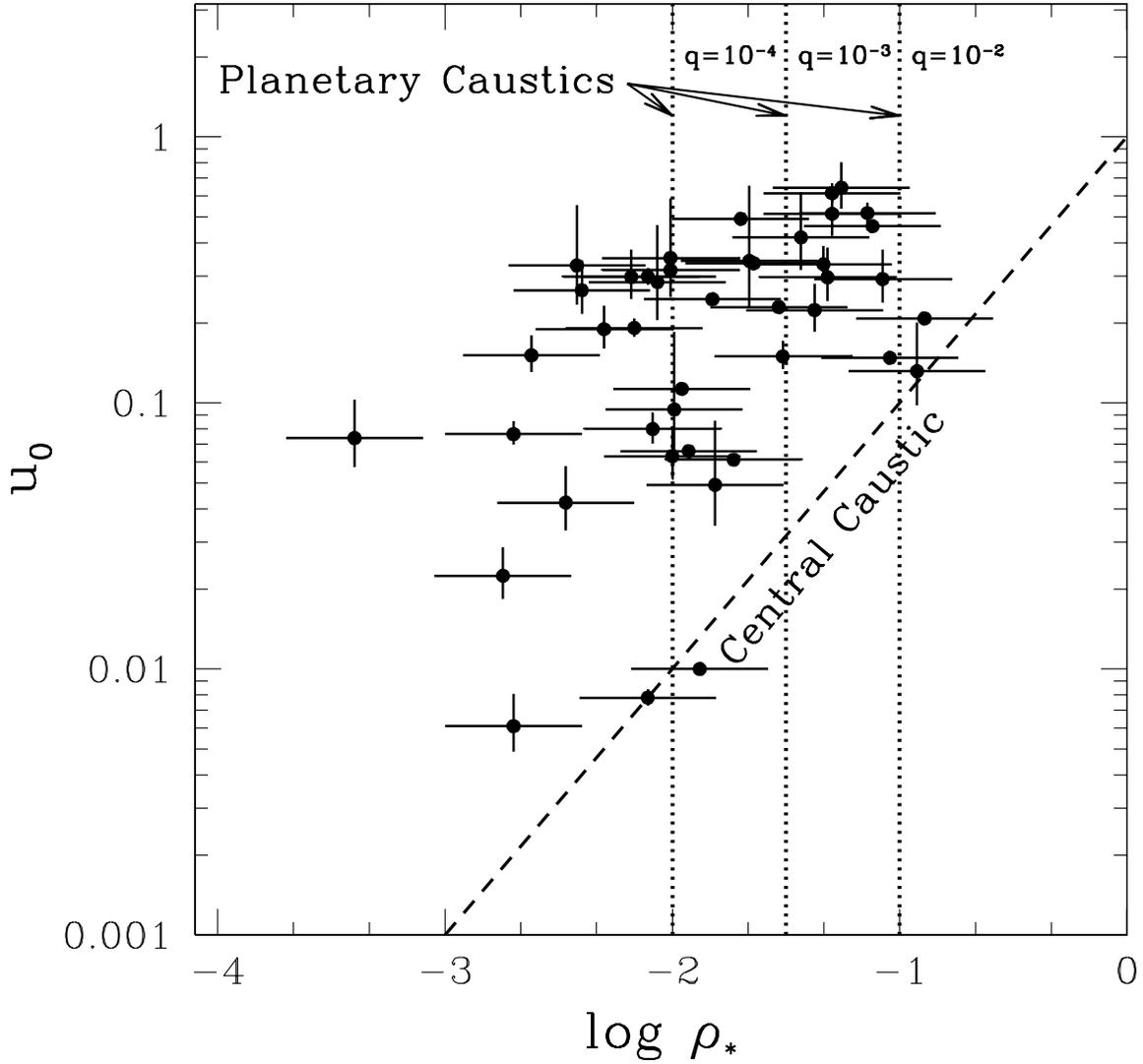}}
\caption{
\footnotesize
The impact parameter $\u0$ versus the logarithm of the source
size $\rhos$ in units of the angular Einstein ring radius $\thetae$.
The dotted lines indicate the boundaries at which finite source effects
become important for the detection of a companion of the indicated mass ratios 
via the planetary caustics; source sizes to the
right of these boundaries significantly affect the amplitude and
duration of the deviation caused by the planetary caustics.  The
dashed line indicates the boundary of the region at which finite source
effects become important for the detection of a companion via the
central caustic.
}
\label{fig:u0vrho}
\end{figure*}

These estimates of $\thetas$ are used to determine $\rhos$ 
under the assumption that all events have the same
relative proper motion $\left<\murel\right>=12.5~\kms \kpc^{-1}$; these values of $\rhos$
are listed in
Table~\ref{tab:tab04}.  For two events, \mb{98}{35} and \ob{99}{35}, the value of $\rhos$
estimated in this way is larger than the fitted $\u0$ of the event.
In both cases, the derived values of $\rhos$ are ruled out by the fact that, despite dense
coverage at the peak, no deviations from
the PSPL form are seen, as would be expected if $\rhos>\u0$ and
the lens was resolving the source \citep{gould1994, neandwick1994,wittandmao1994}.
For these two events, we therefore assume that $\rhos=\u0$.
In Figure~\ref{fig:u0vrho}, we plot $\u0$ versus $\rhos$ for all our
events, along with the boundaries where finite source effects become
important for both the planetary and central caustics
[eqs.~(\ref{eqn:fslimpc}) and (\ref{eqn:fslimcc})].  For the majority of
our events, finite source effects should not alter the results for
companions with $q\ga 10^{-3}$, whereas a large fraction of our
events should be affected for $q\sim 10^{-4}$.

\subsection{Incorporating Finite Sources\label{sec:encorfs}}

In order to incorporate finite sources into the analysis, we repeat the
algorithm presented in \S\ref{sec:algorithm} for all events, but 
fit the events to binary-lens light curves that include the effect of the
finite size of the source.  Evaluating the
finite-source binary-lens magnification for the specific value of
$\rhos$ determined for each event is not computationally
feasible, as finite source magnifications are quite time consuming to
calculate.  We therefore adopt a procedure similar to that described in
\S\ref{sec:maps}: interpolation between a grid of finite-source binary-lens
magnification maps.  We choose the same grid spacing and size
for $(d,q)$, namely $10^{-4} \le q \le 10^{-2}$ at equal
intervals of $0.25$ in $\log{q}$, and $0.1 \le d \le 10$ at $d=0.1,
0.2, ... , 1.0$ and their inverses.  For each
of these $(d,q)$ pairs, we create finite-source magnification maps for
$10^{-4} \le \rhos \le 10^{-1}$ at intervals of $1/3$ dex in
$\log{\rhos}$.  These maps have same the extent and sampling in the source
plane as the point source maps (see \S\ref{sec:maps}).  We evaluate
the finite-source magnification using the Stokes method of integrating over
the boundary of the images \citep{kands1988, gandgauch1997}. 
Our assumption of uniform sources overestimates the size of the finite
source effect relative to limb-darkened sources, and thus is
conservative.  The grid value of $\rhos$
closest to the value estimated for each source is used to calculate
the detection efficiency for that event.  
We have repeated this process for the next-closest value of
$\rhos$ in the grid for all events, and find that there is no appreciable difference
in the conclusions. 

\subsection{Effect of Source Size on Detection Efficiencies}

The distribution of $\dchim$ for the finite-source binary-lens fits 
is shown in Figure~\ref{fig:detdist}, along with the distribution for
the point-source binary-lens fits.  For the most part, the two
distributions are quite similar.  The significance of the
best-fit binary-lens model has increased in some cases
(e.g. \mb{98}{35}), but all of the events that fall below our
detection threshold ($\dchim>-60$) in the point-source case also fall
below this threshold in the finite source case.  We recover the same two anomalies
 in \mb{99}{18} and \ob{99}{36}, but no others.  As argued in \S\ref{sec:detections}, these
two anomalies have explanations other than planetary microlensing for their behavior.  Thus our
conclusions are unchanged: out of a sample of 43 events, we find no
viable planet candidates.

The resulting finite-source lensing zone detection efficiencies [\Eq{elenszone}] 
are shown in Figure~\ref{fig:lze} along with the corresponding
point-source efficiencies.  We find, in agreement with the
expectations in \S\ref{sec:angular}, that the difference between
the point-source and finite-source efficiencies for mass ratios $q \ga
10^{-3}$ is negligible for nearly all events, with the
exception of a few events with very large sources ($\rhos\sim 0.1$).
Finite source effects begin to become appreciable for $q\la
10^{-3}$. For $q=10^{-4}$, the finite-source detection efficiency is
markedly smaller than the point-source efficiency for large
sources.  The finite size of the sources has no appreciable effect on the
detection efficiencies for those
mass ratios where we have significant constraints ($q>10^{-4}$), and conversely for those mass
ratios where finite source effects are appreciable we have no
interesting constraints.  Therefore we conclude that, for this sample
of events, finite source effects are negligible.

\section{Upper Limits on Planetary Companions\label{sec:ulimits}}

The fact that a large fraction of our final sample of 42 microlensing events has significant
detection efficiencies to planetary companions --- despite the fact
that we have detected no viable planetary candidates in these events --- suggests
that the fraction of primary lenses with planetary companions in our range of sensitivity
must be considerably smaller than unity.  To quantify the exact limit
implied by our data, we combine the individual event efficiencies $\epsilon_i (d,q)$ to obtain a
statistical upper limit on the fraction of lenses with companions as a function of
mass ratio $q$ and projected separation $d$. 

Assume that a fraction $f(d,q)$ of primary lenses have planets with parameters
$(d,q)$.  Averaged over a large number of events, the probability that any single
event would harbor such a planet is then also $f(d,q)$.   The
probability that such a planet would be detected in event $i$ is
the detection efficiency, $\epsilon_i(d,q)$.  Therefore, the
probability that any given event has a planet that is detectable with these data is
$f(d,q)\epsilon_i(d,q)$.  The probability that a planet
is not detected is $1-f(d,q)\epsilon_i(d,q)$.  Thus the
probability that a sample of $N$ events would result in at least one detection is simply
\begin{equation}
P(d,q)= 1- \Pi_{i=1}^N \left[ 1- f(d,q) \epsilon_i(d,q)\right].
\label{eqn:pnull}
\end{equation}
The 95\% confidence level (c.l.) upper limit to $f(d,q)$ implied by such a sample of
events is found by setting $P(d,q)=0.05$ and solving for $f(d,q)$.
Note that, in the limit of $f\epsilon_i \ll1$,
equation~(\ref{eqn:pnull}) reduces to the naive formula,
\begin{equation}
P(d,q) \rightarrow 1 - \exp[-N_{\rm exp} (d,q)]\qquad N_{\rm exp}(d,q)=f(d,q)\sum_i\epsilon_i(d,q).
\label{eqn:pnaive}
\end{equation}
We have, however, used the exact expression \eq{pnull} to compute excluded fractions $f(d,q)$.

\begin{figure*}[t]
\epsscale{1.0}
\centerline{\plotone{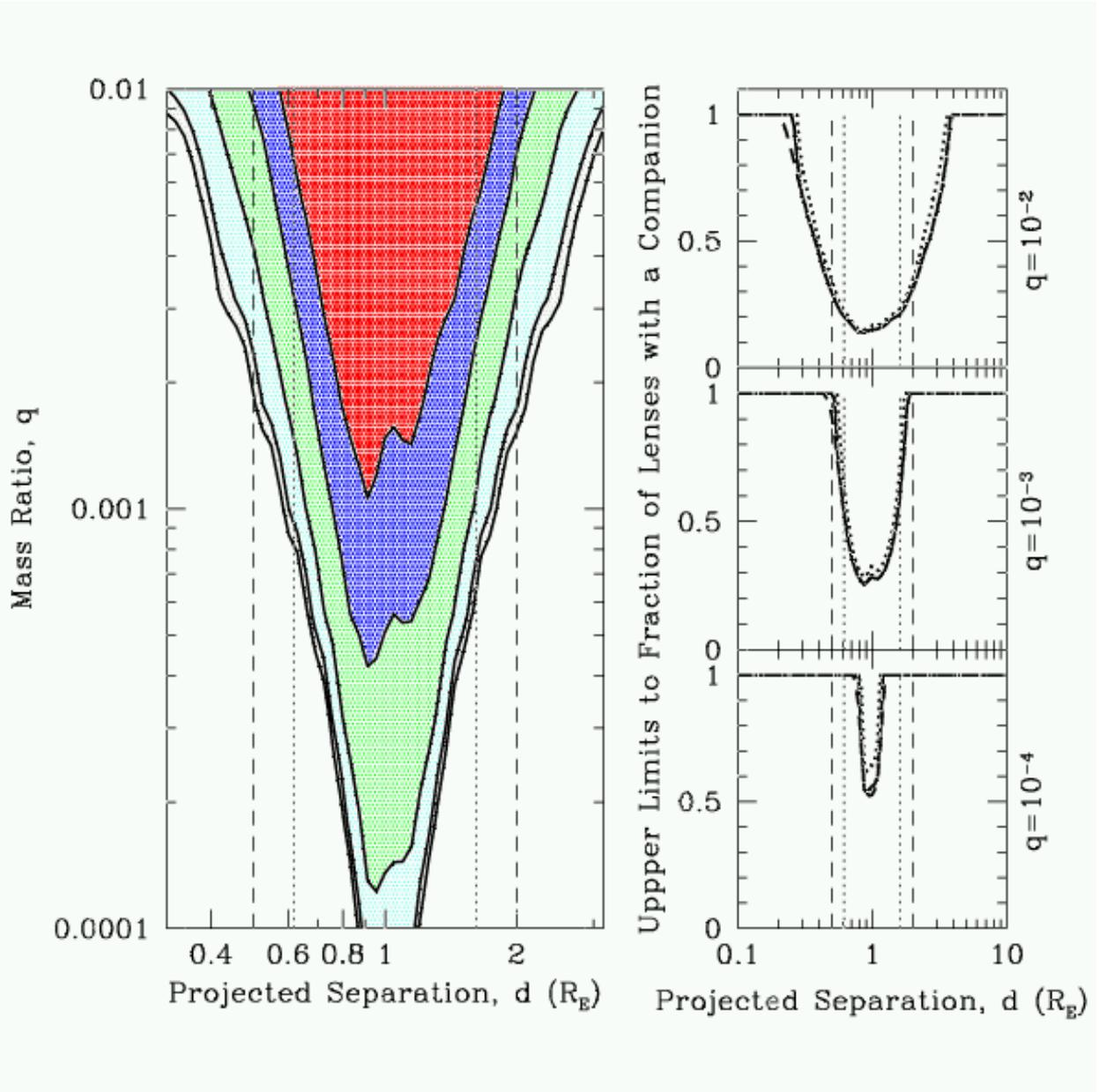}}
\caption{
\footnotesize
{\it Left Panel}: Exclusion contours (95\% c.l.) for the fractions of primary lenses with a
companion derived from our sample of 43 events, 
as a function of the mass ratio and projected separation of the
companion.  Solid black lines show exclusion contours for $f=$75\%, 66\%, 50\%,
33\% and 25\% (outer to inner).  The dotted (dashed) vertical lines indicate
the boundaries of the lensing zone (extended lensing zone). 
{\it Right Panels}:  Cross sections through the
left panel, showing for three different mass ratios the upper limit to the 
fraction of lenses with a companion as a function of
projected separation.  The solid
line is derived from the point-source efficiencies with a threshold of
$\dchit=60$. The dotted line is derived from the point-source efficiencies with a threshold of
$\dchit=100$.  The dashed line is finite-source efficiencies with a threshold of
$\dchit=60$.  The dotted vertical lines indicate
the boundaries of the lensing zone $0.6 \le d \le 1.6$.  The dashed vertical lines
indicate the extended lensing zone, $0.5 \le d \le 2$.}
\label{fig:uldq}
\end{figure*}

\begin{figure*}[t]
\epsscale{1.0}
\centerline{\plotone{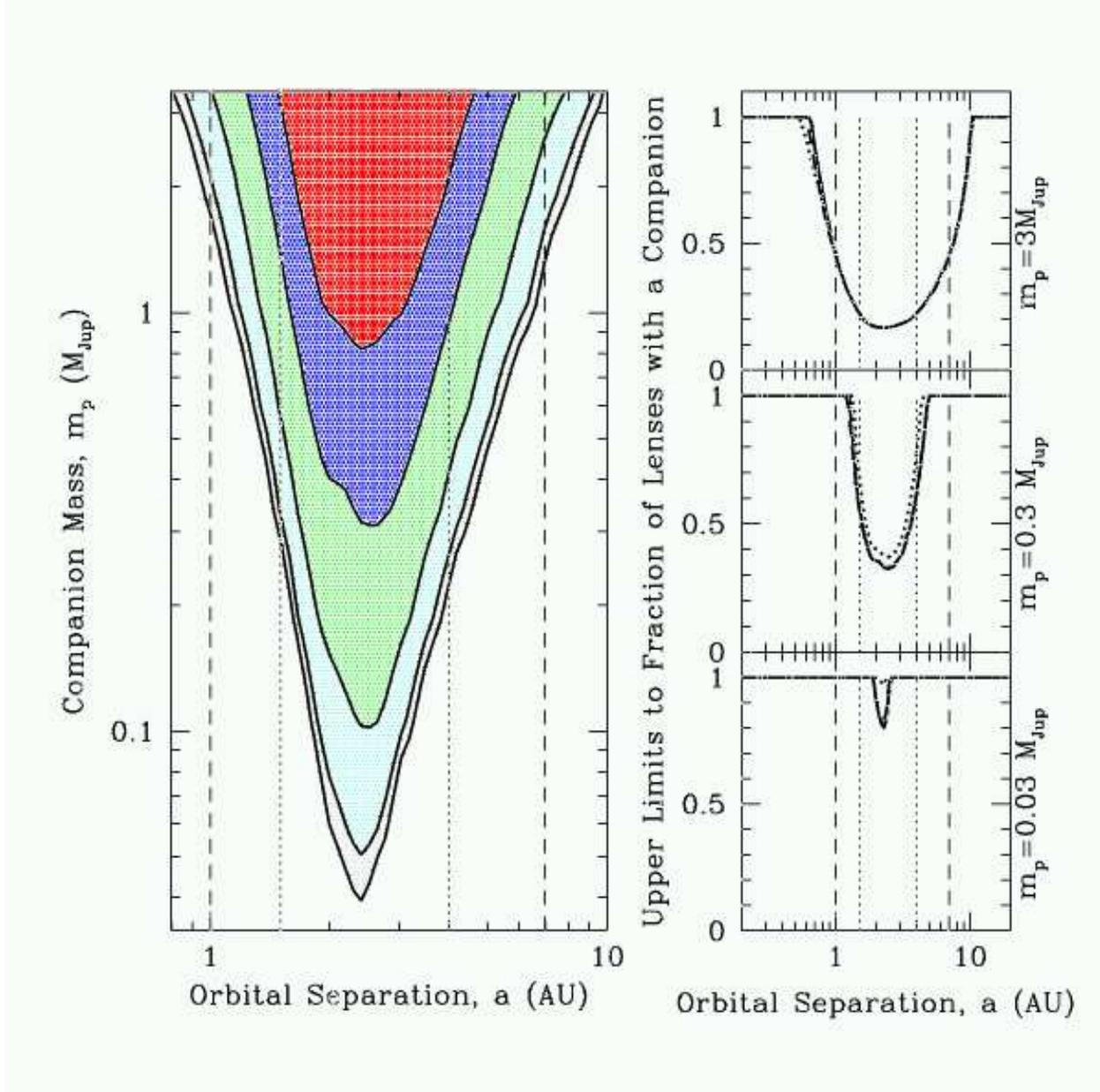}}
\caption{
\footnotesize
Same as Fig.~\ref{fig:uldq}, except we have integrated over all
possible orbital inclinations and phases to convert from projected
separation to orbital separation, and assumed a primary mass of
$M=0.3\msun$ and a primary Einstein ring radius of $\re=2~\au$.
}
\label{fig:ulam}
\end{figure*}

In Figure~\ref{fig:uldq} we show the 95\% c.l.\ upper limit to
$f(d,q)$ as a function of $d,q$ derived from our final sample of 42 events,
assuming $\dchit=60$ and point sources. 
We conclude that $<28\%$ of lenses have a companion of mass ratio
$q\ga10^{-3}$ and projected separation $d\sim1$.  
The hypothesis that more than
one-half of the primary lenses have a companion near $d=1$ for the
full range of mass ratios $10^{-4} \le q \le 10^{-2}$ is excluded with 95\% confidence.
Also shown in Figure~\ref{fig:uldq} are cross
sections of the $(d,q)$ exclusion diagram (95\% c.l.\ upper limits as a
function of $d$) for three different mass ratios, namely $q=10^{-2}$, $10^{-3}$, and $10^{-4}$.
For these cross sections, we also show the 95\% c.l.\ upper limits
derived assuming point sources and $\dchit=100$, and assuming finite
sources and $\dchit=60$.  Clearly finite source effects are negligible
in regions where we have interesting constraints.  

\begin{figure*}[t]
\epsscale{1.0}
\centerline{\plotone{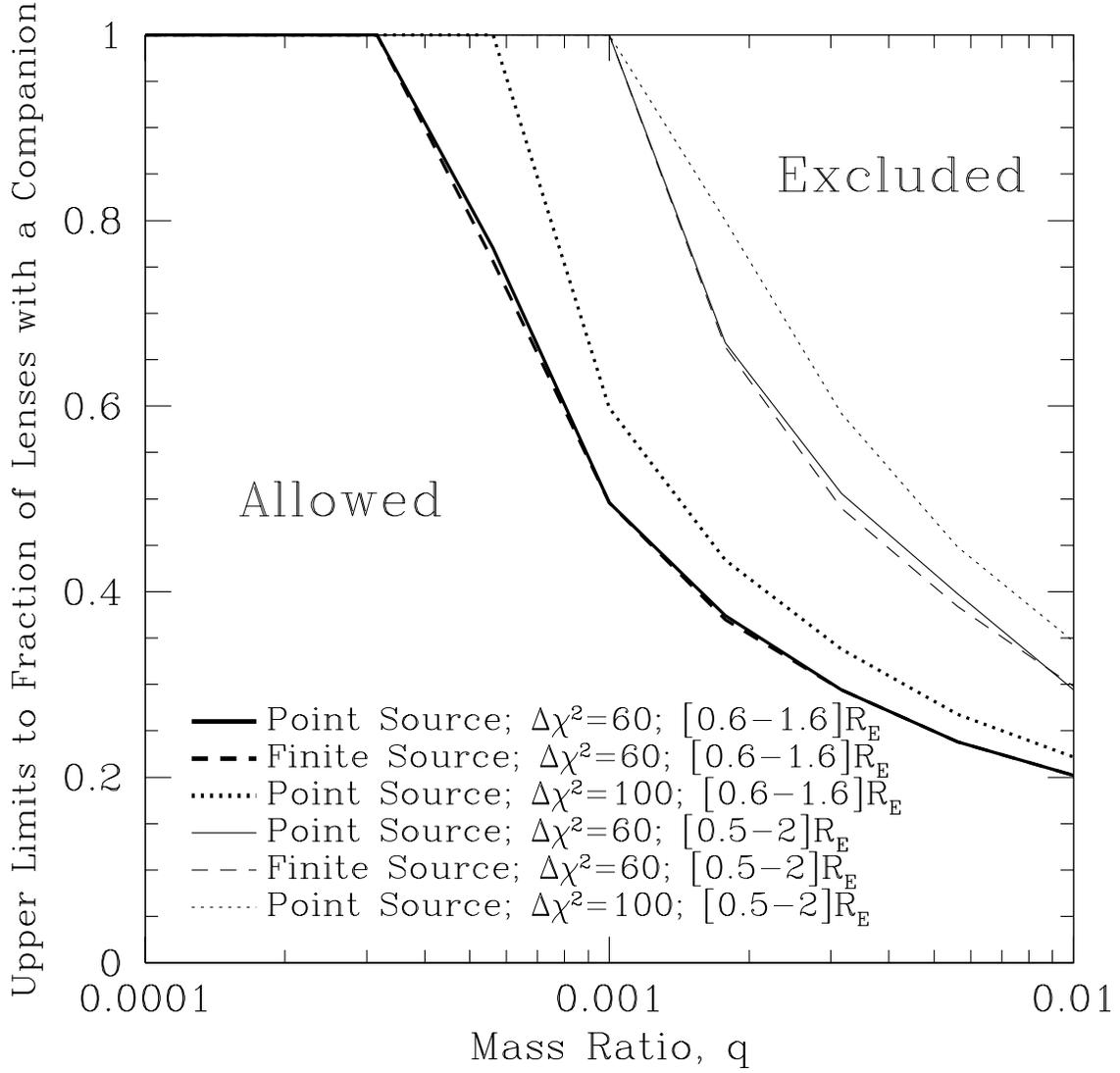}}
\caption{
\footnotesize
Upper limits to the fraction of primary lenses with a companion as a
function of the primary-companion mass ratio. 
Bold lines are for companions with projected separations anywhere in the lensing zone,
$0.6-1.6~\re$. Thinner lines are for projected separations in the extended
lensing zone, $0.5-2~\re$.  
}
\label{fig:lzuldq}
\end{figure*}

\begin{figure*}[t]
\epsscale{1.0}
\centerline{\plotone{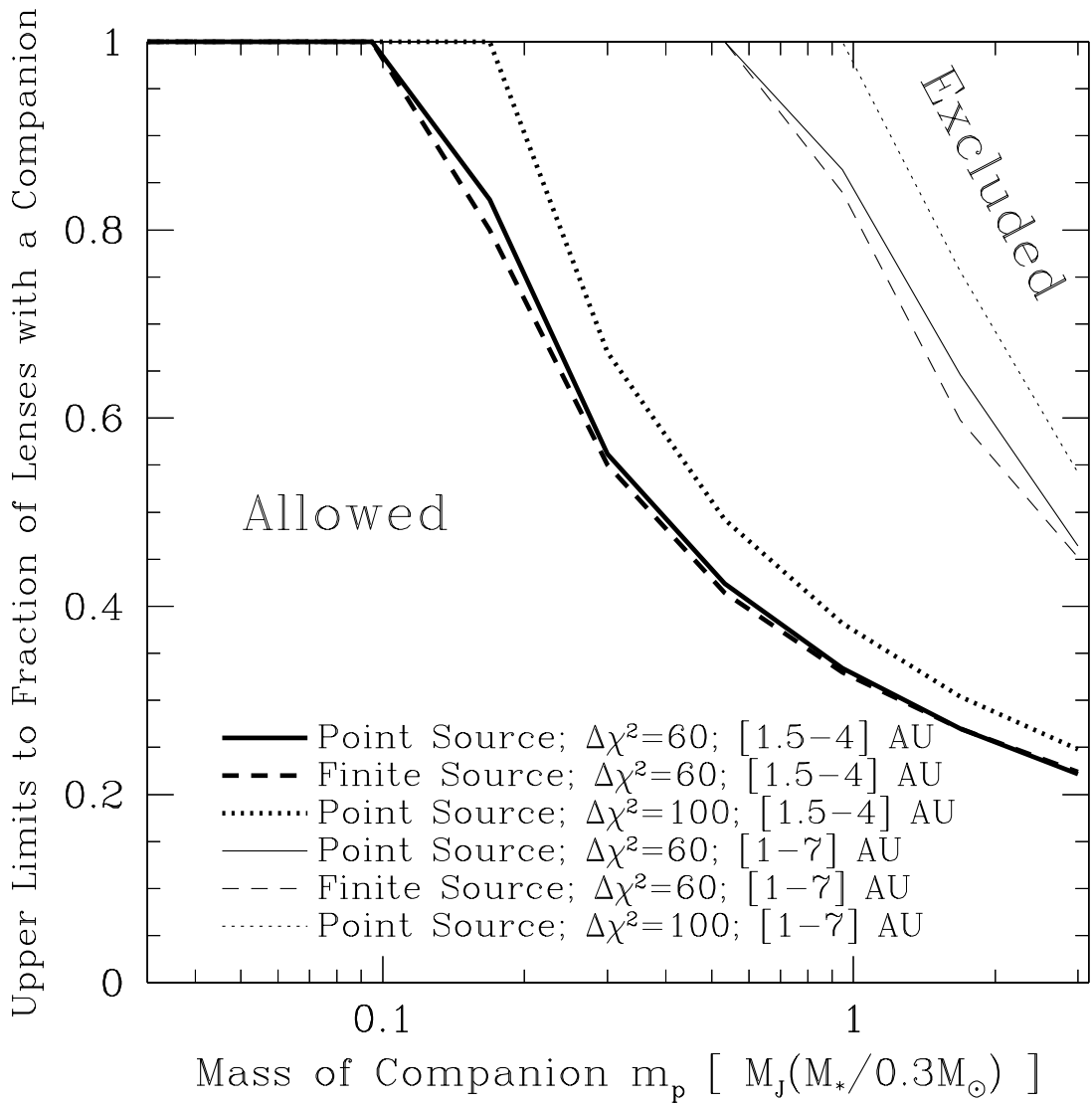}}
\caption{
\footnotesize
Upper limits to the fraction of primary lenses with a companion as a
function of the companion mass. The
bold lines are for companions with orbital separations
$1.5-4~\au$. The light lines are for orbital separations in the extended
lensing zone, $1-7~\au$.  This figure is essentially identical to
Fig.~\ref{fig:lzuldq}, except we have integrated over all
possible orbital inclinations and phases to convert from projected
separation to orbital separation, and assumed a primary mass of
$M=0.3\msun$ and a primary Einstein ring radius of $\re=2~\au$.
}
\label{fig:lzulam}
\end{figure*}

In Figure~\ref{fig:lzuldq} we show the 95\% c.l.\ upper limit as a
function of $q$ for companions anywhere in the lensing zone $0.6 \le d
\le 1.6$, and anywhere in the ``extended'' lensing zone,
$0.5 \le d \le 2.0$.  Statistically, less than $20\%$ of
primaries have a $q=10^{-2}$ mass ratio companion in the lensing zone.  For
$q=10^{-3}$ companions in the lensing zone, the upper limit is $45\%$. 

\section{Converting to Planetary Mass and Orbital Separation\label{sec:iandd}}

The upper limits presented in \S\ref{sec:ulimits} are the most
direct, least model-dependent inferences we can draw from our data.
Unfortunately, they are not the most illuminating, for several
reasons.  First, the nature of primaries
around which we limit planets is not specified.  Second, our results are quoted
in terms of the two natural binary-lens parameters, the mass ratio of
the system $q$ and the instantaneous projected separation $d$ of the companion, rather
than the more common (and more interesting) parameterization of 
planetary mass $\massp$ and orbital separation $a$.  

Unfortunately, it is not possible to directly determine the mass of the
primaries, and hence their nature, because the one
observable parameter containing information about the
lens, the event time scale $\te$, is a degenerate combination of the
mass, distance, and velocity of the lens [Eqs.~\ref{eqn:thetae} and \ref{eqn:te}]. 
Only model-dependent inferences about the nature of the primary
lenses are possible.   The majority
of the microlensing events in our sample are
likely to be due to bulge stars lensing other bulge stars
\citep{kandp1994}.  Following \citet{gould1999}, we adopt the bulge mass function as measured by
\citet{zoccali2000}, and assume a model such that the sources and lenses are distributed
as $r^{-2}$, where $r$ is the Galactocentric distance, and have Gaussian velocity distributions with dispersion
$\sigma=100~\kms$.   This model gives typical parameters for bulge self-lensing
events of $\ave{M} \sim 0.3~\msun$,  $\ave{\pirel}=40~\muas$, and
thus $\ave{\thetae} \sim 320~\muas$.  For the relative proper motion,
this model predicts $\ave{\murel}\sim 25~\kms~\kpc^{-1}$, and thus
$\ave{\te}\sim 20~\days$, which is the median time scale found by OGLE
for events toward the Galactic bulge \citep{udalski2000}.  Taken 
at face value, the fact that the median time scale
of the events in our sample is a factor of two times larger implies that we are
selecting a biased subset of lenses.  From equations~(\ref{eqn:thetae}) and
(\ref{eqn:te}), this bias could be toward higher mass lenses, slower
lenses (smaller $\murel$) or closer lenses (larger $\pirel$), or any combination of
these three factors.  In fact, as demonstrated by \citet{gould1999},
the majority of the dispersion in the expected distribution of
time scales arises from the dispersion in $\murel$, not the dispersion in $\pirel$ or
$M$.  This implies that we are, for the most part, preferentially
selecting slower --- rather than more massive or closer --- lenses,
justifying our assumption of $\ave{\murel}\sim 12.5~\kms~\kpc^{-1}$
for the estimates of $\rhos$ in \S\ref{sec:fs}.  
Thus the typical mass and lens-source relative parallax of the lenses in our sample is likely 
to be close to
those of the complete sample of microlensing events.  
We therefore adopt $\ave{M}= 0.3~\msun$ and  $\ave{\pirel}=40~\muas$, which
for source stars at $\dos\sim 8~\kpc$ implies lens
distances of $\dol\sim 6~\kpc$.  In other words, the majority of our primary lenses are $M$ dwarfs in the
Galactic bulge. 

Some caveats must be noted. \citet{kandp1994} estimate that $\sim
20\%$ of events toward the Galactic bulge are due to lensing of bulge
stars by disk stars.  Of the remaining $\sim 80\%$,
\citet{gould1999} estimates that $\sim 20\%$ are due
to remnants (white dwarfs, neutron stars, and black holes).  Thus, we
would expect $\sim 60\%$ of the events in our sample to be due to normal
stars in the Galactic bulge.  However, we have no idea which events comprise this 60\%.
  Also, some fraction of the events in our sample are likely
members of binary systems with separations that are either too small
or (more often) too large to be distinguishable from single lenses.  We have
no way of determining which events these are, or even what fraction of
our events are in such systems.  
Given our rather small sample of events and the uncertainties in the 
magnitude of these contaminations, we feel it is not appropriate at
this stage to attempt to correct for these effects.

The estimates of $\ave{M}$ and $\ave{\pirel}$ adopted above imply
$\ave{\thetae}=320~\muas$ and thus $\ave{\re}=2~\au$ (for $\dol=6~\kpc$).
We use these values to convert the upper limits derived in
\S\ref{sec:ulimits} from dimensionless units to physical units,
via the relations, 
\begin{equation}
\massp = \left({ q \over 0.003}\right)\mjup , \qquad \ap = \left({ d \over 0.5 }\right)\au,
\label{eqn:convert}
\end{equation}
where $r_{\rm P}$ is the analog of $d$ (the instantaneous projected
separation) in physical units.  To convert from $\ap$ to the
conventional three-dimensional separation $a$, we must convolve with
the distribution function \citep{gandl1992},
\begin{equation}
p(\ap;a)= { \ap \over a}\left( 1- { \ap^2 \over a^2}\right)^{1/2},
\label{eqn:distfunc}
\end{equation}
which is found by integrating over
all random inclinations and orbital phases, assuming circular orbits.  
Thus the detection efficiency of
each event $i$ in the $(a,\massp)$ plane is,
\begin{equation}
\epsilon_i(a,\massp)=\int_0^a {\rm d}\ap p(\ap;a)
\epsilon_i(\ap,\massp)
\label{eqn:deam}
\end{equation}
These individual efficiencies $\epsilon_i(a,\massp)$ can now be
combined in the same manner as in \S\ref{sec:ulimits} to derive
95\% c.l.\ upper limits to the fraction $f(a,\massp)$ of events with
companions as a function of the 
mass $\massp$ and separation $a$ of the companion.

In Figure~\ref{fig:ulam} we show the 95\% c.l.\ upper limit to
$f(a,\massp)$ as a function of $a$ and $\massp$, assuming $\dchit=60$ and point sources. 
This figure is analogous to Figure~\ref{fig:uldq}, except that now
our upper limits are in terms of the physical variables of the
mass of the companion in $\mjup$ and separation of the companion in $\au$, and we have
identified our primaries as M-dwarfs in the Galactic bulge.  
In Figure~\ref{fig:lzulam} we show the 95\% c.l.\ upper limits to the
fraction of lenses with planets in two ranges of orbital separations, 
$(1.5-4)~\au$ and $(1-7)~\au$.  Taking our inference about the nature
of the primary lenses literally, we conclude that less than 33\% of
M-dwarfs in the Galactic bulge have Jupiter mass companions
between 1.5 and 4$~\au$.  Less than 45\% have 3-Jupiter mass companions
between 1 and 7$~\au$.  These are the first significant limits on
planetary companions to M-dwarfs, and are the primary result
of this work.

\section{Discussion\label{sec:discuss}}

The majority of what we know about planetary companions has been
gathered from radial velocity surveys of stars in the local neighborhood. However, these surveys
have told us very little about planetary companions to M-dwarf
primaries, as they have focused on F, G, and K-dwarf
and have only recently begun surveying cooler stars.
To date, the only M-dwarf with known planetary 
companions is Gliese 876 \citep{mbvfl1998,marcy2001b}.  Our
results therefore place interesting limits in an entirely new region of
parameter space.  However, this also means that the comparison between
our results and those of radial velocity surveys is not entirely
straightforward, as we are probing different primaries, and therefore
different regimes of star, disk, and planet formation.  Furthermore,
our primaries are mostly old stars in the bulge, whereas those
studied by radial velocity surveys are relatively young \citep{frs1999,g1999,sim2000}.
Finally, there is evidence that the host stars of local companions
have super-solar metallicity \citep{g1999,sim2000}, whereas stars in the Galactic bulge
likely have solar to sub-solar metallicity. It is not at all 
clear how these differences between the parent samples we probe will affect the various
proposed planet formation mechanisms.

\begin{figure*}[t]
\epsscale{1.0}
\centerline{\plotone{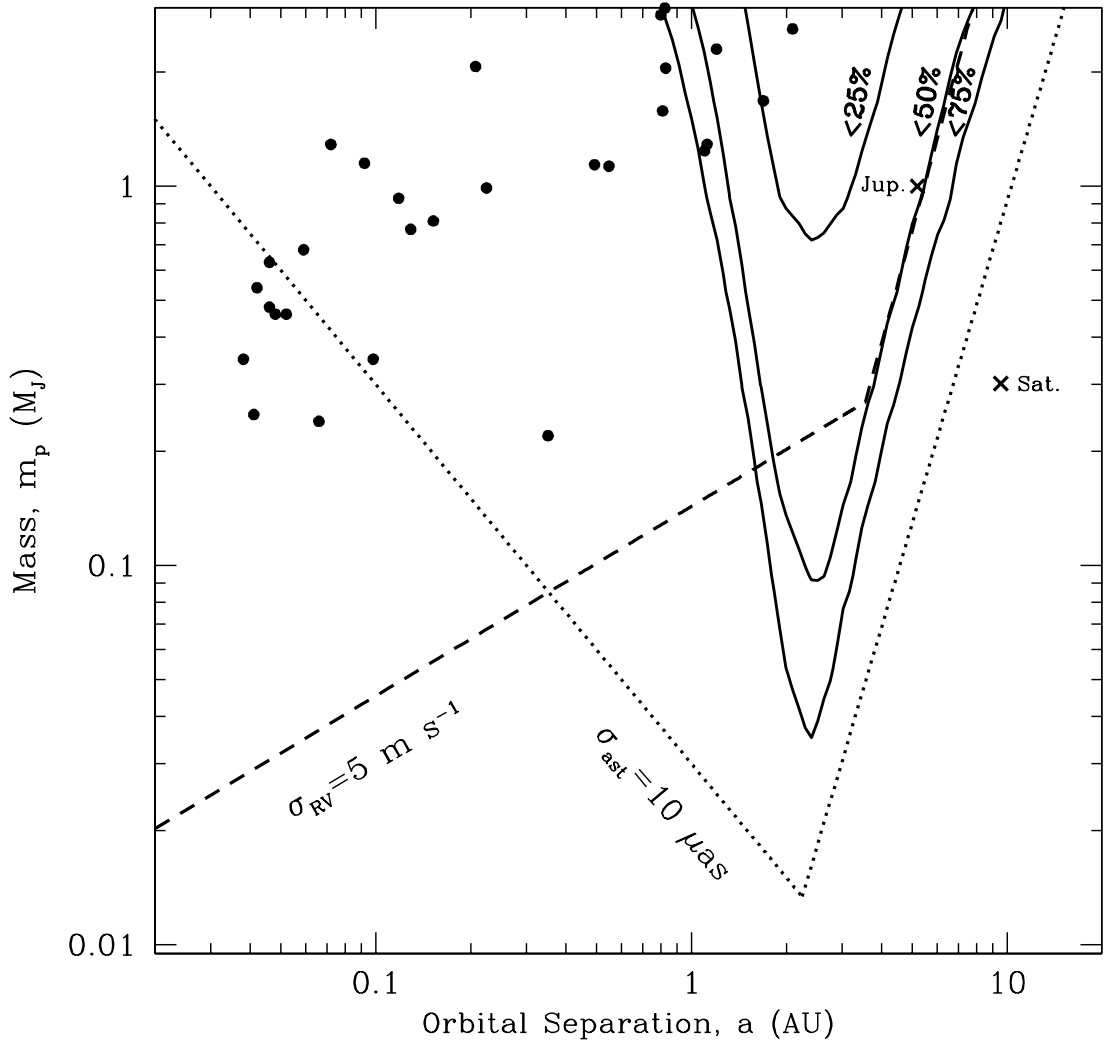}}
\caption{
\footnotesize
Our 95\% c.l.\ upper limit to the fraction of M-dwarf
primaries with a companion as a function of the mass $\massp$ and orbital separation
$a$ of the companion.  The solid black lines show upper limit contours of 75\%, 50\%,
and 25\%.  The points indicate the $\massp \sin{i}$ and $a$  of companions to stars (mostly G-dwarfs)
in the local neighborhood detected by radial velocity surveys. 
Jupiter and Saturn are marked with crosses. The
dashed line shows the radial-velocity detection limit for a precision of $5~{\rm m s^{-1}}$,
a primary mass of $0.3~M_\odot$, and a survey lifetime of
10~years.  The dotted line is the
astrometric detection limit for an accuracy of $10~{\rm \muas}$, a
primary of mass $0.3~M_\odot$ at $10~{\rm pc}$, and a survey lifetime
of 5~years.
}
\label{fig:compare}
\end{figure*}

Rather than attempt to interpret our results in the
context of these various parameters, which may or may not affect planetary formation, 
we simply make a direct comparison between our results
and those of radial velocity surveys.  In Figure~\ref{fig:compare},
we show our 95\% c.l.\ upper limits on the fraction of primaries with a
companion as a function of the mass $\massp$ and orbital separation axis $a$
of the companion, along with the measured $\massp \sin{i}$ and $a$ of
those companions detected by radial velocity surveys. For the most
part, radial velocity surveys are currently sensitive to companions of smaller
$a$ than is microlensing, although there is clearly some overlap.  Also
shown is the radial-velocity detection limit for a precision of $5~{\rm m~s^{-1}}$, 
a primary mass of $0.3~M_\odot$ (typical of our primaries), and a survey lifetime of
10~years. We also show the astrometric detection limit for $0.3~M_\odot$
primaries at $10~{\rm pc}$ expected for SIM, 
which should achieve a precision of $10~\mu{\rm as}$ and have a survey lifetime of five years. 

The results from radial velocity surveys for companions indicate that $f\sim 5\%$ of 
local F, G, and K-dwarfs have companions between $0 \le a \le 3~\au$ \citep{mcm2000}.  
It is interesting to ask how many more events we would need to monitor in
order to limit the fraction of primaries with companions to 5\% in the
range of the separations to which we are sensitive.  From
\eq{pnaive}, we find that, for small $f$, $f \propto N_{\rm exp}^{-1}$.  Given
that our limits are $f \sim 33\%$, we would require $\sim 7$ times more
events of similar quality.  This number could be significantly reduced
if the quality of the alerts could be improved, i.e.\ if a larger
fraction of events we monitor in the future were bright, high-magnification events.
This will likely be possible with the next generation OGLE campaign \citep{udalski2000}.

\section{Summary and Conclusion\label{sec:conclude}}

We have analyzed five years of PLANET photometry of microlensing events
toward the Galactic bulge to search for 
planets.  All of the 126 bulge microlensing events for which PLANET
has acquired data over the last five years can be subdivided into three categories:
events for which the data quality is too poor to determine the nature of the event,
events  that deviate from the single lens in a way not associated with planetary 
companions (roughly equal-mass binaries, parallax, finite source, binary source, etc.), and 
apparently normal point-source point-lens events (PSPL).  We
find no events in a possible fourth category: events that have
short-duration deviations from the single lens light curve that are
indicative of the presence of planetary companions to the primary
microlenses.  This indicates that Jupiter-mass companions to bulge
stars with separations of a few AU are not typical.

In order to justify and quantify this conclusion, we imposed strict
event selection criterion, and derived a well-defined subset of 43
intensively monitored events which we carefully analyzed for the presence of companions.
Using the method of \cite{meands2000}, we searched for the signatures of
planetary companions in these events over a densely sampled, extensive region of
parameter space.  Specifically, we searched for companions with mass
ratios $q$ from $10^{-2} - 10^{-4}$ and instantaneous projected
separations $d$ in units of the angular Einstein ring radius from $0.1
\le d \le 10$.  Based on an analysis of our photometric uncertainties
for constant stars, we required that the difference in $\chi^2$ between the
best-fit binary lens model and the best-fit single lens model be
$<-60$ for a detection candidate.  We found two such candidates, events
\mb{99}{18} and \ob{99}{36}.  Analysis of \mb{99}{18} revealed a significantly better
fit with $q\simeq 0.2$, and was eliminated from
the sample.  \ob{99}{36} displays an overall asymmetry that is
equally-well (in the sense of $\chi^2$) explained by a low-amplitude parallax signal.  Since
we cannot reliably detect planets from global
asymmetries, we explicitly discard this ambiguous anomaly. 
Thus we find no viable planetary candidates out of our original sample of 43
events. 

We then calculated the detection efficiency for our events in $(d,q)$
space.  Of our final sample of 42 events (eliminating \mb{99}{18}), 30 have substantial ($>25\%$)
efficiency for the detection of companions with $q=10^{-2}$ and separations in the
lensing zone $0.6 \le d \le 1.6$.  Had all
of the primary lenses harbored such companions, we should have
detected a planet in at least $\sim 7$ of them.  The fact that we detected no companions
implies that this is not the case.  By combining
our efficiencies, we obtain statistical upper limits on the fraction
of lenses with massive planets in the lensing zone.  At the 95\%
confidence level, we find that $< 25\%$ of lenses can have a companion in the lensing zone
with mass ratio $q=10^{-2}$ . 

Using a model of the mass function, spatial distribution, and velocity distribution
of stars in the Galactic bulge, we infer that the majority of our
lenses are likely due to $M\sim 0.3~\msun$ stars at $6~\kpc$, i.e.\ M
dwarfs in the Galactic bulge.   Using this assumption, we convert our
upper limits from $(q,d)$ space to mass-orbital separation space.  We
conclude that less than 33\% of
M-dwarfs in the Galactic bulge have Jupiter-mass companions
between 1.5 and 4$~\au$, and less than 45\% have 3-Jupiter mass companions
between 1 and 7$~\au$.  These are the first significant limits on
planetary companions to M-dwarfs.

We have also tested the robustness of our conclusions to various assumptions. 
The effect of the finite size of the source
stars was estimated for each event using the color and magnitude of the source
and assuming a mean relative proper motion of the lens.  We find that
the finite source effect becomes important only for mass ratios $q \la
10^{-3}$, where our constraints on companions are already weak.  We therefore
conclude that finite source effects have a negligible effect on our results.
We also tested the effect of changing our detection criterion from
$\dchit=60$ to $\dchit=100$.  As expected, this lowers our sensitivity
somewhat, and increases our upper limits by $\la 20\%$, 
but does not change our conclusions substantially.  Finally,
we have tested the effect of ignoring parallax asymmetries in the
calculation of our detection efficiencies, and find that this changes
our limits by substantially less than our statistical uncertainties.

We find that our median event time scale ($\te=40~\days$) is
a factor of two larger than the median time scale for all events toward
the Galactic bulge, a selection effect that arises from
the manner in which we choose our targets.  We argued that this
primarily biases our events toward slower, rather than closer or more
massive lenses.  Therefore, our assertion of a typical lens mass of
$0.3~\msun$ is justified.  

For the most part our upper limits are for planets with orbital separations that are 
larger than those currently probed by radial velocity techniques, 
since the orbital times are longer than the finite survey lifetimes.  However, the
smallest separations to which we are sensitive overlap with current radial velocity
surveys, and as the radial velocity surveys continue, the degree of overlap will increase.
Thus one will eventually be able to compare the frequency of
companions in the Galactic bulge with that in the solar neighborhood.
We estimate, however, that a sample $\sim 7$ times larger than
that considered here would be needed to probe fractions as small as
those being measured by radial velocity surveys ($\sim 5\%$), assuming
assuming that future microlensing observations are of similar quality to
those analyzed here.  If the number of alerts is increased
substantially, however, more care could be taken to choose
higher-quality (brighter, higher maximum magnification) events.  This
would considerably reduce the number of event needed to probe
companion fractions of $5\%$.   

Our results have implications for theories of planet
formation, as the orbital separations we probe may be closer to
the sites of planet formation than the small separations at
which radial-velocity companions are found, which may be reached via
orbital migration.  In any case, the limits described here provide fundamental
constraints on the frequency and distribution of extrasolar planets orbiting the most
common stars in our Galaxy.

\begin{acknowledgements}

We would like to thank the MACHO, OGLE and EROS collaborations for providing
real-time alerts, without which this work would not be possible, and
MACHO and OGLE for making their data publicly
available.  We single out Andrzej Udalksi and Andy Becker
for the special contributions they have made in this regard.
We are especially
grateful to the observatories that have supported our science
(Canopus, ESO, CTIO, Perth and SAAO) via the generous allocations of time that make this
work possible.  We are indebted to the people that have donated their
time to observe for the PLANET collaboration.  
PLANET acknowledges financial support via award GBE~614-21-009  
from the organization for 
{\sl Nederlands Wetenschappelijk Onderzoek\/} 
(Dutch Scientific Research), 
the Marie Curie Fellowship ERBFMBICT972457 from the European Union, 
a ``coup de pouce 1999'' award from the 
{\sl Minist\`ere de l'\'Education nationale, de la Recherche et de la Technologie, D\'epartement Terre-Univers-Environnement\/}, 
grants AST~97-27520 and AST~95-30619 from the NSF, NASA grant 
NAG5-7589, a Presidential
Fellowship from the Ohio State University, and 
NASA through a Hubble Fellowship grant from the Space Telescope Science Institute, which is operated by the Association of Universities for Research in Astronomy, Inc., under NASA contract NAS5-26555.
\end{acknowledgements}

\appendix

\section{Excluded Anomalous Events\label{sec:app1}}

In \S\ref{sec:esel}, we rejected from the
analysis 19 anomalous events which we asserted were not caused by planetary (i.e.\ small
mass ratio binary) lenses.  Here we list
each of these events, and briefly justify why we believe their anomalies
to be non-planetary in origin.  For those events for which binary-lens fits are available in the
literature, we will simply state the fitted mass ratio(s), and refer the
reader to the paper; for a large fraction of these events, we rely on the
analysis and binary-lens fits of \citet{alcock2000}.  One caveat should be noted.  It is known
\citep{dandh1996,martin1999a, albrow1999b} that binary lens events, even extremely well sampled ones, 
often have degenerate solutions \citep{afonso2000}.  This is due to intrinsic degeneracies in the
binary lens equation \citep{martin1999}.  Finding all of these
degenerate solutions to an observed light curve is highly non-trivial,
due to the extremely sharp variations in $\chi^2$ with respect to the
canonical parameters, although several methods have been proposed to
deal with this difficulty \citep{diandm1996, diandp1998, albrow1999b}.
It is therefore possible, as \citet{alcock2000} allow,
that not all solutions have been found and thus that 
some of the events they analyze actually have
planetary solutions that they missed.  Based
simply on examination of the data we find this unlikely, since the 
deviations from the PSPL form are gross and long duration, contrary to what would be expected
from a small mass ratio binary.   

For caustic-crossing binary-lens events for which the source is
resolved, we can use the following argument to place a lower limit on the
mass ratio $q$.  The maximum magnification obtained when a source of
angular size $\thetas$ crosses a fold caustic is \citep{bible}
\begin{equation}
\acfmax \sim \left( u_{\rm r} \over \theta_* \right)^{1/2},
\label{eqn:acfmax}
\end{equation}
whereas for a cusp caustic, 
\begin{equation}
\accmax \sim \left ( u_{\rm r} \over \theta_* \right).
\label{eqn:accmax}
\end{equation}
Here $u_{\rm r}$ is a factor that describes the characteristic
scale of the caustic.  For caustics originating from binary lenses 
with small $q$, this scale is of order the planetary Einstein ring radius, 
$\thetap$.  Due to possible blending, the observed
maximum magnification, $\acfmobs$  (or $\accmobs$), is a lower limit to the true magnification, and by
combining eqs.~(\ref{eqn:acfmax}), (\ref{eqn:thetae}), and
(\ref{eqn:rhos}), we obtain an approximate lower limit on $q$ for
a fold caustic crossing:
\begin{equation}
\qmin \ga 0.01 \left( {\acfmobs \over 6} \right)^{4} 
\left( { \thetas \over 1~\mu{\rm as}}\right)^{2}
\left( { \murel \over 12.5~\kms~{\rm kpc^{-1}}}\right)^{-2}
\left( { \teobs \over 40~\days}\right)^{-2},
\label{eqn:qminf}
\end{equation}
and combining eqs.~(\ref{eqn:accmax}), (\ref{eqn:thetae}),
and (\ref{eqn:rhos}), we obtain a similar relation for a cusp crossing:
\begin{equation}
\qmin \ga 0.01 \left( {\accmobs \over 30} \right)^{2} 
\left( { \thetas \over 1~\mu{\rm as}}\right)^{2}
\left( { \murel \over 12.5~\kms~{\rm kpc^{-1}}}\right)^{-2}
\left( { \teobs \over 40~\days}\right)^{-2},
\label{eqn:qminc}
\end{equation}
where $\teobs$ is the observed (i.e.\ blended) time scale of the event,
which is always a lower limit to the true time scale.  
Since $\qmin$ is proportional to $\teobs$ squared, while $\qmin$
is proportional to $\acfmobs$ to the fourth power and $\accmobs$ squared,  
the limits in Eqs.\ \ref{eqn:qminc} and \ref{eqn:qminf}  hold
even in the presence of blending.  
The smallest sources in the
Galactic bulge have $\thetas \sim 1~\mu{\rm as}$, and the dispersion
in $\murel$ for bulge-bulge lensing is a factor of $\sim 2$.  Thus, 
an observed fold crossing with $\acfmobs \ga 10$ is almost
certainly due to binary lens with mass ratio $q \ge 0.01$.  
A cusp crossing with
$\accmobs \ga 40$ is almost certainly due to a binary with $q \ge 0.01$. 
In general, for reasonably well-sampled events, a cusp approach can be easily distinguished by eye from caustic crossing events.  
For disk-disk lensing, for which $\murel \sim 5~\kms~{\rm kpc^{-1}}$,
somewhat smaller mass ratios are allowed; however such events are 
generally rare. 

\begin{description}
\item[{\mb{95}{12}}] Both PLANET \citep{albrow1998}, and MACHO/GMAN
\citep{alcock2000} data show a smooth double-peaked event, with both
peaks having comparable duration.  This morphology suggests a weak
binary lens or binary source \citep{gandhu1992}.   However, the
achromaticity of the event favors a binary-lens interpretation, and
we find that a binary-source model provides a poor fit to the PLANET
data.  We cannot uniquely constrain a binary-lens fit, but
\citet{alcock2000} find a binary-lens fit with mass ratio $q=0.47$.
The fact that the peaks are of comparable duration precludes a small
mass ratio binary-lens (i.e.\ planetary) model. 
\item[{\mb{96}{04}}] MACHO/GMAN data show two
nearly equal-duration deviations separated by $\sim 500~\days$ \citep{alcock2000}.  Both
deviations are separately well-fit by a standard PSPL model,
suggesting a widely-separated binary-source or binary-lens
\citep{diandm1996}.  \citet{alcock2000} find $q=0.88$ for their
binary-lens fit.  Regardless of the interpretation, the PLANET data on
this event would not have passed our second cut, due to insufficient data.   
\item[{\mb{97}{28}}] We find only one viable model that fits our data for this event
\citep{albrow1999a}, with $q=0.23$. \citet{alcock2000} find a similar
binary-lens model fit for their dataset, with $q=0.21$.
\item[{\mb{97}{41}}] Our data for this peculiar event is well fit by a rotating binary-lens
model with mass ratio $q=0.34$ \citep{albrow2000a}.
\citet{bennett1999} favor the interpretation that this event is 
a planet orbiting a binary lens.  Our data are clearly
inconsistent with their particular fit, although this does not
preclude the possibility that some fit of this nature would explain
our data.  Regardless of the interpretation, this event is
rejected because of the presence of the binary. 
\item[{\mb{98}{6}}] This is a long-timescale ($>100~\days$) event which
shows global deviations from the PSPL form indicative of parallax.  
\item[{\mb{98}{12}}] MACHO/GMAN data indicate that this event likely 
underwent four caustic crossings, with each pair of crossings separated by $\sim 40~\days$
\citep{alcock2000}.  The MACHO/GMAN data have poor coverage of the first set of
caustic crossings, but constrain the amount of time that the
source was between the second set of crossings to be $\lsim 3~\days$.
Due to its short duration, one might suppose that the second set of crossings
was due to a planetary caustic.  However, the first set of caustic
crossings, combined with the fact that the event exhibits a {\it rise}
toward the second set of crossings, makes this interpretation
impossible.  
Indeed, \citet{alcock2000} find that the event is well fit by an intermediate-topology
binary lens with $q=0.68$.  PLANET acquired a few data points immediately after the
second crossing, and data immediately after the fourth crossing
continuing until the end of the event.  Due to the fact that
the PLANET data did not probe any of the caustic structures, we find
that our dataset is reasonably well fit by a PSPL model.  However, our
data alone fail our $\delta\u0/\u0$ cut.
\item[{\mb{98}{14}}] Both the MACHO/GMAN dataset  \citep{alcock2000}
and the PLANET dataset show a highly asymmetric light curve with a
``shoulder'' and then a peak.  Such a morphology is indicative of a
weakly-perturbed binary-lens event, and as such is prone
to degeneracies.  In fact \citet{alcock2000} find two fits, one
with mass ratio $q=0.09$ and the other with $q=0.22$.  However, the
event deviates from the PSPL form for a large fraction ($\sim 40\%$)
of its apparent duration, making a planetary interpretation unlikely. 
We performed a systematic search of binary-lens fits to this event,
using our data and the MACHO data.  We recover the fits reported by
\citep{alcock2000}, along with a few other fits of similar
significance.  The best-fit binary with $q<0.01$ is ruled out at the
$\Delta\chi^2=50$ level.
\item[{\mb{98}{16}}] MACHO/GMAN data show a short duration peak,
followed by an abrupt rise and a plateau at magnification $\sim 10$ that lasts $\sim
8~\days$. Following the plateau, the event returned to magnification
$\sim 2$ \citep{alcock2000}.  Although the coverage is poorer,  PLANET data
qualitatively confirm this behavior.  This morphology is consistent
with a caustic-crossing binary lens event in which the short-time scale peak
is due to a cusp approach, followed by a pair of fold caustic crossings
with the usual intra-caustic plateau.
MACHO/GMAN data near the peak of the first fold caustic crossing have
$\acfmobs \sim 20$, and thus constrain the event to be non-planetary
by \eq{qminf}.  Indeed, \citet{alcock2000} find a binary-lens fit with $q=0.68$.
\item[{\mb{98}{42}}]  \citet{alcock2000} find $q=0.33$.  PLANET data
cover the second half of the event, including the falling side of a
second caustic crossing.  Our data of the second crossing show no evidence of a
cusp approach, favoring a pure fold caustic crossing. The data near the peak of 
this fold crossing have $\acfmobs \sim 40$; therefore the
event must be non-planetary in origin [eq.(\ref{eqn:qminf})]. 
\item[{\ob{98}{28}}]  This event displays a double-peaked structure
indicative of a weak binary-lens or binary source.  This is seen in
both OGLE and PLANET data for the event.  Using the combined dataset, we find the best-fit
binary-lens model has $q=0.34$ and $b=0.42$.  Normalizing the errors
to this model, the best model in the range $q=10^{-4}-10^{-2}$ has
$\Delta\chi^2 \sim 19$, and thus is excluded.  
\item[{\ob{98}{29}}]  PLANET data for this high-magnification ($\amax \sim 50$) event
show deviations from the PSPL form near the peak of the event 
that are indicative of source resolution effects.  We find that a point-lens finite-source model fits the
data quite well.  In contrast, we find that the best-fit point-source planetary model in the range $q=10^{-4}-10^{-2}$ is a significantly worse fit ($\Delta\chi^2>100$). 
\item[{\mb{99}{8}}] Similar to \mb{98}{6}, this long-time scale event
shows severe parallax effects.  We also find short time scale
variability in the source.
\item[{\mb{99}{22}}] Although the PLANET, MACHO, and OGLE data show no
obvious anomalous behavior, our PSPL fit to the combined datasets
yielded a time scale of $\te\sim 900~\days$, leading us to suspect
parallax effects might be present.  In fact, we find
that a fit with parallax improves $\chi^2$ significantly, and results
in a much more reasonable time scale.  This interpretation is confirmed
by the analysis of \citet{mao2001}.  
This event is excluded since our algorithm
does not currently allow the search for planets atop other microlensing anomalies.
\item[{\mb{99}{25}}]MACHO data for MACHO 99-BLG-25 show a clear deviation from
PSPL at early times, in the form a smaller amplitude, but
nearly equal duration peak occurring before PLANET began monitoring the
event.  The fact that both peaks are of similar duration suggests that
this event is likely due to a binary
source, and exclude the possibility that it is due to a planet.  
Our data  only cover the rise and fall of the second
peak and are perfectly consistent with a PSPL model.  In fact, we find that this event does not
have a significant planetary signal, nor does it have a large detection
efficiency to planetary companions.  Therefore excluding this event
has no significant impact on our conclusions. 
\item[{\mb{99}{47}}] PLANET data show a departure from the PSPL form lasting
$\sim 3~\days$ near the peak.  Detailed analysis of this event shows
that the deviation is caused by a small separation binary with $q\sim
0.4$ \citep{albrow2001b}.
\item[{\mb{99}{57}}] MACHO data show a large, long duration deviation
from the PSPL form that is likely due to a binary-source or
binary-lens.  PLANET has very little data on this event, and so cannot
confirm or clarify the nature of this anomaly.  
\item[{\ob{99}{11}}] A caustic-crossing binary lens; PLANET data
resolve the second crossing.  The full dataset indicate a pair of pure
fold caustic crossings. The second (fold) crossing
has $\acfmobs \sim 10$; thus the event must be non-planetary in
origin [eq.(\ref{eqn:qminf})].
\item[{\ob{99}{23}}] We find only one viable fit to this event, with
$q=0.39$ \citep{albrow2000c}.
\item[{\ob{99}{25}}] PLANET data show a large positive deviation lasting $<
1~\day$ superposed atop an extremely noisy light curve.  As we
see a sharp change in the slope of the light curve immediately after this deviation, we
conclude that it is due a caustic crossing of some kind.  It is not clear whether
this deviation is due to a cusp or fold caustic crossing.  We therefore conservatively
assume that it is due to a cusp. The observed
magnification at the peak of this deviation is $\accmobs \sim 40$, and
thus from \eq{qminc}, the deviation cannot be due to planet.
\item[{\ob{99}{42}}] OGLE data indicate a double peaked structure to the light curve, which is 
likely due to an nearly equal-mass binary lens or a binary source.  The PLANET data cover the rise and fall
of the second peak, and are consistent with a single lens model.  Regardless of the 
nature of the anomaly, the light curve would not pass the cut on the uncertainty in $\u0$.

\end{description}

\section{\ob{99}{36} and Parallax Contamination\label{sec:parallax}}

\begin{figure*}[t]
\epsscale{1.0}
\centerline{\plotone{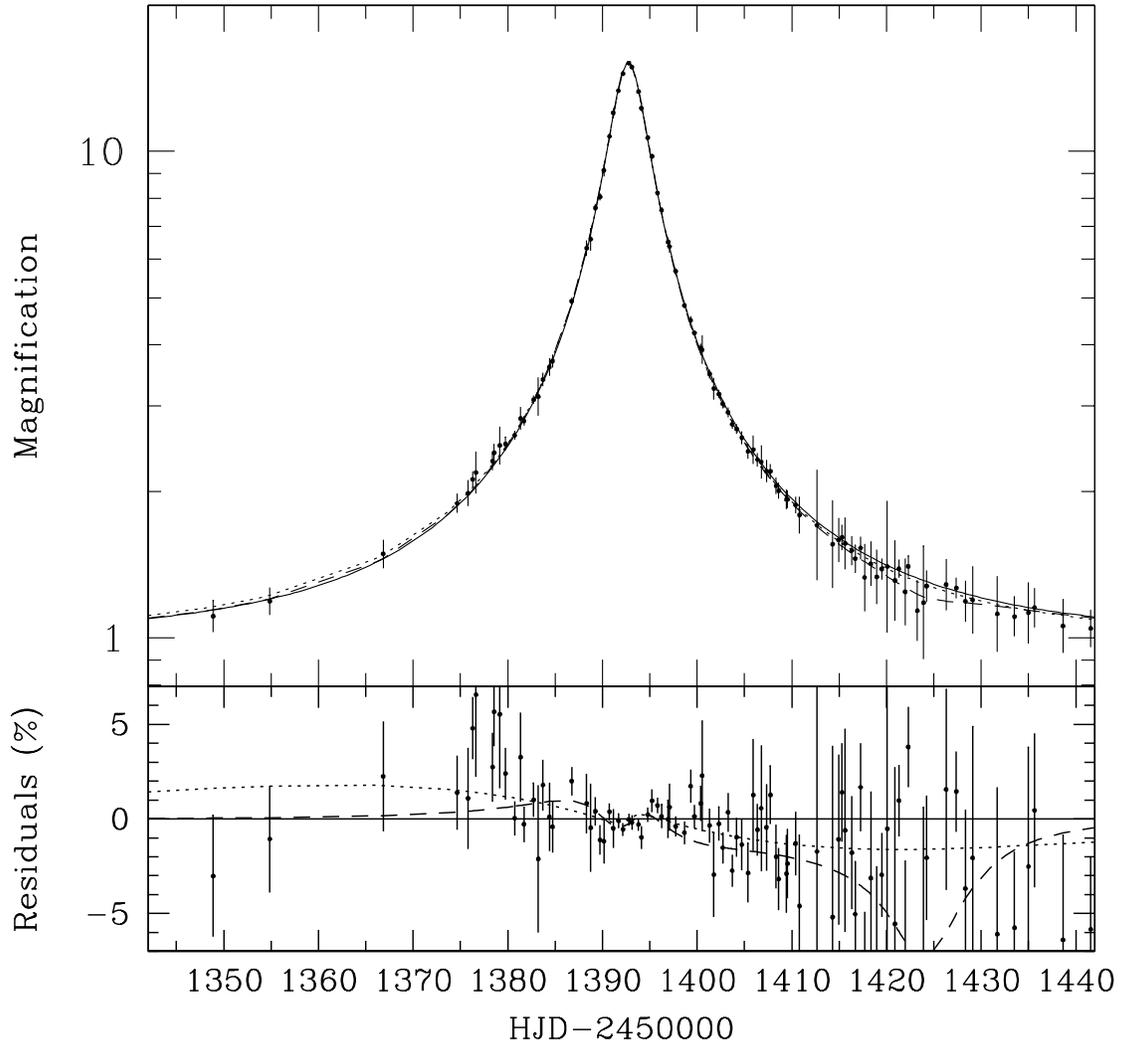}}
\caption{
\footnotesize
{\it Top Panel}: Points show the magnification as a function of
time for PLANET and OGLE data of event \ob{99}{36}, binned into 1 day
intervals.  The solid line shows the best-fit point-source point-lens (PSPL)
model, the dotted line the best-fit parallax asymmetry
model, and the dashed line the best-fit binary model.  {\it
Bottom Panel}:  The residuals from the best-fit PSPL model (in \%) as
a function of time.  The dotted (dashed) line shows the deviation of the
parallax asymmetry (binary-lens) model from the PSPL model.
}
\label{fig:ob36}
\end{figure*}

Figure~\ref{fig:ob36} shows the PLANET and OGLE data for event \ob{99}{36},
binned into 1 day intervals, along with the best-fit single-lens,
binary-lens, and parallax asymmetry models to the unbinned data. The
difference in $\chi^2$ between the binary-lens and parallax models is
$4$ (with the binary-lens model giving the worse fit); both models are
favored over the PSPL model by $\Delta\chi^2 \sim 80$.  The best-fit
binary lens model has $q=0.0028$, $d=0.60$, and $\alpha =1^\circ$.  We also find
fits for other mass ratios and separations that are nearly as good.
The parallax asymmetry fit (see \citealt{gbm1994} for the exact form)
yields a measurement of the asymmetry factor, $\kappa$, given by,
\begin{equation}
\kappa = \Omega_{\oplus} {v_{\oplus}\over \vtilde} \sin\lambda \sin\phi,
\label{eqn:kappa}
\end{equation}
where $\vtilde = \vperp (\dos/\dls)$ is the transverse velocity of the lens projected on the
observer plane, $\Omega_{\oplus}=2\pi~{\rm yr}^{-1}$, $v_{\oplus}\simeq 30~\kms$
is the speed of the Earth, and $\lambda$ is the angle between the source
and Sun at the time of maximum magnification.  In the case of \ob{99}{36},
$\sin\lambda\sim0.6$.  We find $\kappa=0.0021 \pm 0.0001$, which
implies,
\begin{equation}
{\vtilde\over \sin{\phi}}=143 \pm 7~\kms.
\label{eqn:vtocos}
\end{equation}
Combining this constraint with $\te$, we find an upper limit to the
mass of the lens as a function of the distance to the lens,
\begin{equation}
M \la 0.1~\msun { {1-x} \over x},
\label{eqn:mlimits}
\end{equation}
where $x=\dol/\dos$.  Thus, if the parallax interpretation is correct,
the lens must be closer to us than $\sim 4~\kpc$ in order to be above the hydrogen
burning limit.  

\begin{figure*}[t]
\epsscale{1.0}
\centerline{\plotone{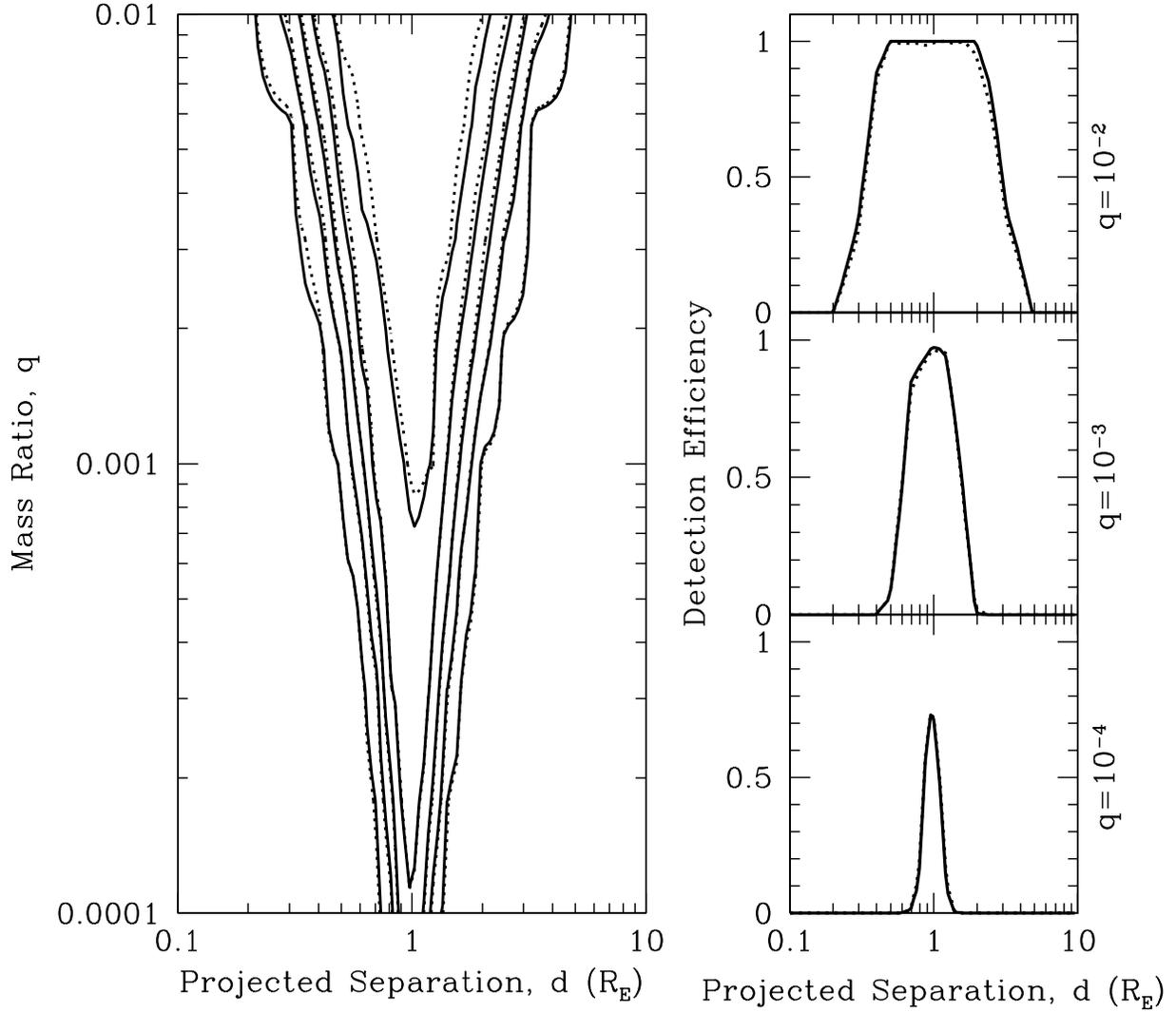}}
\caption{
\footnotesize
{\it Left Panel}: Contours of constant detection efficiency, $\epsilon$, as a
function of mass ratio and projected separation for event \ob{98}{14}.
The contours are $\epsilon=5\%, 25\%, 50\%, 75$\%, and $95\%$ (outer
to inner).  The solid contours are the efficiencies under the
assumption that the parallax asymmetry parameter is zero, while the dotted
contours are calculated with the asymmetry as a free parameter.
{\it Right Panels}: Detection efficiencies as a function of projected
separation for three mass ratios $q$.  The solid lines are for no parallax
asymmetry, and the dashed lines are with asymmetry as a free parameter.
}
\label{fig:ob14}
\end{figure*}

The primary lesson learned from the analysis of \ob{99}{36} is that we
cannot robustly detect planetary companions based on global asymmetries, since
they cannot be distinguished from low-level parallax.
However, when calculating our detection efficiencies
(\S\ref{sec:des}), we excluded all deviations that produced
$\dchi>60$, including asymmetries.  Therefore, our efficiencies are
overestimated.  In order to estimate by how much, we choose a well-sampled,
high-quality event, \ob{98}{14}, that contains data on both rising and falling
sides.  We repeat the algorithm in \S\ref{sec:algorithm}
to calculate the efficiency of this event but simultaneously
fit for both the binary-lens magnification and parallax asymmetry.
This procedure removes all detections based on asymmetry alone.  In
Figure~\ref{fig:ob14} we show the detection efficiency both with and
without excluding such detections. The difference is quite small, a
few percent, because a very small range of angles produce
deviations consistent with asymmetries.  The majority of our events have
sampling and photometric accuracy that is poorer than \ob{98}{14}, in
which case they will be less sensitive to asymmetries.  We
therefore conclude that this effect is negligible.
  
{}

\clearpage

\newpage

\begin{table}
\begin{tabular}{cc|cc}
\hline\hline
Official Alert Name & Abbreviated Name & Status  & Notes \\ 
\hline
 MACHO 95-BLG-10& MB95010 & Cut & Uncertainty in $u_0$ \\
 MACHO 95-BLG-12& MB95012 & Cut & Binary\tablenotemark{1,2} \\	
 MACHO 95-BLG-13& MB95013 & Passed & -- \\
 MACHO 95-BLG-17& MB95017 & Cut & Uncertainty in $u_0$ \\
 MACHO 95-BLG-18& MB95018 & Cut & Uncertainty in $u_0$ \\
 MACHO 95-BLG-19& MB95019 & Passed & -- \\
 MACHO 95-BLG-21& MB95021 & Cut & Insufficient Data \\
 MACHO 95-BLG-24& MB95024 & Cut & Insufficient Data \\
 MACHO 95-BLG-25& MB95025 & Cut & Insufficient Data \\
 MACHO 95-BLG-30& MB95030 & Cut & Uncertainty in $u_0$ \\
 OGLE-1995-BUL-04& OB95004 & Cut & Uncertainty in $u_0$ \\
\hline
 MACHO 96-BLG-1& MB96001 & Cut & Insufficient Data \\
 MACHO 96-BLG-4& MB96004 & Cut & Binary\tablenotemark{2}  \\
 MACHO 96-BLG-5& MB96005 & Cut & Uncertainty in $u_0$ \\
 MACHO 96-BLG-6& MB96006 & Cut & Uncertainty in $u_0$ \\
 MACHO 96-BLG-7& MB96007 & Cut & Uncertainty in $u_0$ \\
 MACHO 96-BLG-8& MB96008 & Cut & Uncertainty in $u_0$ \\
 MACHO 96-BLG-9& MB96009 & Cut & Uncertainty in $u_0$ \\
 MACHO 96-BLG-10& MB96010 & Cut & Uncertainty in $u_0$ \\
 MACHO 96-BLG-11& MB96011 & Passed & -- \\
 MACHO 96-BLG-12& MB96012 & Cut & Uncertainty in $u_0$ \\
 MACHO 96-BLG-13& MB96013 & Cut & Insufficient Data \\
 MACHO 96-BLG-14& MB96014 & Cut & Uncertainty in $u_0$ \\
 MACHO 96-BLG-15& MB96015 & Cut & Insufficient Data \\
 MACHO 96-BLG-16& MB96016 & Passed & -- \\
 MACHO 96-BLG-17& MB96017 & Cut & Insufficient Data \\
 MACHO 96-BLG-18& MB96018 & Passed & -- \\
 MACHO 96-BLG-19& MB96019 & Passed & -- \\
 MACHO 96-BLG-20& MB96020 & Cut & Uncertainty in $u_0$ \\
 MACHO 96-BLG-21& MB96021 & Cut & Uncertainty in $u_0$ \\
 MACHO 96-BLG-23& MB96023 & Cut & Uncertainty in $u_0$ \\
 MACHO 96-BLG-24& MB96024 & Cut & Insufficient Data \\
 MACHO 96-BLG-25& MB96025 & Cut & Uncertainty in $u_0$ \\
 MACHO 96-BLG-26& MB96026 & Cut & Uncertainty in $u_0$ \\
 MACHO 96-BLG-27& MB96027 & Cut & Insufficient Data \\
\hline
\end{tabular}
\caption{All Events from 1995-1999 with PLANET data.}\label{tab:tab01}
\end{table}

\begin{table}
\begin{tabular}{cc|cc}
\hline\hline
Official Alert Name & Abbreviated Name & Status  & Notes \\ 
\hline
 MACHO 97-BLG-10& MB97010 & Cut & Insufficient Data \\
 MACHO 97-BLG-18& MB97018 & Passed & -- \\
 MACHO 97-BLG-25& MB97025 & Passed & -- \\
 MACHO 97-BLG-26& MB97026 & Passed & -- \\
 MACHO 97-BLG-28& MB97028 & Cut & Binary\tablenotemark{2,3} \\
 MACHO 97-BLG-30& MB97030 & Passed & -- \\
 MACHO 97-BLG-31& MB97031 & Passed & -- \\
 MACHO 97-BLG-36& MB97036 & Cut & Insufficient Data \\
 MACHO 97-BLG-37& MB97037 & Cut & Uncertainty in $u_0$ \\
 MACHO 97-BLG-41& MB97041 & Cut & Binary\tablenotemark{4} \\
 MACHO 97-BLG-49& MB97049 & Cut & Insufficient Data \\
 MACHO 97-BLG-50& MB97050 & Cut & Uncertainty in $u_0$ \\
 MACHO 97-BLG-52& MB97052 & Cut & Uncertainty in $u_0$ \\
 MACHO 97-BLG-54& MB97054 & Cut & Insufficient Data \\
 MACHO 97-BLG-56& MB97056 & Cut & Insufficient Data \\
 MACHO 97-BLG-58& MB97058 & Cut & Insufficient Data \\
 MACHO 97-BLG-59& MB97059 & Cut & Insufficient Data \\
\hline
\end{tabular}
\\
\\
Table 1: Continued\\
\end{table}

\begin{table}
\begin{tabular}{cc|cc}
\hline\hline
Official Alert Name & Abbreviated Name & Status  & Notes \\ 
\hline
 EROS BLG-1998-2& EB98002 & Passed & -- \\
 EROS BLG-1998-4& EB98004 & Cut & Insufficient Data \\
 MACHO 98-BLG-1& MB98001 & Cut & Insufficient Data \\
 MACHO 98-BLG-5& MB98005 & Cut & Insufficient Data \\
 MACHO 98-BLG-6& MB98006 & Cut & Parallax \\
 MACHO 98-BLG-12& MB98005 & Cut & Binary \\
 MACHO 98-BLG-13& MB98013 & Passed & -- \\
 MACHO 98-BLG-14& MB98013 & Cut & Binary\\
 MACHO 98-BLG-16& MB98016 & Cut & Binary\\
 MACHO 98-BLG-17& MB98017 & Cut & Uncertainty in $u_0$ \\
 MACHO 98-BLG-18& MB98018 & Cut & Uncertainty in $u_0$ \\
 MACHO 98-BLG-19& MB98019 & Cut & Insufficient Data \\
 MACHO 98-BLG-22& MB98022 & Cut & Insufficient Data \\
 MACHO 98-BLG-24& MB98024 & Cut & Insufficient Data \\
 MACHO 98-BLG-26& MB98026 & Passed & -- \\
 MACHO 98-BLG-27& MB98027 & Cut & Uncertainty in $u_0$\tablenotemark{a} \\
 MACHO 98-BLG-28& MB98028 & Cut & Insufficient Data \\
 MACHO 98-BLG-30& MB98030 & Passed & -- \\
 MACHO 98-BLG-31& MB98031 & Cut & Insufficient Data \\
 MACHO 98-BLG-33& MB98033 & Passed & -- \\
 MACHO 98-BLG-35& MB98035 & Passed & -- \\
 MACHO 98-BLG-37& MB98037 & Cut & Uncertainty in $u_0$ \\
 MACHO 98-BLG-38& MB98038 & Cut & Uncertainty in $u_0$ \\
 MACHO 98-BLG-39& MB98039 & Cut & Uncertainty in $u_0$ \\
 MACHO 98-BLG-40& MB98040 & Cut & Uncertainty in $u_0$ \\
 MACHO 98-BLG-42& MB98042 & Cut & Binary\tablenotemark{2}\\
 $*$OGLE-1998-BUL-13& OB98013 & Passed & -- \\
 $*$OGLE-1998-BUL-14& OB98014 & Passed & -- \\
 $*$OGLE-1998-BUL-15& OB98015 & Passed & -- \\
 $*$OGLE-1998-BUL-18& OB98018 & Passed & -- \\
 $*$OGLE-1998-BUL-20& OB98020 & Cut & Insufficient Data \\
 $*$OGLE-1998-BUL-21& OB98021 & Passed & -- \\
 $*$OGLE-1998-BUL-23& OB98023 & Passed & -- \\
 $*$OGLE-1998-BUL-25& OB98025 & Passed & -- \\
 $*$OGLE-1998-BUL-26& OB98026 & Cut & Uncertainty in $u_0$ \\
 $*$OGLE-1998-BUL-27& OB98027 & Cut & Uncertainty in $u_0$ \\
 $*$OGLE-1998-BUL-28& OB98028 & Cut & Binary \\
 $*$OGLE-1998-BUL-29& OB98029 & Cut & Finite Source \\
 $*$OGLE-1998-BUL-30& OB98030 & Passed & -- \\
\hline
\end{tabular}
\\
\\
Table 1: Continued\\
\end{table}

\begin{table}
\begin{tabular}{cc|cc}
\hline\hline
Official Alert Name & Abbreviated Name & Status  & Notes \\ 
\hline
 EROS BLG-1999-1& EB99001 & Passed & -- \\
 EROS BLG-1999-2& EB99002 & Cut & Insufficient Data \\
 $*$MACHO 99-BLG-6 & MB99006 & Passed & -- \\
 $*$MACHO 99-BLG-8& MB99008 & Cut & Parallax \\
 $*$MACHO 99-BLG-11& MB99011 & Passed & -- \\
 $*$MACHO 99-BLG-18& MB99018 & Passed & Deviation near peak\tablenotemark{b}\\
 $*$MACHO 99-BLG-22& MB99022 & Cut & Parallax \\
 $*$MACHO 99-BLG-24& MB99024 & Passed & -- \\
 $*$MACHO 99-BLG-25& MB99025 & Cut & Binary Source? \\
 $*$MACHO 99-BLG-34& MB99034 & Passed & -- \\
 $*$MACHO 99-BLG-37& MB99037 & Passed & -- \\
 $*$MACHO 99-BLG-42& MB99042 & Cut & Insufficient Data \\
 $*$MACHO 99-BLG-45& MB99045 & Cut & Insufficient Data \\
 $*$MACHO 99-BLG-47& MB99047 & Cut & Binary \\
 $*$MACHO 99-BLG-57& MB99057 & Cut & Binary Lens/Binary Source? \\
 $*$OGLE-1998-BUL-5& OB99005 & Passed & -- \\
 $*$OGLE-1999-BUL-7& OB99007 & Passed & -- \\
 $*$OGLE-1999-BUL-8& OB99008 & Passed & -- \\
 $*$OGLE-1999-BUL-11& OB99011 & Cut & Binary \\
 $*$OGLE-1999-BUL-13& OB99013 & Passed & -- \\
 $*$OGLE-1999-BUL-14& OB99014 & Cut & Uncertainty in $u_0$ \\
 $*$OGLE-1999-BUL-16& OB99016 & Passed & -- \\
 $*$OGLE-1999-BUL-17& OB99017 & Cut & Insufficient Data \\
 $*$OGLE-1999-BUL-19& OB99019 & Cut & Insufficient Data \\
 $*$OGLE-1999-BUL-22& OB99022 & Passed & -- \\
 $*$OGLE-1999-BUL-23& OB99023 & Cut & Binary\tablenotemark{5} \\
 $*$OGLE-1999-BUL-25& OB99025 & Cut & Binary\\
 $*$OGLE-1999-BUL-27& OB99027 & Passed & -- \\
 $*$OGLE-1999-BUL-33& OB99033 & Passed & -- \\
 $*$OGLE-1999-BUL-35& OB99035 & Passed & -- \\
 $*$OGLE-1999-BUL-36& OB99036 & Passed & Global Asymmetry\tablenotemark{c} \\
 $*$OGLE-1999-BUL-39& OB99039 & Passed & -- \\
 $*$OGLE-1999-BUL-40& OB99040 & Cut & Insufficient Data \\
 $*$OGLE-1999-BUL-42& OB99042 & Cut & Uncertainty in $u_0$\tablenotemark{a}\\
 $*$OGLE-1999-BUL-43& OB99043 & Cut & Insufficient Data \\
\hline
\end{tabular}
\tablenotetext{a}{Also shows evidence for binarity}
\tablenotetext{b}{The lightcurve of MACHO 99-BLG-18 has a small deviation near the peak of the
event that is fit by a nearly equal mass binary lens.  It is therefore excluded from the
final event sample.}
\tablenotetext{c}{The lightcurve of OGLE-1999-BUL-36 has a global asymmetry that is equally 
well-fit by a planetary model and a parallax asymmetry model.  See \S6.3.}
\tablenotetext{*}{MACHO and/or OGLE data included in the PSPL fit.}
\tablerefs{ (1)\citet{albrow1998}; (2) \citet{alcock2000};
(3)\citet{albrow1999a}; (4)\citet{albrow2000a};
(5)\citet{albrow2000c}.}
\\
\\
Table 1: Continued\\
\end{table}

\begin{table}
\begin{tabular}{c|cccc}
\hline\hline
Event Name & $t_0$\tablenotemark{a} & $t_E$\tablenotemark{a} & $u_0$\tablenotemark{a} & $\delta u_0/u_0 $  \\ 
&(HJD-2450000) & (days) & & (\%) \\
\hline
MB95013 & $-$101.169 $\pm$  0.034 &  80.85 $\pm$   2.71 &  0.245 $\pm$  0.010 &          3  \\
MB95019 & $-$93.573 $\pm$  0.030 &  38.22 $\pm$   6.41 &  0.189 $\pm$  0.035 &         18  \\
MB96011 & 241.405 $\pm$  0.058 &  10.59 $\pm$   1.66 &  0.223 $\pm$  0.046 &         20  \\
MB96016 & 259.777 $\pm$  0.265 &  57.19 $\pm$  24.70 &  0.094 $\pm$  0.046 &         48  \\
MB96018 & 259.391 $\pm$  0.041 &   7.07 $\pm$   1.98 &  0.132 $\pm$  0.045 &         34  \\
MB96019 & 266.944 $\pm$  0.129 &  12.03 $\pm$   1.95 &  0.292 $\pm$  0.065 &         22  \\
MB97018 & 609.529 $\pm$  0.497 & 100.19 $\pm$  30.14 &  0.329 $\pm$  0.134 &         40  \\
MB97025 & 598.011 $\pm$  1.423 &  20.75 $\pm$   6.18 &  0.342 $\pm$  0.164 &         47  \\
MB97026 & 636.624 $\pm$  0.007 &  68.17 $\pm$   2.70 &  0.113 $\pm$  0.005 &          4  \\
MB97030 & 601.243 $\pm$  0.050 &  23.43 $\pm$   2.69 &  0.080 $\pm$  0.011 &         13  \\
MB97031 & 593.439 $\pm$  1.772 &  41.05 $\pm$   4.33 &  0.645 $\pm$  0.128 &         19  \\
EB98002 & 964.024 $\pm$  0.041 &  23.97 $\pm$   1.12 &  0.335 $\pm$  0.021 &          6  \\
MB98013 & 930.529 $\pm$  0.027 &  18.24 $\pm$   3.47 &  0.063 $\pm$  0.014 &         21  \\
MB98026 & 986.377 $\pm$  0.022 &  33.73 $\pm$   1.15 &  0.229 $\pm$  0.009 &          4  \\
MB98030 & 992.117 $\pm$  0.093 &  26.95 $\pm$   8.95 &  0.285 $\pm$  0.111 &         38  \\
MB98033 & 990.463 $\pm$  0.002 &   7.33 $\pm$   0.15 &  0.148 $\pm$  0.004 &          2  \\
MB98035 & 999.157 $\pm$  0.001 &  27.46 $\pm$   1.17 &  0.0100 $\pm$  0.0005 &          4  \\
$*$OB98013 & 945.081 $\pm$  0.157 &  55.30 $\pm$   3.03 &  0.299 $\pm$  0.022 &          7  \\
$*$OB98014 & 956.033 $\pm$  0.005 &  41.52 $\pm$   0.78 &  0.061 $\pm$  0.001 &          2  \\
$*$OB98015 & 943.840 $\pm$  0.007 &  52.24 $\pm$  10.72 &  0.006 $\pm$  0.001 &         24  \\
$*$OB98018 & 971.078 $\pm$  0.006 &   7.64 $\pm$   0.16 &  0.208 $\pm$  0.006 &          2  \\
$*$OB98021 & 992.190 $\pm$  0.410 &  26.64 $\pm$   5.75 &  0.419 $\pm$  0.135 &         32  \\
$*$OB98023 & 998.735 $\pm$  0.151 &  18.60 $\pm$   2.92 &  0.514 $\pm$  0.107 &         20  \\
$*$OB98025 & 1041.701 $\pm$  0.250 &  50.49 $\pm$   7.88 &  0.298 $\pm$  0.063 &         21  \\
$*$OB98030 & 1043.417 $\pm$  0.069 &  54.26 $\pm$  21.85 &  0.049 $\pm$  0.021 &         42  \\
EB99001 & 1415.023 $\pm$  0.021 &  20.19 $\pm$   1.39 &  0.517 $\pm$  0.045 &          8  \\
$*$MB99006 & 1247.546 $\pm$  0.130 &  27.24 $\pm$   1.69 &  0.150 $\pm$  0.017 &         11  \\
$*$MB99011 & 1286.711 $\pm$  0.117 &  45.17 $\pm$   2.42 &  0.191 $\pm$  0.015 &          7  \\
$*$MB99018 & 1301.897 $\pm$  0.019 &  21.69 $\pm$   0.52 &  0.462 $\pm$  0.016 &          3  \\
$*$MB99024 & 1304.710 $\pm$  0.192 &  59.74 $\pm$   7.37 &  0.151 $\pm$  0.024 &         15  \\
$*$MB99034 & 1326.699 $\pm$  0.152 &   7.04 $\pm$   0.59 &  0.332 $\pm$  0.049 &         14  \\
$*$MB99037 & 1354.220 $\pm$  0.031 &  63.59 $\pm$   5.63 &  0.076 $\pm$  0.007 &          9  \\
\hline
\end{tabular}
\caption{\footnotesize
Point-Source Point-Lens Fit Parameters for the final event sample.\label{tab:tab02}}
\end{table}

\begin{table}
\begin{tabular}{c|cccc}
\hline\hline
Event Name & $t_0$\tablenotemark{a} & $t_E$\tablenotemark{a} & $u_0$\tablenotemark{a} & $\delta u_0/u_0 $ \\ 
&(HJD-2450000) & (days) & &(\%) \\
\hline
$*$OB99005 & 1275.168 $\pm$  0.009 &  72.24 $\pm$  15.48 &  0.022 $\pm$  0.005 &         21  \\
$*$OB99007 & 1316.100 $\pm$  0.048 &  36.87 $\pm$   1.05 &  0.492 $\pm$  0.020 &          4  \\
$*$OB99008 & 1287.546 $\pm$  0.154 &  41.94 $\pm$   6.06 &  0.042 $\pm$  0.011 &         26  \\
$*$OB99013 & 1318.005 $\pm$  0.053 &  19.43 $\pm$   1.09 &  0.614 $\pm$  0.052 &          8  \\
$*$OB99016 & 1334.421 $\pm$  0.615 &  44.02 $\pm$  11.91 &  0.351 $\pm$  0.142 &         40  \\
$*$OB99022 & 1323.514 $\pm$  0.066 &   7.68 $\pm$   1.23 &  0.297 $\pm$  0.067 &         22  \\
$*$OB99027 & 1365.833 $\pm$  0.280 &  50.59 $\pm$   7.82 &  0.265 $\pm$  0.060 &         22  \\
$*$OB99033 & 1434.789 $\pm$  0.099 &  58.67 $\pm$   2.18 &  0.316 $\pm$  0.016 &          5  \\
$*$OB99035 & 1392.552 $\pm$  0.001 &  48.97 $\pm$   3.32 &  0.008 $\pm$  0.001 &          6  \\
$*$OB99036 & 1392.730 $\pm$  0.004 &  29.84 $\pm$   0.55 &  0.066 $\pm$  0.001 &          2  \\
$*$OB99039 & 1436.605 $\pm$  0.395 & 219.90 $\pm$  55.16 &  0.074 $\pm$  0.021 &         28  \\
\hline
\end{tabular}
\tablenotetext{*}{Indicates those events for which MACHO and/or OGLE
data were included in the PSPL fit.}
\tablenotetext{a}{$t_0$=time of maximum magnification; $t_E$=Einstein
ring radius crossing time; $u_0$=minimum impact parameter.}
\\
\\
Table 2: Continued \\
\end{table}

\begin{table}
\begin{tabular}{c|cccc}
\hline\hline
Event Name & \# Points & ${\sigma_{\rm med}}$\tablenotemark{a} &
$\Delta t_{\rm med}$\tablenotemark{a} & $\Delta t_{\rm med}/t_E$\\ 
& & (\%)&(hrs)& \\ 
\hline
MB95013 & 266 & 0.7 &  1.21 & 6.23$\times 10^{-4}$ \\
MB95019 & 163 & 1.4 &  1.54 & 1.68$\times 10^{-3}$ \\
MB96011 & 40 & 2.5 &  3.62 & 1.43$\times 10^{-2}$ \\
$\dagger$MB96016 & 169 & 6.3 &  0.20 & 1.46$\times 10^{-4}$ \\
MB96018 & 21 & 4.4 &  2.34 & 1.38$\times 10^{-2}$ \\
MB96019 & 95 & 1.5 &  1.47 & 5.10$\times 10^{-3}$ \\
MB97018 & 257 & 2.6 &  2.08 & 8.66$\times 10^{-4}$ \\
MB97025 & 78 & 1.5 &  3.09 & 6.21$\times 10^{-3}$ \\
MB97026 & 556 & 1.1 &  0.71 & 4.34$\times 10^{-4}$ \\
$\dagger$MB97030 & 106 & 2.2 &  1.91 & 3.39$\times 10^{-3}$ \\
MB97031 & 328 & 0.8 &  1.08 & 1.09$\times 10^{-3}$ \\
EB98002 & 160 & 1.2 &  1.81 & 3.15$\times 10^{-3}$ \\
$\dagger$MB98013 & 80 & 5.0 &  1.93 & 4.42$\times 10^{-3}$ \\
MB98026 & 253 & 1.3 &  2.34 & 2.90$\times 10^{-3}$ \\
MB98030 & 82 & 2.2 &  4.86 & 7.51$\times 10^{-3}$ \\
MB98033 & 278 & 1.0 &  0.19 & 1.10$\times 10^{-3}$ \\
$\dagger$MB98035 & 356 & 4.6 &  1.02 & 1.55$\times 10^{-3}$ \\
OB98013 & 147 & 1.2 &  2.71 & 2.04$\times 10^{-3}$ \\
$\dagger$OB98014 & 619 & 1.9 &  1.02 & 1.02$\times 10^{-3}$ \\
$\dagger$OB98015 & 121 & 7.0 &  1.19 & 9.49$\times 10^{-4}$ \\
OB98018 & 404 & 1.4 &  0.25 & 1.39$\times 10^{-3}$ \\
OB98021 & 115 & 6.3 &  1.44 & 2.25$\times 10^{-3}$ \\
OB98023 & 128 & 1.7 &  2.35 & 5.25$\times 10^{-3}$ \\
OB98025 & 148 & 3.9 &  1.99 & 1.64$\times 10^{-3}$ \\
$\dagger$OB98030 & 65 & 9.9 &  2.31 & 1.77$\times 10^{-3}$ \\
EB99001 & 333 & 0.8 &  0.84 & 1.74$\times 10^{-3}$ \\
MB99006 & 38 & 0.9 &  1.15 & 1.76$\times 10^{-3}$ \\
MB99011 & 118 & 2.9 &  0.16 & 1.43$\times 10^{-4}$ \\
MB99018 & 407 & 0.8 &  0.28 & 5.29$\times 10^{-4}$ \\
MB99024 & 74 & 4.3 &  7.59 & 5.30$\times 10^{-3}$ \\
MB99034 & 88 & 1.7 &  1.70 & 1.01$\times 10^{-2}$ \\
$\dagger$MB99037 & 301 & 2.7 &  0.89 & 5.84$\times 10^{-4}$ \\
\hline
\end{tabular}
\caption{Data Characteristics for the Final Event Sample.\label{tab:tab03}}
\end{table}

\begin{table}
\begin{tabular}{c|cccc}
\hline\hline
Event Name & \# Points & ${\sigma_{\rm med}}$\tablenotemark{a} &
$\Delta t_{\rm med}$\tablenotemark{a} & $\Delta t_{\rm med}/t_E$\\ 
& & (\%)&(hrs)& \\ 
\hline
$\dagger$OB99005 & 229 & 2.6 &  0.27 & 1.55$\times 10^{-4}$ \\
OB99007 & 388 & 1.4 &  2.35 & 2.66$\times 10^{-3}$ \\
$\dagger$OB99008 & 31 & 9.1 &  3.50 & 3.48$\times 10^{-3}$ \\
OB99013 & 256 & 1.6 &  2.53 & 5.43$\times 10^{-3}$ \\
OB99016 & 75 & 2.9 &  1.43 & 1.35$\times 10^{-3}$ \\
OB99022 & 59 & 5.7 &  1.53 & 8.28$\times 10^{-3}$ \\
OB99027 & 94 & 3.4 &  2.58 & 2.13$\times 10^{-3}$ \\
OB99033 & 162 & 2.3 &  1.81 & 1.29$\times 10^{-3}$ \\
$\dagger$OB99035 & 316 & 3.4 &  1.46 & 1.24$\times 10^{-3}$ \\
$\dagger$OB99036 & 501 & 2.1 &  1.05 & 1.47$\times 10^{-3}$ \\
$\dagger$OB99039 & 77 & 3.9 & 23.89 & 4.53$\times 10^{-3}$ \\
\hline
\end{tabular}
\tablenotetext{\dagger}{Indicates high-magnification ($u_0\le 0.1; A_{\rm
max}\ge 10$) events.}
\tablenotetext{a}{${\sigma_{\rm med}}$: the median photometric error;
$\Delta t_{\rm med}$: the median sampling interval.}
\\
\\
Table 3: Continued \\
\end{table}

\begin{table}
\begin{tabular}{c|cccc}
\hline\hline
Event Name & $(V-I)_0$ & $I_0$ & $\theta_*$\tablenotemark{a} & $\rho_*$\tablenotemark{a} \\ 
& & &$(\mu{\rm as})$& \\
\hline
MB95013 & 1.137$\pm$0.014 & 13.88$\pm$ 0.05 &  8.71 & 1.61$\times 10^{-2}$ \\
MB95019 & 0.681$\pm$0.014 & 17.04$\pm$ 0.22 &  1.38 & 5.40$\times 10^{-3}$ \\
MB96011\tablenotemark{b}& 1.014       & 15.86$\pm$ 0.27 &  3.21 & 4.54$\times 10^{-2}$ \\
MB96016\tablenotemark{b}& 1.060       & 15.35$\pm$ 0.70 &  4.20 & 1.10$\times 10^{-2}$ \\
MB96018\tablenotemark{c}& --          & --              &  6.00 & 1.28$\times 10^{-1}$ \\
MB96019\tablenotemark{b}& 1.297       & 14.42$\pm$ 0.32 &  7.27 & 9.04$\times 10^{-2}$ \\
MB97018 & 0.963$\pm$0.075 & 16.11$\pm$ 0.61 &  2.74 & 4.09$\times 10^{-3}$ \\
MB97025 & 1.085$\pm$0.028 & 15.94$\pm$ 0.76 &  3.27 & 2.36$\times 10^{-2}$ \\
MB97026 & 1.352$\pm$0.008 & 15.12$\pm$ 0.05 &  5.36 & 1.18$\times 10^{-2}$ \\
MB97030 & 1.101$\pm$0.057 & 17.83$\pm$ 0.15 &  1.38 & 8.79$\times 10^{-3}$ \\
MB97031 & 1.343$\pm$0.010 & 12.69$\pm$ 0.39 & 16.38 & 5.97$\times 10^{-2}$ \\
EB98002 & 1.078$\pm$0.008 & 15.51$\pm$ 0.09 &  3.95 & 2.47$\times 10^{-2}$ \\
MB98013 & 0.809$\pm$0.013 & 17.35$\pm$ 0.24 &  1.31 & 1.07$\times 10^{-2}$ \\
MB98026 & 1.313$\pm$0.006 & 14.47$\pm$ 0.05 &  7.15 & 3.17$\times 10^{-2}$ \\
MB98030 & 1.103$\pm$0.029 & 17.41$\pm$ 0.51 &  1.68 & 9.31$\times 10^{-3}$ \\
MB98033 & 1.074$\pm$0.002 & 15.09$\pm$ 0.03 &  4.78 & 9.76$\times 10^{-2}$ \\
MB98035 & 1.022$\pm$0.002 & 16.31$\pm$ 0.05 &  2.62 & 1.43$\times 10^{-2}$ \\
OB98013 & 0.936$\pm$0.012 & 15.77$\pm$ 0.10 &  3.12 & 8.43$\times 10^{-3}$ \\
OB98014 & 1.092$\pm$0.004 & 14.80$\pm$ 0.02 &  5.55 & 2.00$\times 10^{-2}$ \\
OB98015 & 0.911$\pm$0.057 & 18.77$\pm$ 0.24 &  0.76 & 2.18$\times 10^{-3}$ \\
OB98018 & 1.120$\pm$0.030 & 14.31$\pm$ 0.04 &  7.07 & 1.38$\times 10^{-1}$ \\
OB98021 & 1.145$\pm$0.026 & 14.35$\pm$ 0.49 &  7.07 & 3.97$\times 10^{-2}$ \\
OB98023 & 1.323$\pm$0.014 & 14.60$\pm$ 0.33 &  6.75 & 5.43$\times 10^{-2}$ \\
OB98025 & 0.791$\pm$0.208 & 16.02$\pm$ 0.30 &  2.39 & 7.09$\times 10^{-3}$ \\
OB98030\tablenotemark{c}& --          & --              &  6.00 & 1.65$\times 10^{-2}$ \\
EB99001 & 1.380$\pm$0.005 & 13.69$\pm$ 0.14 & 10.50 & 7.78$\times 10^{-2}$ \\
MB99006\tablenotemark{c}& --          & --              &  6.00 & 3.30$\times 10^{-2}$ \\
MB99011 & 0.961$\pm$0.020 & 16.54$\pm$ 0.09 &  2.24 & 7.43$\times 10^{-3}$ \\
MB99018 & 1.320$\pm$0.006 & 13.37$\pm$ 0.05 & 11.86 & 8.18$\times 10^{-2}$ \\
MB99024 & 0.653$\pm$0.028 & 17.55$\pm$ 0.18 &  1.07 & 2.68$\times 10^{-3}$ \\
MB99034 & 0.906$\pm$0.020 & 16.31$\pm$ 0.23 &  2.34 & 4.98$\times 10^{-2}$ \\
MB99037 & 0.831$\pm$0.010 & 18.27$\pm$ 0.11 &  0.88 & 2.06$\times 10^{-3}$ \\
\hline
\end{tabular}
\caption{Source Characteristics for the Final Event Sample.\label{tab:tab04}}
\end{table}

\begin{table}
\begin{tabular}{c|cccc}
\hline\hline
Event Name & $(V-I)_0$ & $I_0$ & $\theta_*$\tablenotemark{a} & $\rho_*$\tablenotemark{a} \\ 
& & &$(\mu{\rm as})$& \\
\hline
OB99005 & 0.699$\pm$0.006 & 17.99$\pm$ 0.24 &  0.91 & 1.87$\times 10^{-3}$ \\
OB99007 & 1.100$\pm$0.008 & 14.91$\pm$ 0.07 &  5.29 & 2.15$\times 10^{-2}$ \\
OB99008\tablenotemark{b}& 0.895& 18.14$\pm$ 0.19 &  1.00 & 3.56$\times 10^{-3}$ \\
OB99013 & 1.112$\pm$0.012 & 14.31$\pm$ 0.15 &  7.04 & 5.42$\times 10^{-2}$ \\
OB99016\tablenotemark{b}& 1.012& 15.90$\pm$ 0.73 &  3.14 & 1.07$\times 10^{-2}$ \\
OB99022 & 1.021$\pm$0.050 & 16.27$\pm$ 0.32 &  2.66 & 5.19$\times 10^{-2}$ \\
OB99027\tablenotemark{b}& 0.890& 17.26$\pm$ 0.36 &  1.49 & 4.40$\times 10^{-3}$ \\
OB99033 & 0.987$\pm$0.021 & 15.24$\pm$ 0.07 &  4.17 & 1.06$\times 10^{-2}$ \\
OB99035\tablenotemark{c}& -- &  -- &  6.00 & 7.80$\times 10^{-3}$ \\
OB99036 & 0.938$\pm$0.005 & 16.21$\pm$ 0.02 &  2.55 & 1.28$\times 10^{-2}$ \\
OB99039 & 0.870$\pm$0.412 & 19.45$\pm$ 0.32 &  0.53 & 3.62$\times 10^{-4}$ \\
\hline
\end{tabular}
\tablenotetext{a}{$\theta_*$= angular size of the source;
$\rho_*$=estimated angular size of the source in units of the angular Einstein
ring radius of the lens. See \S7.1.}

\tablenotetext{b}{Insufficient $V$-band data to determine the color of
the source; the source is assumed to have the typical $(V-I)$ for its magnitude.}
\tablenotetext{c}{No CMD available, or CMD inconclusive.  The source
is assumed to be a clump giant.}
Table 4: Continued \\
\end{table}

\newpage

\clearpage

\clearpage

\clearpage

\clearpage

\clearpage

\clearpage

\clearpage

\clearpage

\clearpage

\clearpage

\clearpage

\clearpage

\clearpage

\clearpage

\end{document}